\documentclass[12pt, preprint]{aastex}

\begin{document}

\title{A Catalog of Broad Absorption Line Quasars in Sloan Digital Sky Survey Data Release 5}

\author{Robert R. Gibson\altaffilmark{1}, Linhua Jiang\altaffilmark{2}, W.~N. Brandt\altaffilmark{1}, Patrick B. Hall\altaffilmark{3}, Yue Shen\altaffilmark{4}, Jianfeng Wu\altaffilmark{1}, Scott F. Anderson\altaffilmark{5}, Donald P. Schneider\altaffilmark{1}, Daniel Vanden Berk\altaffilmark{1}, S.~C. Gallagher\altaffilmark{6,7}, Xiaohui Fan\altaffilmark{2}, and Donald G. York\altaffilmark{8}}
\email{rgibson@astro.psu.edu}

\altaffiltext{1}{Department of Astronomy and Astrophysics, Pennsylvania State University, 525 Davey Laboratory, University Park, PA 16802}
\altaffiltext{2}{Steward Observatory, The University of Arizona, 933 North Cherry Avenue, Tucson, AZ 85721--0065}
\altaffiltext{3}{Department of Physics and Astronomy, York University, 4700 Keele St., Toronto, ON, M3J 1P3, Canada}
\altaffiltext{4}{Princeton University Observatory, Princeton, NJ 08544}
\altaffiltext{5}{Astronomy Department, University of Washington, Seattle, WA 98195}
\altaffiltext{6}{Department of Physics and Astronomy, The University of Western Ontario, 1151 Richmond Street, London, ON N6A 3K7, Canada}
\altaffiltext{7}{Department of Physics and Astronomy, University of California -- Los Angeles, 430 Portola Plaza, Los Angeles, CA 90095--1547}
\altaffiltext{8}{Department of Astronomy and  Astrophysics and The Enrico Fermi Institute, The University of  Chicago, 5640 So. Ellis Ave., Chicago, IL 60637}

\shorttitle{SDSS DR5 BAL Catalog}
\shortauthors{Gibson et al.}



\begin{abstract}
We present a catalog of 5039 broad absorption line (BAL) quasars (QSOs) in the Sloan Digital Sky Survey (SDSS) Data Release 5 (DR5) QSO catalog that have absorption troughs covering a continuous velocity range $\ge 2000$~km~s$^{-1}$.  We have fit ultraviolet (UV) continua and line emission in each case, enabling us to report common diagnostics of BAL strengths and velocities in the range $-25,000$ to $0$~km~s$^{-1}$ for \ion{Si}{4}~$\lambda$1400, \ion{C}{4}~$\lambda$1549, \ion{Al}{3}~$\lambda$1857, and \ion{Mg}{2}~$\lambda$2799.  We calculate these diagnostics using the spectrum listed in the DR5 QSO catalog, and also for spectra from additional SDSS observing epochs when available.  In cases where BAL QSOs have been observed with {\it Chandra} or {\it XMM-Newton}, we report the \mbox{X-ray} monochromatic luminosities of these sources.

We confirm and extend previous findings that BAL QSOs are more strongly reddened in the rest-frame UV than non-BAL QSOs and that BAL QSOs are relatively \mbox{X-ray} weak compared to non-BAL QSOs.  The observed BAL fraction is dependent on the spectral signal-to-noise (S/N); for higher-S/N sources, we find an observed BAL fraction of $\approx$15\%.  BAL QSOs show a similar Baldwin effect as for non-BAL QSOs, in that their \ion{C}{4} emission equivalent widths decrease with increasing continuum luminosity.  However, BAL QSOs have weaker \ion{C}{4} emission in general than do non-BAL QSOs.  Sources with higher UV luminosities are more likely to have higher-velocity outflows, and the BAL outflow velocity and UV absorption strength are correlated with relative \mbox{X-ray} weakness.  These results are in qualitative agreement with models that depend on strong \mbox{X-ray} absorption to shield the outflow from over-ionization and enable radiative acceleration.  In a scenario in which BAL trough shapes are primarily determined by outflow geometry, observed differences in \ion{Si}{4} and \ion{C}{4} trough shapes would suggest that some outflows have ion-dependent structure.
\end{abstract}


{Accepted to ApJ. (c) Copyright 2008.  The American Astronomical Society.  All rights reserved.  Printed in U.S.A.}

\section{INTRODUCTION\label{introSec}}

The study of broad absorption lines (BALs) commonly observed in the ultraviolet (UV) spectra of optically-selected quasars \citep[QSOs; e.g.,][and references therein]{thrrsvkafbkn06} significantly advances our understanding of the structure and emission/absorption physics of Active Galactic Nuclei (AGN).  BALs are at least 2000~km~s$^{-1}$ wide by definition \citep{wmfh91}, and can be accelerated up to and beyond velocities of $0.1c$.  The depth of absorption at a given outflow velocity is largely determined by the fraction of the UV emission region covered by the BAL outflow \citep[e.g.,][and see also Appendix A of \citet{h+02}]{ablgwbd99}, and is therefore an indicator of the outflow structure.  In some cases, BAL troughs have been observed to change dramatically over several (rest-frame) years, indicating that the structure of the BAL outflow evolves on observable time scales \citep[e.g.,][]{gbsg08}.  It is thought that an \mbox{X-ray} absorber shields the BAL outflow from overionization by the soft \mbox{X-rays} generated in the heart of the QSO so that UV radiation can accelerate the outflow to high velocities \citep[e.g.,][]{mcgv95}.  The shielding \mbox{X-ray} absorber may also influence other BAL properties, such as the maximum outflow velocity \citep[e.g.,][]{gbcpgs06}.

BAL characteristics such as outflow velocities, velocity widths, and absorption depths vary widely among individual sources; large samples are therefore required in order to obtain a general understanding of QSO physics.  Variability and multi-wavelength studies of BAL QSOs can be performed with dedicated observations, but are most economically conducted using publicly-archived, serendipitous observations of optically-identified BAL QSOs.  Such studies have become possible due to large QSO surveys with well-understood selection properties that have greatly increased the number of known, representative BAL QSOs.  \citet{hf03} identified a sample of 67 BAL QSOs in the Large Bright Quasar Survey \citep[LBQS;][and references therein]{hfc95}.  \citet{tkt02} and \citet{rrhsvfykb03} studied 116 and 224 BAL QSOs, respectively, in the Sloan Digital Sky Survey \citep[SDSS;][]{y+00} Early Data Release (EDR).  \citet{gbcssv07} have identified $\approx$600 QSOs with broad absorption in SDSS Data Relase 2 (DR2), while \citet{thrrsvkafbkn06} have recently identified 1986 traditional BAL QSOs in the SDSS Data Release 3 (DR3) QSO catalog \citep{s+05}.  As survey technology progresses (e.g., from the LBQS to the SDSS), fainter sources are catalogued, increasing our coverage of the QSO luminosity/redshift plane.  As a result, we are not only increasing sample sizes over time, but also increasing the physical parameter space for QSO studies.

In the current study, we identify and quantify the BAL absorption in sources in the SDSS DR5 QSO catalog \citep{s+07}.  Of the 77,429 DR5 QSOs in the catalog, we identify 4204 traditionally-defined BAL QSOs, 481 of which have BALs from low ionization stages of \ion{Al}{3} and/or \ion{Mg}{2}.  Our sample size increases to 5039 BAL QSOs if we extend the traditional definition to include BALs that have absorption at outflow velocities within 3000~km~s$^{-1}$ of the emission line redshift.  (The ``traditional'' and ``extended'' BAL definitions are formally detailed in \S\ref{obsDataSec}.)  In cases where multiple SDSS spectra are publicly available for a QSO, we identify BALs present in the duplicate spectra as well as in the primary spectrum listed in the DR5 QSO catalog.  We also determine the monochromatic \mbox{X-ray} luminosities of BAL QSOs that were observed with {\it Chandra} and/or {\it XMM-Newton} and test for any correlations between BAL properties and relative \mbox{X-ray} brightness.

Because conventions vary among studies, we briefly describe the terminology used in this catalog.  BAL QSOs are broadly classified by ionization into low-ionization BAL QSOs (``LoBALs'') and high-ionization BAL QSOs (``HiBALs'').  LoBALs are QSOs that have BALs from ions at lower ionization states such as \ion{Al}{3} or \ion{Mg}{2}.  They may also have BALs from higher ionization stages such as \ion{Si}{4} or \ion{C}{4}.  ``FeLoBALs,'' which show broad absorption from iron blueward of 2800~\AA, are included in the category of LoBALs for this work.  HiBALs have BALs {\it only} from high ionization stages.  While most BAL QSOs do not have broad \ion{Al}{3} or \ion{Mg}{2} absorption lines, the unambiguous classification of HiBALs requires spectra that extend to the \ion{Al}{3} and \ion{Mg}{2} regions in order to verify that no BALs are present for these ions.

We define velocities flowing outward (with respect to QSO emission rest frames) to be negative.  Positive velocities indicate features that are at {\it longer} wavelengths than the wavelength corresponding to (rest-frame) zero velocity.  However, we use the terms ``greater'' and ``smaller'' velocities to refer to the {\it magnitude} of the velocity, so that an outflow velocity of \mbox{--10,000~km~s$^{-1}$} is ``greater'' than a velocity of \mbox{--5000~km~s$^{-1}$}.  We adopt this convention to match common terminology.

Throughout this work we use a cosmology in which $H_0 = 70$~km~s$^{-1}$~Mpc$^{-1}$, $\Omega_M = 0.3,$ and $\Omega_{\Lambda} = 0.7$.


\section{OBSERVATIONS AND DATA REDUCTION\label{obsDataSec}}

This BAL catalog is constructed from SDSS Data Release 5 \citep{a-m+07}.  The SDSS uses a CCD camera \citep{gcrsbehiklpshswyhabbbhkhhpprdfsohsmbhslb98} on a dedicated 2.5~m telescope \citep{gunn+06} at Apache Point Observatory, New Mexico, to obtain images in five broad optical bands \citep[$ugriz$;][]{figdss96} over approximately 10,000~deg$^2$ of the high Galactic latitude sky.  The survey data-processing software measures the properties of each detected object in the imaging data in all five bands, and determines and applies both astrometric and photometric
calibrations \citep{stkrfiijughwthapmbchsy02, pmhkli03, ilsbayksfprgksrenkknstsmstl04}.  The SDSS quasar target selection algorithm is presented in \citet{rfnsvsybbfghikllrsssy02}.  See \citet{s+02} for a detailed description of many of the technical aspects of the SDSS.  The SDSS photometry is reported on the asinh magnitude scale \citep{lgs99}.

In order to account for the loss of flux due to ``fiber leak,'' we multiply each spectra by a constant that matches the photometric $g$, $r$, and $i$ PSF magnitudes to those synthesized from the spectra.  In cases where the synthetic magnitudes are lower (brighter) than the PSF magnitudes (e.g., because the source luminosity has varied), we leave the spectrum unchanged.  We also correct the spectra for Galactic extinction using the reddening curve of \citet{ccm89} with the near-UV extension of \citet{o94}.  We obtain $E(B-V)$ from the NASA Extragalactic Database (NED)\footnote{http://nedwww.ipac.caltech.edu/}, which uses the dust maps of \citet{sfd98}.  The redshifts for our QSOs are taken from the DR5 QSO catalog, with the exception of two sources, J$100424.88+122922.2$ and J$153029.05+553247.9$, which we consider to be misclassified.  We assign these sources redshifts of 4.66 and 1.73, respectively.

We do not rebin SDSS spectra except in a few special cases for which we note the rebinning method directly.  A ``spectral bin'' therefore refers to the standard wavelength binning of SDSS spectra, which is $\approx$69~km~s$^{-1}$ per bin in the observing frame.


We fit SDSS spectra using the algorithm of \citet{gbsg08}, which we summarize here.  For QSOs at $z \ge 1.7$, our continuum model is a power law reddened using the Small Magellanic Cloud reddening curve of \citet{p92}.  For QSOs at lower redshifts, we use a fourth- or sixth-degree polynomial; in our experience this non-physical model is able to reproduce well the complex continuum at longer wavelengths.  We initially fit regions that are generally free from strong absorption or emission features:  1250--1350, 1700--1800, 1950--2200, 2650--2710, 2950--3700, 3950--4050, 4140--4270, 4400--4800, 5100--6400, and $>$6900~\AA.  We then iteratively fit the continuum, ignoring at each step wavelength bins that deviate by $>$3$\sigma$ from the current fit in order to exclude strong absorption and emission features.  We fit Voigt profiles to the strongest emission lines expected in the spectrum:  \ion{Si}{4}~$\lambda$1400, \ion{C}{4}~$\lambda$1549, \ion{Al}{3}~$\lambda$1857, \ion{C}{3}]~$\lambda$1909, and \ion{Mg}{2}~$\lambda$2799.  These wavelengths are taken from the SDSS vacuum wavelength list used by the SDSS pipeline to determine emission-line redshifts.\footnote{See http://www.sdss.org/dr6/algorithms/linestable.html and http://www.sdss.org/dr6/algorithms/redshift\_type.html}  The wavelengths of the emission lines are allowed to vary in the fit.  We ascribe no physical significance to the Voigt profile; it simply enables us to model emission line cores and wings using a small number of parameters.  We fit emission lines iteratively as well, ignoring at each step bins that are absorbed by more than $2.5\sigma$ from the continuum $+$ emission fit.  Due to the degeneracy between the UV emission continuum shape and the magnitude of intrinsic reddening, we do not attach physical significance to the values of $E(B-V)$ obtained from our fits.  The UV luminosities we report are therefore corrected for Galactic, but not intrinsic, reddening.  After the initial automatic fits were performed, RRG inspected and manually adjusted the automatic fits.  Other reviewers (LJ, YS, WNB, JW, PBH, SFA, DVB, DPS) then examined this adjusted sample of spectra in order to ensure that BALs were not missed or mis-characterized.

We smoothed each spectrum using a boxcar three bins wide\footnote{Smoothing is performed to reduce scatter in the data that could obscure broad absorption features.  We use the same smoothing width as \citet{thrrsvkafbkn06} did for the DR3 BAL QSO catalog.  Although the smoothing width we use is much smaller than the minimum width of a BAL, we note that the BAL detection process is technically dependent on the smoothing width, as even a single bin can ``break up'' a potential BAL.} and then searched the smoothed spectra for BALs by calculating $BI$, the traditional ``balnicity index'' of \citet{wmfh91}.  The balnicity index is defined as
\begin{eqnarray}
BI &\equiv& \int^{25,000}_{3000} \biggl(1 - \frac{f(-v)}{0.9}\biggr) C~dv,\label{bIDefnEqn}
\end{eqnarray}
where $f(v)$ is the continuum-normalized spectral flux at a velocity $v$ (in km~s$^{-1}$) from the line rest wavelength (in the system frame).  The dimensionless value $C$ is $0$ unless the observed spectrum has fallen at least 10\% below the continuum for a velocity width of at least 2000~km~s$^{-1}$ on the red side of the absorption trough, at which point $C$ is set to $1$.  Traditional BALs are defined to have $BI > 0$.  Because this metric does not include a large number of broad absorption features at lower outflow velocities, we also define and calculate $BI_0$ in a similar fashion to $BI$, but integrate to $0$~km~s$^{-1}$ instead of $-3000$~km~s$^{-1}$.  We measure $BI$ and $BI_0$ for \ion{Si}{4}, \ion{C}{4}, \ion{Al}{3}, and \ion{Mg}{2}, using the wavelength for the red component of the doublets given in the theoretical line list of \citet{vfky95} to define zero velocity for each ion.  These wavelengths are 1402.77~\AA\ (\ion{Si}{4}), 1550.77~\AA\ (\ion{C}{4}), 1862.79~\AA\ (\ion{Al}{3}), and 2803.53~\AA\ (\ion{Mg}{2}).  In some cases, BAL troughs may be contaminated by absorption from other ions, such as \ion{Fe}{2}~$\lambda$2632 and \ion{C}{2}~$\lambda$1335.  Such contamination is difficult to identify except when strong absorption begins at the known wavelengths of these lines.  In general, we do not account for contamination, but ascribe all broad absorption to \ion{Si}{4}, \ion{C}{4}, \ion{Al}{3}, or \ion{Mg}{2}.  In \S\ref{catDataSec}, we list 92 sources in our BAL catalog that appear to have contaminated BALs, based on visual inspection.

\begin{figure} [ht]
  \begin{center}
      \includegraphics[width=4in, angle=270]{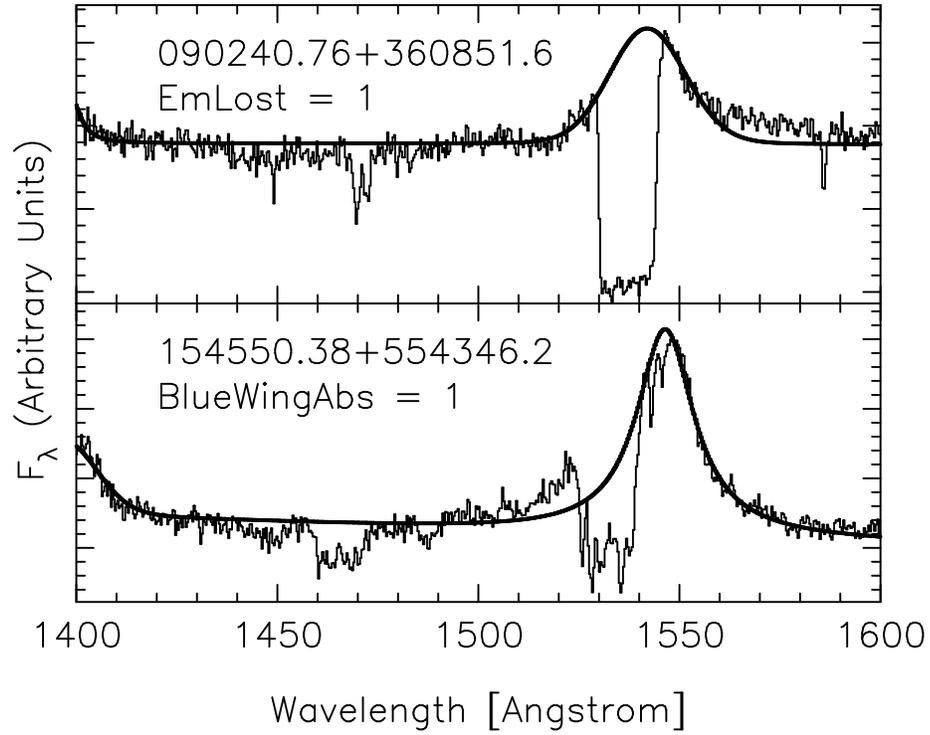}
      \caption{\label{bALCatFlagExamplesFig}Examples of \ion{C}{4} absorption regions for BAL QSOs that have the \ion{C}{4} {\tt EmLost} ({\it top}) and {\tt BlueWingAbs} ({\it bottom}) flags set.  The $x$-axis shows rest-frame wavelength, while the $y$-axis shows the (unsmoothed) flux density scaled arbitrarily.  Our model fit is plotted with a thick line.}
   \end{center}
\end{figure}

Overall, our models are able to accurately reproduce the observed continuum and emission line features in the SDSS QSO spectra.  However, there are cases where reproducing emission is problematic.  This could occur, for example, if an emission line profile is highly asymmetric.  There are also cases, such as strongly-absorbed emission line profiles, for which we were unable to accurately determine the underlying emission.  We define a set of flags that we use to indicate when the fit model is particularly uncertain due to such effects.
\begin{enumerate}
\item{{\tt EmLost}:  The profile of the emission line at (nearly) zero velocity has been strongly absorbed and may not have been accurately reconstructed.  The EmLost flag is set if a BAL is found for this ion and either:  the center of the (best-guess) emission line fit is in the BAL region, the FWHM of our (best-guess) emission line fit is $<$10~\AA, or there is evidence that a broad range of spectral bins in the line center are absorbed.  Quantitatively, for the last criterion, we require that at least 15 adjacent bins (20 for \ion{Mg}{2}) fall $\ge$5\% below our emission model fit, and these absorbed bins lie in the wavelength region where the emission is above the line half-maximum.  Our fitting experience indicates that characterizing emission lines is particularly difficult in cases meeting these criteria.  Low-velocity absorption is very common; nearly half of the sources with \ion{C}{4} $BI_0 > 0$ have this flag set for the \ion{C}{4} emission line.  Setting this flag indicates that measurements of BAL troughs are less certain in cases where the BAL extends to low velocities.}
\item{{\tt BlueWingAbs}:  Additional ``blue wing'' emission occurs on the blue side of the \ion{C}{4} emission line, and absorption is apparent between the blue wing and the line core.\footnote{For example, some asymmetry can be seen on the blue side of \ion{C}{4}~$\lambda$1549 in the SDSS QSO composite spectrum \citep{v+01}.  The strength of this emission varies widely among individual cases, making it difficult to model with our current methods.  Some of the emission may also be due to lines of other ions, such as \ion{Si}{2}~$\lambda$1527 \citep{vvf96}.}  Since we fit only the line core and cannot accurately reconstruct the ``blue wing'' emission, we could potentially miss BAL absorption in these cases.  The {\tt BlueWingAbs} flag is set to ``1'' when 15 consecutive bins redward of 1500~\AA\ lie above the model fit, and 15 more consecutive bins further redward lie below the model, and all these bins are shortward of the wavelength at which the blue side of the \ion{C}{4} emission line reaches its half-maximum.}
\item{{\tt BALManyBadBins}:  Some putative BALs consist of a large number of spectral bins that have been flagged as ``bad'' in the SDSS pipeline.\footnote{See, e.g., the masks reported at http://www.sdss.org/dr5/dm/flatFiles/spPlate.html\#andmask}  Unfortunately, the pipeline occasionally flags bins that are acceptable, and the low flux density and signal-to-noise (S/N) in BAL troughs make them especially susceptible to flagging.  We report the {\tt BALManyBadBins} flag when $>$25\% of the bins in a BAL have been flagged by the SDSS.  However, we ignore the flags {\tt BRIGHTSKY} and {\tt EMLINE}, because these warnings can be triggered by the low flux levels in strong BAL troughs.}
\end{enumerate}
Figure ~\ref{bALCatFlagExamplesFig} shows examples of spectra that have the {\tt EmLost} and {\tt BlueWingAbs} flags set for the \ion{C}{4} absorption region.

The 2200--3000~\AA\ region surrounding the \ion{Mg}{2} emission line is especially difficult to model due to poorly-constrained emission line structure and blended emission from ions such as \ion{Fe}{2}.  Modeling the full emission structure in this region is beyond the scope of this study.  For \ion{Mg}{2} absorption, we have fit the emission line core and estimated the continuum model from the fit to the surrounding region.  We therefore do not claim that our \ion{Mg}{2} absorption list is complete, although visual inspections indicate that we have not missed a large number of strong, broad absorption features.

In some cases, we report BAL absorption detected in the observed spectrum even though the entire outflow velocity range (defined to be $-25,000$ to $0$~km~s$^{-1}$) does not fall in the SDSS spectral bandpass.  As a result, the reported BAL properties may not accurately represent the measurements that would be obtained if the full velocity range were observed.  If a representative sample of BAL properties is required, redshift restrictions should be placed on QSO samples to ensure that the entire potential absorption region falls in the SDSS bandpass.  We discuss this point further in \S\ref{analysisSec}.

Small, systematic misestimates of the true continuum level could cause us to miss or mistakenly identify broad, shallow BALs.  This could occur, for example, if the true emission continuum is more structured than our continuum models.  While detailed studies of BAL morphology are beyond the scope of the current work, we did implement a statistic to search the catalog for broad ($>$50~\AA\ wide in the rest frame), shallow absorption features with relatively little absorption structure and found that such features are rare.  (Only 7 of 4664 \ion{C}{4} BALs matched our criterion.)  Nonetheless, users of this catalog should be aware of such limitations in BAL identification.

We have identified several DR5 QSOs that have spectra that are strongly contaminated by superposed stellar spectra.  The stellar features can trigger false BAL identifications because they do not match the typical QSO spectrum.  We have omitted these sources (J$012412.46-010049.8$, J$014349.15+002128.3$, J$024804.70+002357.0$, and J$073405.24+320315.2$) from the catalog.  Additionally, we are aware of one source, J$114240.94+601550.5$, that would be classified as a LoBAL QSO based on the spectra provided in SDSS DR3 and DR5.  However, DR6 calibration has adjusted this spectrum and removed the BAL feature.  We have omitted this source from the catalog and caution that, while such instances are rare, users should be aware of the latest calibration data for individual objects.

\section{CATALOG DATA\label{catDataSec}}

\begin{figure} [ht]
  \begin{center}
      \includegraphics[width=4in, angle=270]{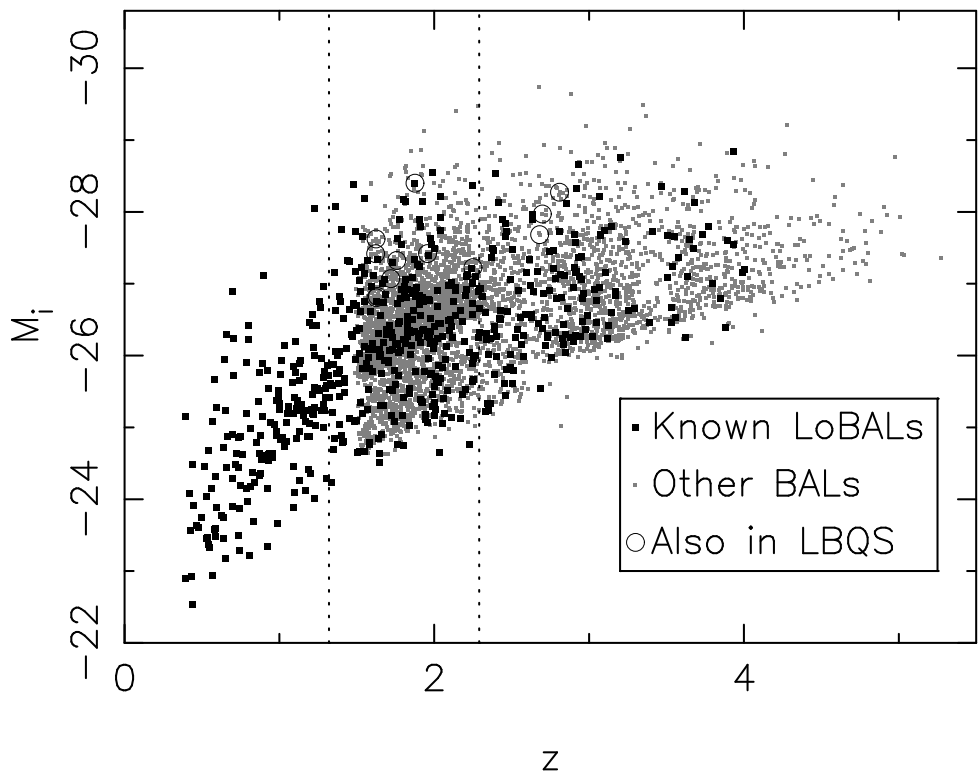}
      \caption{\label{bALCatLzPlaneFig}Large, dark squares show the absolute $i$-magnitude and redshift of the LoBALs (with \ion{Al}{3} or \ion{Mg}{2} $BI_0 > 0$) in this catalog.  Smaller, grey squares show the same for HiBALs or sources that did not have low-ionization regions in the SDSS bandpass and could therefore not be classified as HiBALs or LoBALs with certainty (see \S\ref{introSec}).  For the region between the vertical dotted lines, the full \ion{Al}{3} and \ion{Mg}{2} BAL absorption regions are in the SDSS bandpass.  BAL QSOs which are also in the LBQS are circled.}
   \end{center}
\end{figure}

The redshift and absolute $i$-magnitude (obtained from the DR5 catalog) are plotted for our BAL sources in Figure~\ref{bALCatLzPlaneFig}.  In Table~\ref{bALCatTab}, we list the UV spectral properties for all QSOs in the DR5 catalog that show \ion{Si}{4}, \ion{C}{4}, \ion{Al}{3}, or \ion{Mg}{2} BALs identified by the procedures described in \S\ref{obsDataSec}.  Table~\ref{bALCatTab} contains 5039 QSOs in all, with $BI_0 > 0$ for 1083 BALs attributed to \ion{Si}{4}, 4664 BALs attributed to \ion{C}{4}, 391 BALs attributed to \ion{Al}{3}, and 332 BALs attributed to \ion{Mg}{2}.  For each source with $BI_0 > 0$ for at least one of these ions, we list the name, right ascension (RA), declination (using J2000 coordinates), and redshift given in the DR5 catalog.  For each ion (\ion{Si}{4}, \ion{C}{4}, \ion{Al}{3}, and \ion{Mg}{2}), we list the measurements for $BI$, $BI_0$, $EW_0$, $v_{min}$, $v_{max}$, and $f_{deep}$ (see below).  Although it is, in principle, possible to calculate formal errors for $BI$ and $EW$ measurements, we do not report formal errors in the tables.  The true error in these measurements is dominated by the continuum placement and the absorption trough shape, which in turn determines which spectral bins are included in BAL troughs.  We therefore do not report formal errors, because they are an inaccurate measure of the complex factors involved.

The quantity $EW_0$ is the rest-frame absorption equivalent width measured in BAL regions, i.e., in spectral bins which are part of a continuous absorption region $\ge$10\% below the continuum and $>$2000~km~s$^{-1}$ wide.  Narrower absorption troughs are not included in this measurement.  We define $v_{max}$ and $v_{min}$ to be the outflow velocities corresponding to the shortest and longest wavelengths, respectively, in the BAL trough(s) of a given spectrum, where a BAL trough is any region for which the flux drops $\ge$10\% below the continuum for a velocity range of $>$2000~km~s$^{-1}$.  For example, if a given spectrum has two \ion{C}{4} BALs extending from --20,000 to --15,000~km~s$^{-1}$ and from --8000 to --5000~km~s$^{-1}$, then for this source the \ion{C}{4} $v_{max} = -20,000$ and $v_{min} = -5000$~km~s$^{-1}$.  The quantity $f_{deep}$ is defined to be the fraction of BAL bins that fall at least 50\% below the continuum.  (The definition of ``spectral bin'' is given in \S\ref{obsDataSec}.)

We also report, for each ion, the {\tt EmLost} and {\tt BALManyBadBins} flags.  For \ion{C}{4} we also report the {\tt BlueWingAbs} flag.  Finally, we report measured values of the monochromatic luminosities $L_{\nu}(1400~\mathring{\rm A})$ and $L_{\nu}(2500~\mathring{\rm A})$ (corrected for Galactic reddening), as well as $SN_{1700}$, which we use to estimate the S/N associated with the measurement of \ion{C}{4} absorption.  $SN_{1700}$ is defined as the median of the flux divided by the noise (reported by the SDSS pipeline) for all spectral bins in the 1650--1750~\AA\ region, using the intrinsic binning provided in the SDSS spectra.  This wavelength region was selected to be (typically) free of strong absorption, but still near the \ion{C}{4} BAL region.  $SN_{1700}$ depends on many factors, including the integration time and the source redshift and luminosity; we use it simply to construct samples of QSOs that are less influenced by selection effects associated with low S/N.

We have used the same procedures to analyze additional spectra of DR5 QSOs that are available in the DR6 public database.  These spectra provide one or more additional epochs of SDSS spectroscopy that can be used to study the variability of BALs on (rest-frame) time scales up to $\approx$1~yr.  As previous studies of samples of BAL QSOs have shown, BALs commonly vary significantly on such time scales \citep[e.g.][]{b93, lwbhsyvb07, gbsg08}.  However, we caution that $BI$ measures can change dramatically due to small changes in a few spectral bins and, for this reason, $BI$ and $BI_0$ alone are not necessarily good metrics of variability.  A full variability study should account for these issues as well as differences in S/N and calibration issues between epochs; such a study is beyond the scope of this work.  In Table~\ref{bALCatDupTab}, we report the UV spectral properties measured for each available spectrum of any QSO that has at least one BAL with $BI_0 > 0$ for any epoch.  Additionally, we report the MJD, plate, and fiber numbers used to uniquely identify each spectrum.  The MJD field is the modified Julian date of each SDSS observation.

In Table~\ref{nonBALCIVBWATab}, we list 118 DR5 QSOs that have the \ion{C}{4} {\tt BlueWingAbs} flag set, but are not considered to have \ion{C}{4} BALs in this catalog.  Because these QSOs have emission features that are not well-matched by a single, symmetric Voigt profile, the absorption in these cases is more difficult to accurately determine.  Small (and somewhat subjective) changes to the continuum placement for these QSOs can affect whether or not we measure $BI_0 > 0$ in some cases.

During the visual inspection process, we have identified BAL troughs that are likely associated with ions other than \ion{Si}{4}, \ion{C}{4}, \ion{Al}{3}, or \ion{Mg}{2}.  In particular, \ion{Si}{4} BALs may be contaminated by \ion{C}{2}~$\lambda$1335~\AA, \ion{Al}{3} BALs by \ion{Si}{2}~$\lambda$1808~\AA, and \ion{Mg}{2} BALs by \ion{Fe}{2} absorption at rest wavelengths such as $2750$, $2632$, and $2414$~\AA.  In Table~\ref{contamBALTab}, we list 92 BAL QSOs in our catalog that show visual evidence of probable contamination.  In the table, we identify the ion for which BAL contamination is likely.

We have searched the {\it Chandra} and {\it XMM-Newton} archives for targeted or serendipitous observations of BAL QSOs in this catalog.  The data analysis for the \mbox{X-ray} observations is described in \S\ref{xRayBALQSOSec}.  We report in Table~\ref{xInfoTab} the relevant \mbox{X-ray} properties for the 73 radio-quiet sources at $z < 2.68$ described in that section.  The redshift cut ensures that the 2500~\AA\ continuum flux can be accurately determined.

\section{ANALYSIS AND DISCUSSION\label{analysisSec}}

In the following sections, we briefly analyze the absorption and emission properties of BAL QSOs and discuss some physical implications of these properties.  Because BALs can appear at a range of velocities, we consider only QSOs at redshifts sufficiently high that the entire velocity range (defined to be --25,000 to 0~km~s$^{-1}$) is visible in the SDSS spectra.  For the various ions we consider, these redshift ranges are: $1.96 \le z \le 5.55$ (\ion{Si}{4}), $ 1.68 \le z \le 4.93$ (\ion{C}{4}), $1.23 \le z \le 3.93$ (\ion{Al}{3}), and $0.48 \le z \le 2.28$ (\ion{Mg}{2}).  Catalog users who wish to conduct statistical studies of BAL properties should, in most cases, apply similar redshift criteria when selecting their samples to ensure that additional BAL absorption is not missed because it falls outside the SDSS spectral bandpass.  Similar bandpass constraints are necessary when assembling a sample of QSOs that are required to be free of BAL features.

\subsection{BAL QSO Reddening\label{bALQSOReddeningSec}}

\begin{figure} [ht]
  \begin{center}
      \includegraphics[width=4in, angle=270]{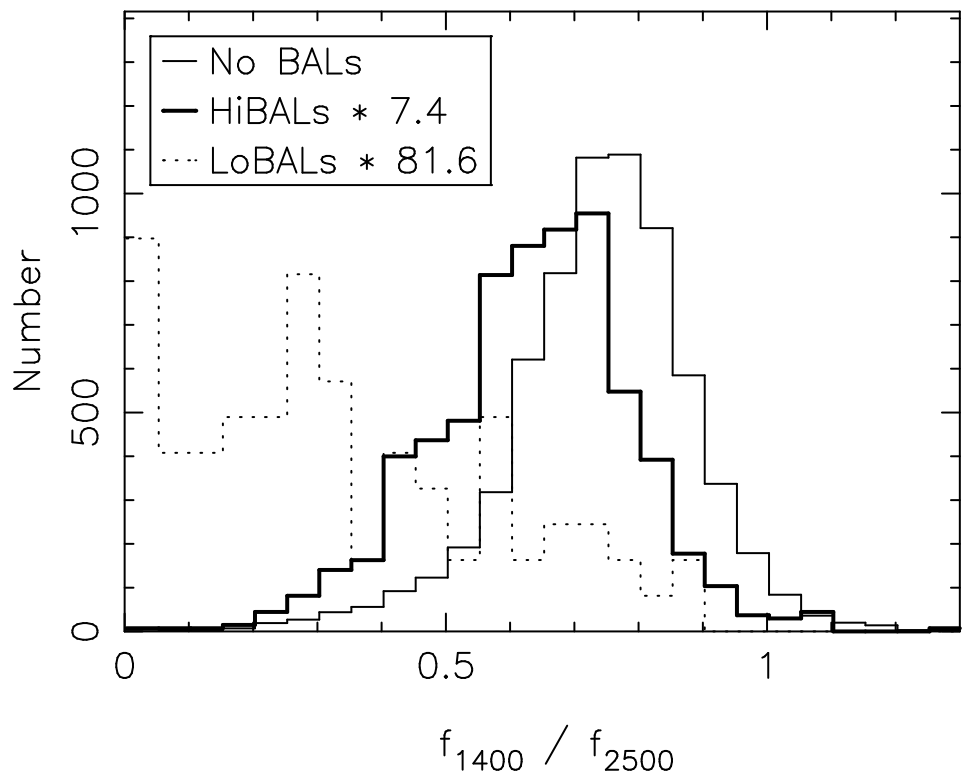}
      \caption{\label{bALFluxReddeningFig}The distribution of the ratio $R_{red} \equiv F_{\nu}(1400~\mathring{\rm A}) / F_{\nu}(2500~\mathring{\rm A})$ for QSOs with no BALs (thin solid line), for HiBAL QSOs (thick solid line), and for LoBAL QSOs (thick dotted line).  The distributions for the BAL QSOs have been renormalized to match the area under the non-BAL QSO distribution for easy comparison.}
   \end{center}
\end{figure}

Previous studies have found that BAL QSOs are redder than non-BAL QSOs, with LoBALs even redder than HiBALs \citep[e.g.,][]{wmfh91, btbglw01, rrhsvfykb03, thrrsvkafbkn06}.  In order to investigate reddening in DR5 BAL QSOs, we measure the ratio of the continuum flux densities (in ergs~s$^{-1}$~cm$^{-2}$~Hz$^{-1}$) at 1400~\AA\ and 2500~\AA.  (The continuum flux densities are calculated from the continuum fit alone with emission lines excluded from the model.  The continuum-fitting procedures described in \S\ref{obsDataSec} should make continuum placement relatively independent of line emission and/or absorption.)  This ratio, $R_{red} \equiv F_{\nu}(1400~\mathring{\rm A}) / F_{\nu}(2500~\mathring{\rm A})$, is plotted in Figure~\ref{bALFluxReddeningFig} for HiBAL, LoBAL, and non-BAL QSOs at $1.96 \le z \le 2.28$.  The distribution of $R_{red}$ for HiBALs is similar to that of non-BALs, but shifted to lower $R_{red}$.  The distribution for LoBALs is much broader and shifted to even lower $R_{red}$.

Lower $R_{red}$ values could be due to increased dust reddening, to a different shape for the underlying continuum emission, or to a combination of both factors.  The median value of $R_{red}$ is 0.75 for non-BAL QSOs, 0.64 for HiBAL QSOs, and 0.30 for LoBAL QSOs.  If we assume that all QSOs have similar emission continuum slopes and the differences are attributed completely to SMC-like dust reddening, HiBAL QSOs have a typical $E(B-V) \approx 0.023$ and LoBALs have $E(B-V) \approx 0.14$ compared to non-BAL QSOs.  The $E(B-V)$ value for HiBALs agrees with that found by \citet{rrhsvfykb03}, while the LoBAL reddening is somewhat stronger in our sample than the value of 0.077 reported in that study.  The difference between HiBAL and LoBAL reddening is $\Delta E(B-V) \approx 0.117$, similar to the value $E(B-V) \approx 0.1$ found by \citet{sf92}.

\subsection{Fraction of BAL QSOs in SDSS\label{bALFracSec}}

The {\it observed} fraction of BAL QSOs in optical surveys differs from the {\it intrinsic} fraction due to selection effects \citep[e.g.,][]{hfc95, kv98, rrhsvfykb03, dss08, ksgc08}.  Because BAL QSOs are redder than non-BAL QSOs (\S\ref{bALQSOReddeningSec}), they are rejected in magnitude-limited surveys at a higher rate than less-reddened QSOs of the same intrinsic luminosity.  Furthermore, color-selection criteria may incorrectly classify BAL QSOs due to the influence of absorption on BAL QSO colors.  BAL identification is also dependent on the S/N ratio of the spectrum, because bins in the middle of a broad absorption feature can randomly rise above the 90\% continuum level and prevent the formal identification of a BAL.  At lower S/N, this occurs more frequently.  False identifications can also occur when the S/N level is low, due to continuum mis-placement or flux levels randomly falling below the required threshhold for BAL identification.

\begin{figure} [ht]
  \begin{center}
      \includegraphics[width=4in, angle=270]{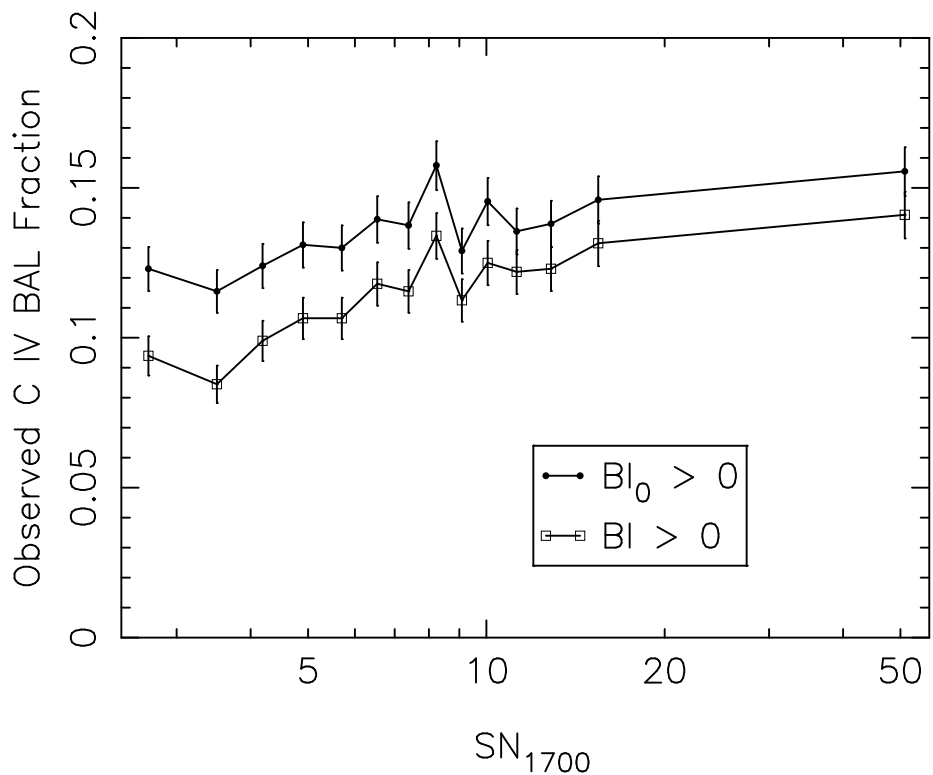}
      \caption{\label{bALFracsVsSNFig}The fraction of BALs observed in samples of 2000 QSOs as a function of $SN_{1700}$, the median S/N for spectral bins in the 1650--1700~\AA\ range.  Filled circles indicate \ion{C}{4} BALs with $BI_0 > 0$ and open squares indicate \ion{C}{4} BALs with $BI > 0$.  The error bars represent 1$\sigma$ confidence regions generated using binomial statistics.  Note that $SN_{1700}$ is not independent of the source luminosity (\S\ref{bALFracSec}).}
   \end{center}
\end{figure}

Figure~\ref{bALFracsVsSNFig} shows the fraction of QSOs with \ion{C}{4} BALs as a function of $SN_{1700}$.  Each point in the plot represents the fraction of BAL QSOs in a bin of 2000 QSOs arranged in order of increasing $SN_{1700}$.  The $x$ coordinate of each point represents the center of the bin.  The ``error bars'' represent 1$\sigma$ limits on the fractions calculated using a binomial distribution.  We show the observed fractions of cases with $BI > 0$ and also for the larger class with $BI_0 > 0$.  The observed fraction of cases with $BI_0 > 0$ rises with $SN_{1700}$ to $\approx$15\% at higher S/N levels.  The difference in BAL fractions may not be attributable to S/N alone, as $SN_{1700}$ is correlated with the observed flux and also the QSO luminosity.

At redshifts $z < 2.3$, the SDSS color selection criteria are relatively unaffected by BAL absorption, and the typical level of reddening for BAL QSOs causes them to be $\sim$0.15 to 0.2~mag fainter than non-BAL QSOs \citep{rrhsvfykb03, ksgc08}.  In order to estimate the intrinsic fraction of BAL QSOs, we have identified all optically-selected QSOs with $SN_{1700} \ge 9$, $i \le 18.5$, and $1.68 \le z \le 2.3$.  We call this set of QSOs subsample $S_0$.  The S/N cut minimizes the tendency of BALs to go undetected in low-S/N spectra.  The $i$-magnitude cut is brighter than the SDSS limit to allow us to examine BAL QSOs that are fainter than the parent sample due to excess reddening.  The lower limit on redshift ensures that the full \ion{C}{4} absorption region is in the SDSS spectral bandpass, and the upper redshift limit represents the point at which SDSS selection effects begin to significantly bias the observed BAL fraction \citep{rrhsvfykb03}.  We have additionally required that the QSOs in subsample $S_0$ be selected for SDSS spectroscopy purely on optical grounds, in order to remove any bias due to specially targeting radio- or \mbox{X-ray}-bright QSOs.

Of the 3427 QSOs in $S_0$, 455 have \ion{C}{4} $BI > 0$ and 519 have $BI_0 > 0$.  This corresponds to BAL fractions of 13.3\% ($\pm 0.6$\%) and 15.1\% ($\pm 0.6$\%), respectively.  The stated errors are 1$\sigma$ ranges calculated from the binomial distribution, but do not include additional error due to inaccuracies in the estimates of BAL QSO selection effects in the SDSS.

These numbers do not account for the fact that BAL QSOs tend to be redder than non-BAL QSOs and therefore more likely to be excluded in a flux-limited sample.  If we make the assumption that BAL QSOs are drawn from a parent population $\Delta m~i$-magnitudes brighter, we can estimate the intrinsic BAL fraction by comparing the number of BAL QSOs with $18.7 \ge i \ge 17.7$ to non-BAL QSOs with $18.7-\Delta m \ge i \ge 17.7 - \Delta m$.  For $\Delta m = 0.16$ (corresponding to $E(B-V)=0.023$ at 2500~\AA), the intrinsic \ion{C}{4} BAL fractions are 16.4 $\pm 0.6$\% ($BI > 0$) and 18.5 $\pm 0.7$\% ($BI_0 > 0$), respectively.  However, because the number of DR5 QSOs per unit magnitude is a rapidly-increasing function of $i$-magnitude, these fractions are highly dependent on the idealized assumption that we can directly compare BAL QSOs to a (presumed) parent population $\Delta m~i$-magnitudes brighter.  A similar method was used by \citet{hf03} to obtain a fraction of 22 $\pm 4$\% for the LBQS BAL sample.  Our estimated BAL fraction agrees reasonably well with those from surveys of QSOs identified in other wavebands such as the radio \citep[18 $\pm 4$\%;][]{bwgbla00} and infrared \citep[20 $\pm 2$\%; see \S6 of][]{dss08}.

It has been previously suggested that more-luminous QSOs are more likely to show BALs in their spectra \citep{gbcssv07}.  In order to test for luminosity dependence of the intrinsic BAL fraction, we have divided subsample $S_0$ into two groups depending on whether the monochromatic luminosity at 2500~\AA\ is greater or less than the median value of $1.81\times 10^{31}$~erg~s$^{-1}$~Hz$^{-1}$.  We find 231 (264) of 1713 lower-luminosity QSOs have $BI > 0$ ($BI_0 > 0$), compared to 224 (255) of 1714 QSOs in the higher-luminosity group.  A Kolmogorov-Smirnov test finds no indication of differing distributions of $BI_0$ for the higher- and lower-luminosity subsamples.  We therefore do not find any significant dependence of the BAL fraction on luminosity in sources with $SN_{1700} \ge 9$.  However, we note that the luminosity range of our $S_0$ sample is rather small, with $10^{31} \la L_{\nu}(2500~\mathring{\rm A}) \la 6\times10^{31}$~erg~s$^{-1}$~Hz$^{-1}$ in most cases, and a larger luminosity baseline is needed to test this result more sensitively.  In particular, we note that \citet{blw00} reported (in their \S8.1) an observed BAL fraction of only 6 $\pm 2$\% for the 87 QSOs at $z < 0.5$ in the Bright Quasar Survey \citep[BQS;][]{sg83}.  These QSOs are generally less luminous than the SDSS QSOs in our sample, suggesting that the BAL fraction may be luminosity-dependent over luminosity ranges wider than that of our SDSS sample.

\subsection{Distribution of BI for \ion{Si}{4} and \ion{C}{4}\label{bIDistSec}}

\begin{figure} [ht]
  \begin{center}
      \includegraphics[width=4in, angle=270]{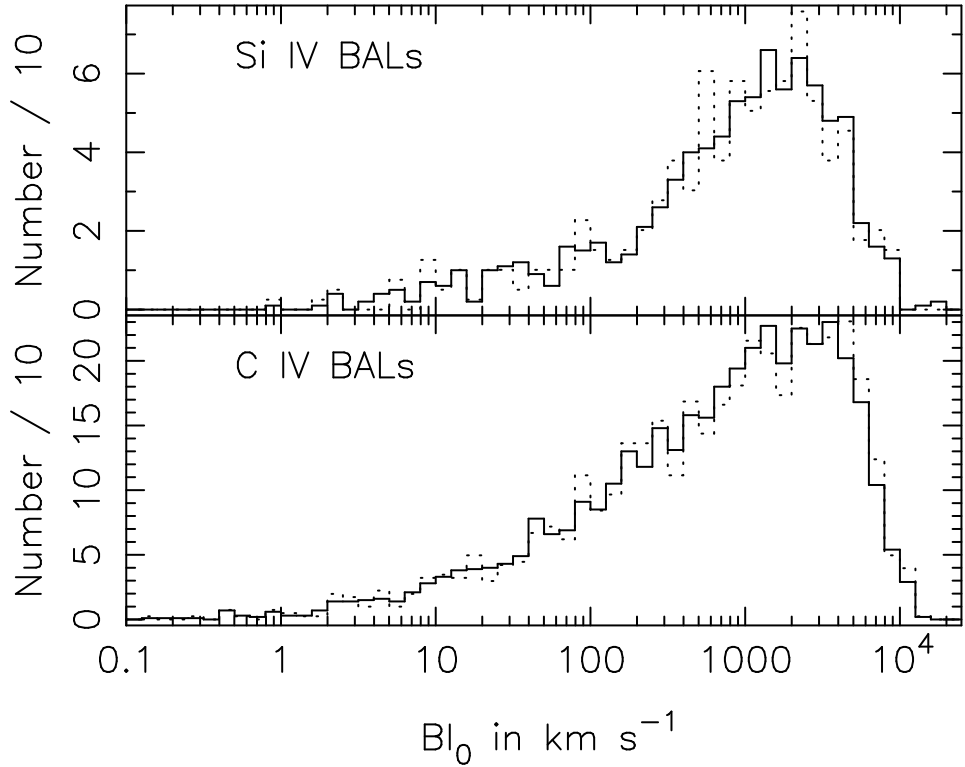}
      \caption{\label{bALCatBI0DistFig}({\it Top panel:})  The distribution of $BI_0$ for \ion{Si}{4} BALs at $z > 1.96$.  The solid line shows the distribution for all BALs, while the dotted line shows the distribution for QSOs with $SN_{1700} > 9$, scaled to match the area of the full distribution.  ({\it Bottom panel:})  Same as the top panel, but for \ion{C}{4} BALs at $z > 1.68$.}
   \end{center}
\end{figure}

Figure~\ref{bALCatBI0DistFig} shows the distributions of $BI_0$ for 345 BALs of \ion{Si}{4} (top panel) and 1603 of \ion{C}{4} (bottom panel).  In each case, we have considered only QSOs at redshifts sufficiently large that the entire potential absorption region (out to --25,000~km~s$^{-1}$) is in the SDSS bandpass (\S\ref{analysisSec}).  The distributions are not strongly affected by spectral S/N, in the sense that a Kolmogorov-Smirnov test finds no significant difference between the distributions of $BI_0$ for the full sample and for a high-S/N subsample (scaled and plotted with a dotted line in Figure~\ref{bALCatBI0DistFig}), which has $SN_{1700} > 9$.

\begin{figure} [ht]
  \begin{center}
      \includegraphics[width=4in, angle=270]{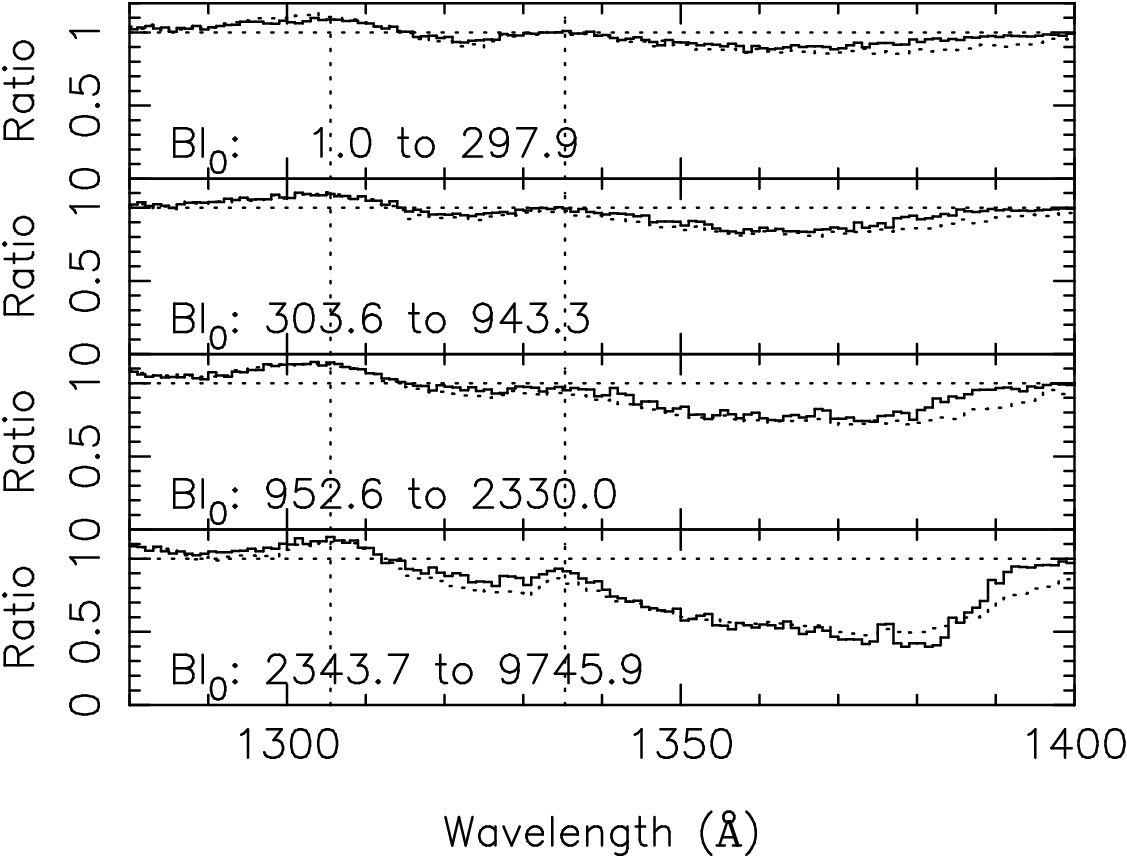}
      \caption{\label{bALCatBI0DistSiIVMedSpecsFig}Median (solid line) and mean (dotted line) rest-frame spectra of the absorber transmission for \ion{Si}{4} BALs with $BI_0 > 0$.  Spectra with \ion{Si}{4} BALs were sorted by $BI_0$ and divided into four groups of 111 objects.  Each spectrum was divided by its emission model and rebinned to a common wavelength grid.  The spectra for each group are plotted in order of increasing $BI_0$.  The local maxima at shorter wavelengths are attributable to unmodeled emission lines such as \ion{O}{1}~$\lambda$1306 and \ion{C}{2}~$\lambda$1335.  Vertical dotted lines mark the wavelengths of these two lines.}
   \end{center}
\end{figure}

Stacking BAL spectra in groups ordered by $BI_0$ can provide a qualitative picture of trends that distinguish strong BALs from weaker BALs.  For illustrative purposes, we have constructed median and mean spectra of BAL QSOs with $SN_{1700} \ge 9$ at a range of $BI_0$ values.  First, we arranged all BAL spectra with $SN_{1700} \ge 9$ in order of increasing $BI_0$.  We divided this ordered set of spectra into four groups of equal size (111 per group for \ion{Si}{4}, 400 per group for \ion{C}{4}), so that each group represents a unique range of $BI_0$.  We divided each individual spectrum by its emission model to construct an absorber ``transmission spectrum,'' and rebinned these spectra to a common grid.  Finally, for each wavelength bin at wavelength $\lambda$ in the grid, we determined the median, $M(\lambda)$, of all the transmission spectra in a given group.  Plots of the median and mean spectra $M(\lambda)$ are shown for each group in Figure~\ref{bALCatBI0DistSiIVMedSpecsFig} (for \ion{Si}{4} BALs) and Figure~\ref{bALCatBI0DistCIVMedSpecsFig} (for \ion{C}{4} BALs).

Two local maxima seen at $\approx$1305~\AA\ and $\approx$1335~\AA\ in the median spectra for \ion{Si}{4} BALs (Figure~\ref{bALCatBI0DistSiIVMedSpecsFig}) are likely due to unmodeled emission from lines such as \ion{O}{1}~$\lambda$1306 and \ion{C}{2}~$\lambda$1335.  Apart from the effects of unmodeled emission, the median spectra for \ion{Si}{4} and \ion{C}{4} BALs are qualitatively similar.  The broad, shallow features in the median spectra for the low-$BI_0$ groups reflect a tendency for weaker BALs to appear at a wide range of velocities.  Stronger BAL absorption tends to appear at lower outflow velocities.



\subsection{Distribution of $v_{max}$ and $v_{min}$\label{vDistSec}}

\begin{figure} [ht]
  \begin{center}
      \includegraphics[width=4in, angle=270]{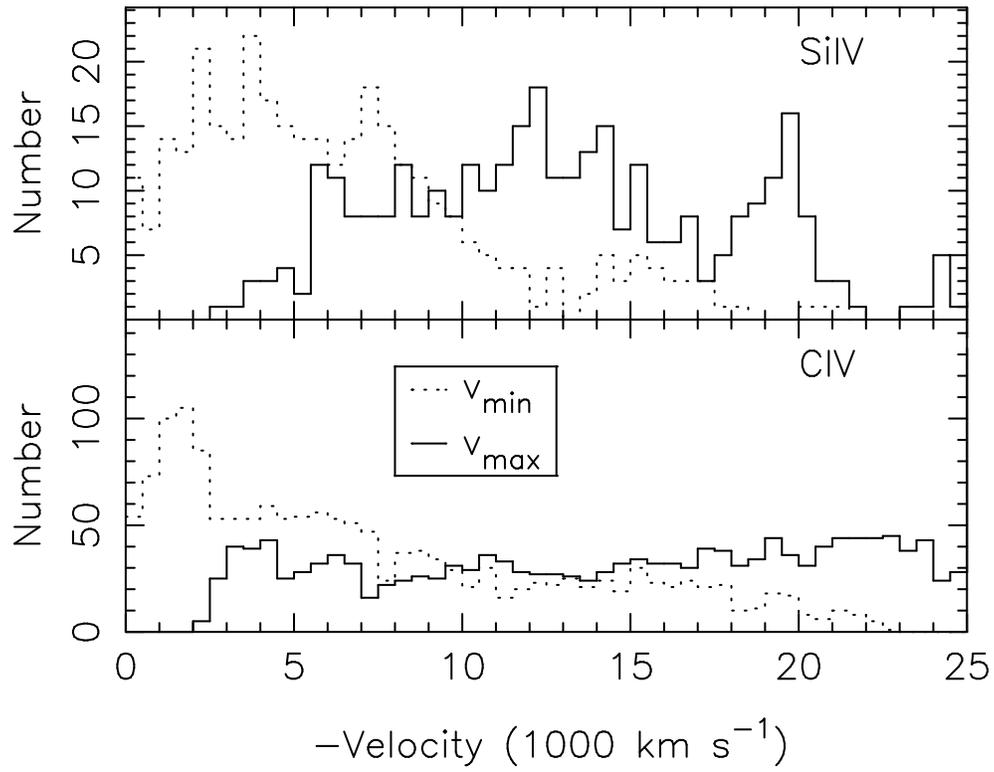}
      \caption{\label{plotBALCatVDistFig}({\it Top panel:})  The distribution of $v_{max}$ (solid line) and $v_{min}$ (dotted line) for \ion{Si}{4} BALs with $SN_{1700} > 9$.  ({\it Bottom panel:})  Same as top panel, but for \ion{C}{4}.}
   \end{center}
\end{figure}

\begin{figure} [ht]
  \begin{center}
      \includegraphics[width=4in, angle=270]{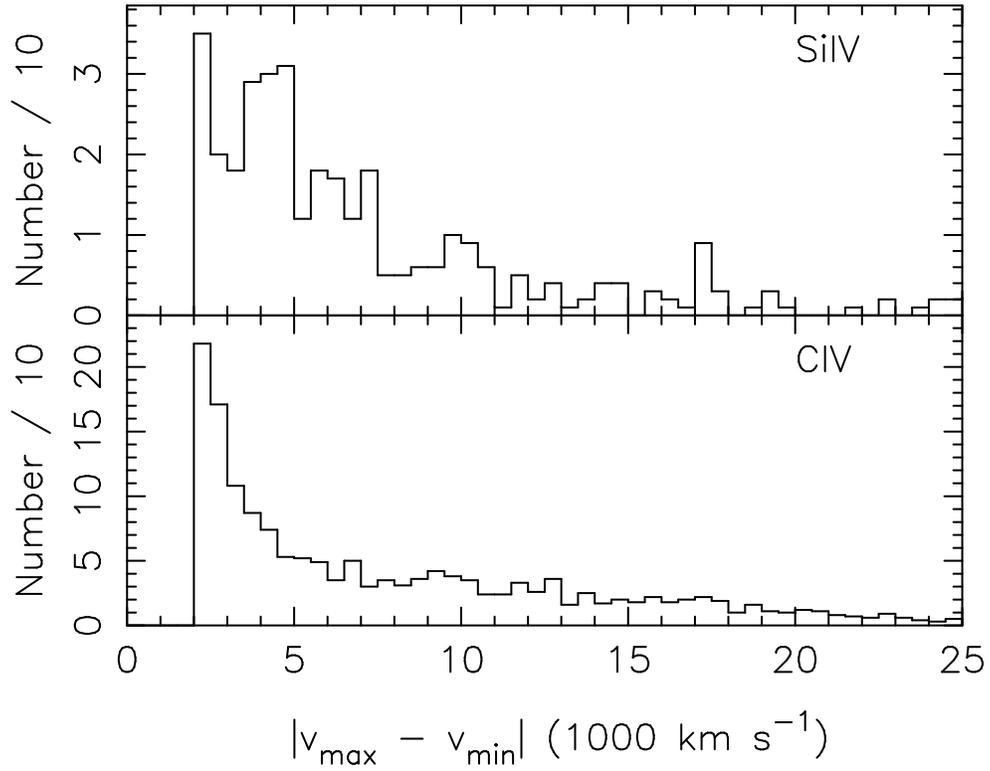}
      \caption{\label{plotBALCatdVDistFig}({\it Top panel:})  The distribution of $|v_{max} - v_{min}|$ for \ion{Si}{4} BALs with $SN_{1700} > 9$.  ({\it Bottom panel:})  Same as top panel, but for \ion{C}{4}.}
   \end{center}
\end{figure}

Figure~\ref{plotBALCatVDistFig} shows the distribution of $v_{max}$ and $v_{min}$ values for 344 \ion{Si}{4} and 1603 \ion{C}{4} BALs with $SN_{1700} \ge 9$ and $BI_0 > 0$.  The distribution for \ion{Si}{4} drops at outflow velocities of about --15,000 and --20,000~km~s$^{-1}$.  These drops are attributable to BAL troughs that would otherwise terminate near these velocities being filled in by the emission features mentioned in \S\ref{bIDistSec}.  We also show the distribution of $|v_{max} - v_{min}|$ for the same BALs in Figure~\ref{plotBALCatdVDistFig}.  Note that, by definition (\S\ref{catDataSec}), the velocity range between $v_{max}$ and $v_{min}$ may include relatively unabsorbed regions in cases with multiple BAL troughs for a single ion.  Although BAL absorption can cover a wide velocity range, narrower troughs ($<$5000~km~s$^{-1}$) are more common.

Previous studies have found that BAL maximum outflow velocities are correlated with QSO UV luminosities, as might be expected if BAL outflows are accelerated by radiation pressure \citep{lb02, gbcssv07}.  In order to test this conclusion, we have taken the set of 1239 QSOs with $1.68 \le z \le 2.68$, \ion{C}{4} $BI_0 > 0$, and $SN_{1700} \ge 9$, and divided it into two subsamples, depending on whether the monochromatic luminosity at 2500~\AA\ falls above or below the median $L_{\nu}(2500~\mathring{\rm A}) \approx 1.38\times 10^{31}$~erg~s$^{-1}$~Hz$^{-1}$.\footnote{Because our estimates of the flux at 2500~\AA\ are based on the continuum fit, $L_{\nu}(2500~\mathring{\rm A})$ does not include the additional emission in this region associated with blended lines of ions such as \ion{Fe}{2}.}  The median (mean) value of $L_{\nu}(2500~\mathring{\rm A})$ for the higher-luminosity subsample is $2.05\times 10^{31}$ ($2.53\times 10^{31}$)~erg~s$^{-1}$~Hz$^{-1}$, and for the lower-luminosity subsample it is $1.06\times 10^{31}$ ($1.04\times 10^{31}$)~erg~s$^{-1}$~Hz$^{-1}$.

\begin{figure} [ht]
  \begin{center}
      \includegraphics[width=4in, angle=270]{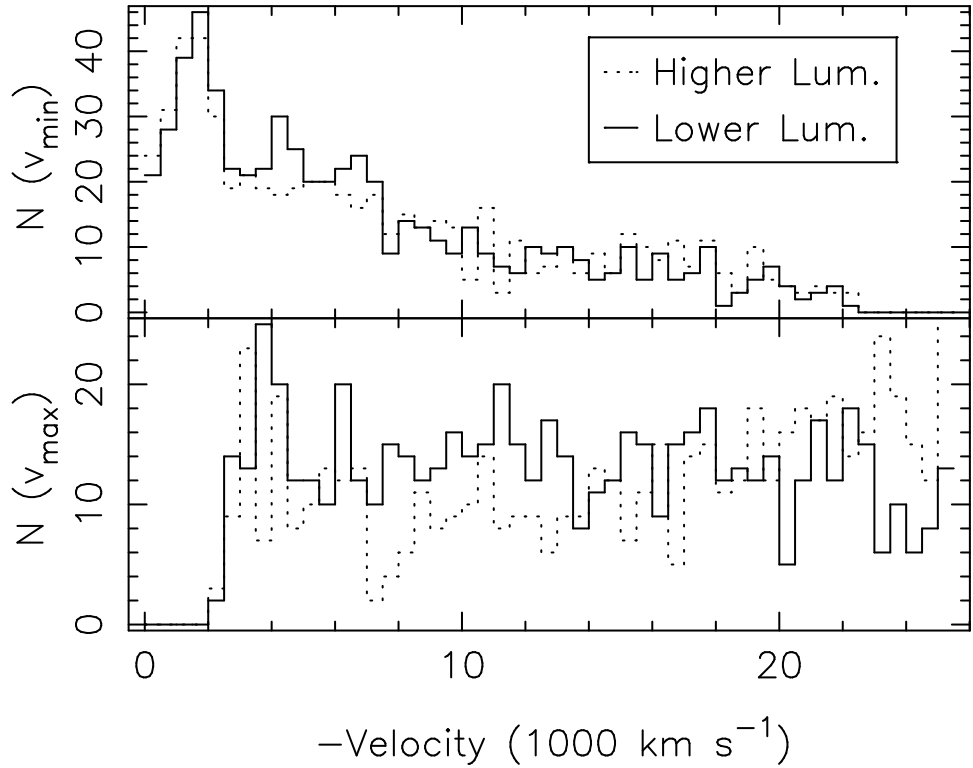}
      \caption{\label{pBCVDistByLumFig}({\it Top panel:})  The distribution of \ion{C}{4} BAL $v_{min}$ for lower-luminosity (solid line) and higher-luminosity (dotted line) QSOs with $SN_{1700} > 9$.  ({\it Bottom panel:})  Same as top panel, but showing the distribution of $v_{max}$.}
   \end{center}
\end{figure}

We have plotted the distributions for $v_{min}$ and $v_{max}$ for the lower- and higher-luminosity samples in Figure~\ref{pBCVDistByLumFig}.  While the distribution of $v_{min}$ appears to be unrelated to the source luminosity, the distribution of $v_{max}$ differs significantly (at $>$99.99\% confidence) between the lower- and higher-luminosity subsamples.  Compared to the lower-luminosity subsample, the higher-luminosity QSOs have fewer BALs with $|v_{max}| < 15,000$~km~s$^{-1}$ and more BALs with $|v_{max}| > 15,000$~km~s$^{-1}$.  In fact, a Spearman rank correlation test finds a highly significant correlation (at $>$99.99\% confidence) between $L_{\nu}(2500~\mathring{\rm A})$ and $v_{max}$ for the full sample.  If radiation pressure contributes significantly to accelerating the BAL outflow \citep[e.g.,][]{mcgv95}, we would expect to see such trends.  As noted in \S\ref{bALFracSec}, the luminosity range covered by our sample is limited; stronger effects may be observable with samples covering a wider range of $L_{2500}$.

We caution that low-S/N sources can have BALs obscured due to statistical noise.  This effect could lead to a false trend in which lower-luminosity (lower-S/N) sources have lower values of $v_{max}$ due to obscuration of the higher-velocity region of the BAL.  We have applied a simple cut to our sample ($SN_{1700} \ge 9$) to limit the effects of low S/N.

\begin{figure} [ht]
  \begin{center}
      \includegraphics[width=4in, angle=270]{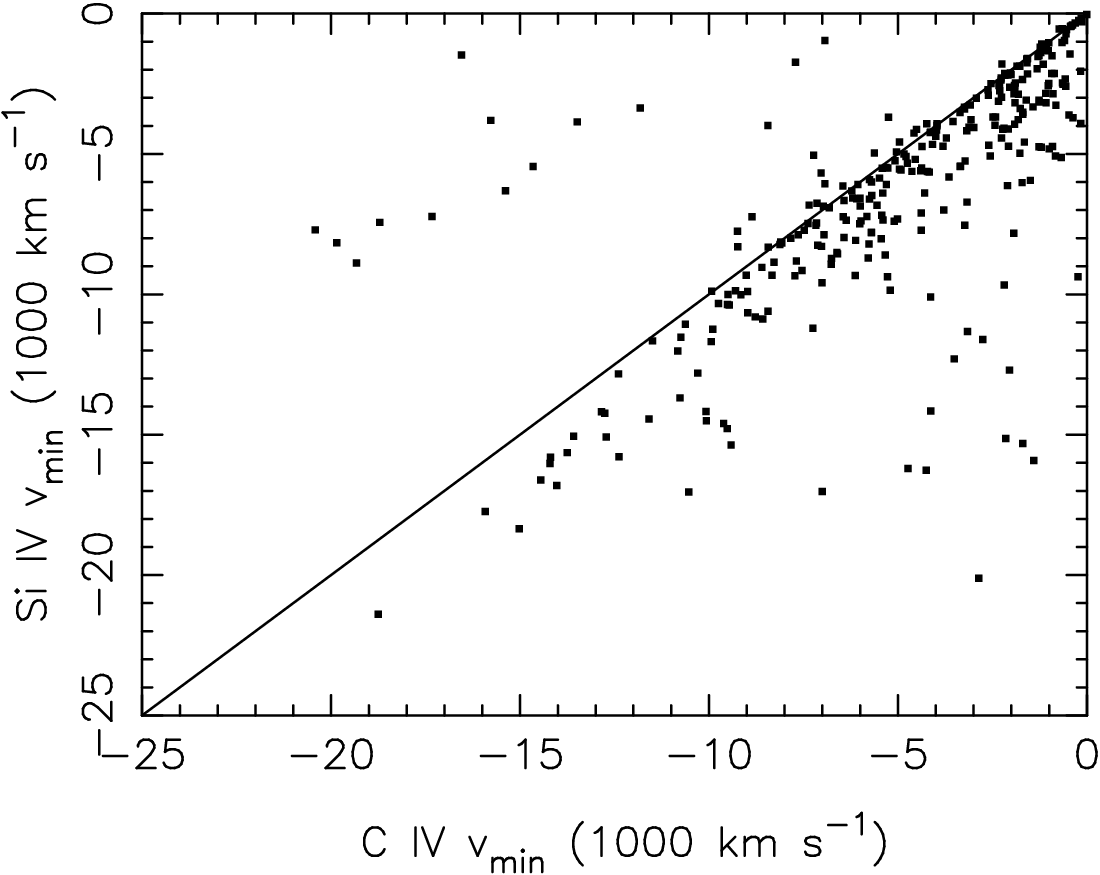}
      \caption{\label{siIVVsCIVVminFig}\ion{Si}{4} BAL $v_{min}$ plotted against \ion{C}{4} BAL $v_{min}$ for QSOs with $SN_{1700} > 9$ and $BI_0 > 0$ for both \ion{Si}{4} and \ion{C}{4}.  The solid line indicates where $v_{CIV} = v_{SiIV}$.}
   \end{center}
\end{figure}
\begin{figure} [ht]
  \begin{center}
      \includegraphics[width=4in, angle=270]{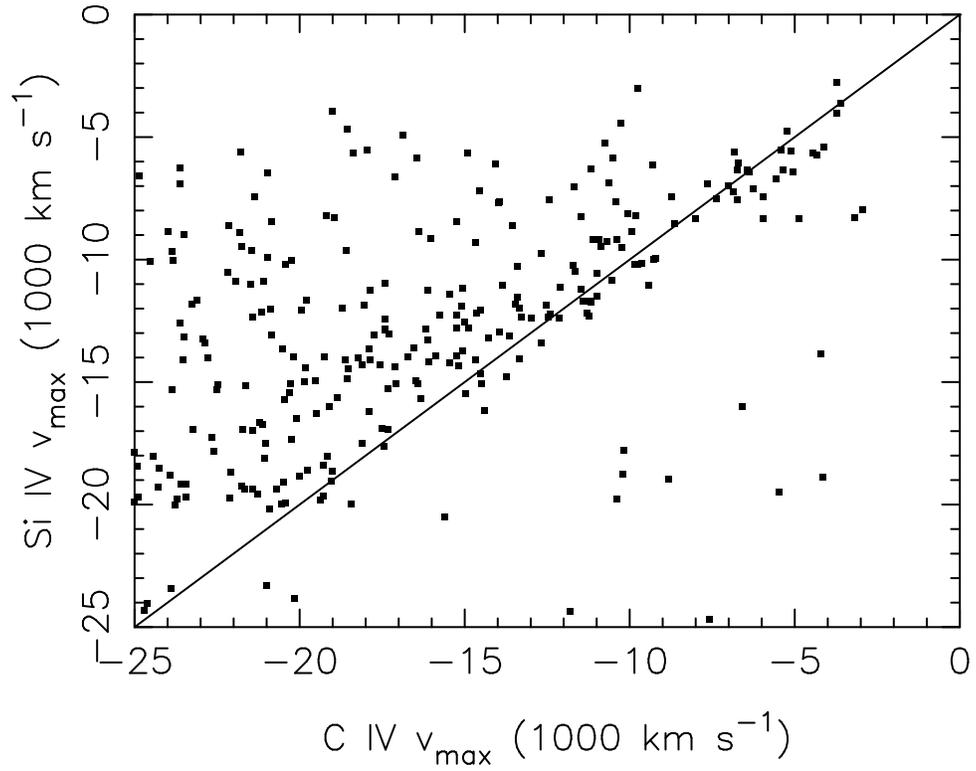}
      \caption{\label{siIVVsCIVVmaxFig}\ion{Si}{4} BAL $v_{max}$ plotted against \ion{C}{4} BAL $v_{max}$ for QSOs with $SN_{1700} > 9$ and $BI_0 > 0$ for both \ion{Si}{4} and \ion{C}{4}.  The solid line indicates where $v_{CIV} = v_{SiIV}$.}
   \end{center}
\end{figure}

It has previously been observed that BALs associated with different ions can have different absorption profiles \citep[e.g.,][]{vwk93}.  Here, we briefly compare \ion{Si}{4} and \ion{C}{4} BAL properties using a large sample of SDSS BAL QSOs.  There are 798 QSOs in our catalog that have $BI_0 > 0$ for both \ion{Si}{4} and \ion{C}{4} absorption and are also at high enough redshift ($z > 1.96$) so that \ion{Si}{4} absorption could be detected up to $|v_{max}| \approx$25,000~km~s$^{-1}$.  In Figure~\ref{siIVVsCIVVminFig} we plot \ion{Si}{4} $v_{min}$ against \ion{C}{4} $v_{min}$ for sources with $SN_{1700} > 9$, and in Figure~\ref{siIVVsCIVVmaxFig} we do the same for $v_{max}$.  From these figures, we see that $v_{min}$ is well correlated between the two ions.  This supports the general identification of BAL absorption blueward of 1400~\AA\ with \ion{Si}{4}, rather than extremely high-velocity \ion{C}{4} absorption.  \ion{Si}{4} tends (with some exceptions) to have equal or greater values of $|v_{min}|$ than \ion{C}{4} does, and also equal or lesser values of $|v_{max}|$.  At outflow velocities $|v| \ga 15,000$~km~s$^{-1}$, the \ion{Si}{4} BAL trough can be truncated by unmodeled emission (such as \ion{C}{2}~$\lambda$1335); however, the effect is also seen for cases with lower values of \ion{C}{4} $|v_{max}|$.  Additional unmodeled emission may be affecting the \ion{Si}{4} trough even at lower outflow velocities, but strong lines are not typically seen at these wavelengths in AGN spectra, and according to theoretical line lists \citep[e.g.,][]{vvf96}, even stronger line emission would be noticable from the same ions in other spectral regions.

Visual inspection indicates that the narrower \ion{Si}{4} troughs can, in some cases, be attributed to the fact that \ion{Si}{4} troughs are shallower and may be categorized as narrower because their high- and low-velocity regions lie above the 90\% continuum line due to a sloping trough shape or spectral noise.  If the trough depth is governed by the geometric covering factor, this points to geometric differences between the \ion{C}{4} and \ion{Si}{4} absorbers.  For stronger \ion{Si}{4} BALs, differences in outflow velocities also suggest differences in the outflow structure.  Studies of individual sources have found that the outflow geometry can differ for ions of various elements \citep[e.g.,][]{ablgwbd99, akdjb99}.  Our results for a large sample of BAL QSOs indicate that BAL outflows do, in general, show some ion-dependent structure, if we assume that trough depths are dominated by outflow structure.

We have visually inspected the less-common cases in which the \ion{Si}{4} $|v_{min}|$ is less than that of \ion{C}{4}.  In some cases, there is broad absorption at lower velocities for \ion{C}{4} that was not formally detected as a BAL; a less-restrictive measure of identifying broad absorption features would not classify these sources as outliers.  In other cases, the BAL troughs are indeed unusual; in particular, we cannot rule out the possibility that extremely high-velocity \ion{C}{4} BALs are being mis-classified as low-velocity \ion{Si}{4} absorption.  We have also visually inspected cases where the \ion{Si}{4} $|v_{max}|$ is much larger than that of \ion{C}{4}.  The high-velocity absorption attributed to \ion{Si}{4} is typically in the 1330--1340~\AA\ range, and is therefore likely due to contamination by \ion{C}{2}~$\lambda$1335, as discussed in \S\ref{obsDataSec}.


\subsection{Distribution of \ion{C}{4} Emission\label{cIVEmSec}}

\citet{thrrsvkafbkn06} found that the BAL QSOs in their catalog tended to match emission templates with broader line emission than for non-BAL QSOs, and \citet{r06} has suggested that BAL QSOs may be drawn from a parent sample of QSOs with larger \ion{C}{4} blueshifts and weaker \ion{C}{4} emission.  To test the hypothesis that \ion{C}{4} emission is different in BAL QSOs, we constructed two QSO subsamples.  For each subsample, we required that $z \ge 1.68$ so that the entire \ion{C}{4} absorption region (out to --25,000~km~s$^{-1}$) is visible in the SDSS bandpass, and also that $SN_{1700} > 9$, in order to ensure that the spectra have high S/N.  We also required that the {\tt EmLost} flag be false for \ion{C}{4} and that $v_{min}$ for any BALs present be $< -10,000$~km~s$^{-1}$.  These samples consist of 369 \ion{C}{4} BAL QSOs and 9597 non-BAL QSOs.  While these sources may not, in principle, be completely representative of all BAL QSOs, selecting BALs detached from the \ion{C}{4} emission lines is necessary to ensure that BALs are not noticably affecting \ion{C}{4} line emission.  However, we are only using 23\% of the available BAL QSO population, and if QSOs having BALs with higher values of $|v_{min}|$ are somehow physically different from the rest of the BAL QSO population, our result will not represent the entire BAL QSO population.  We note that \citet{tfgw88} identified a small sample of three BAL QSOs having strong, smooth \ion{C}{4} BALs with $v_{min} \approx 0$~km~s$^{-1}$ and found that they had emission line characteristics differing from those of QSOs with significantly detached BALs.  Further study of this subject will clearly be valuable, but is outside the scope of the current work.

\begin{figure} [ht]
  \begin{center}
      \includegraphics[width=4in, angle=270]{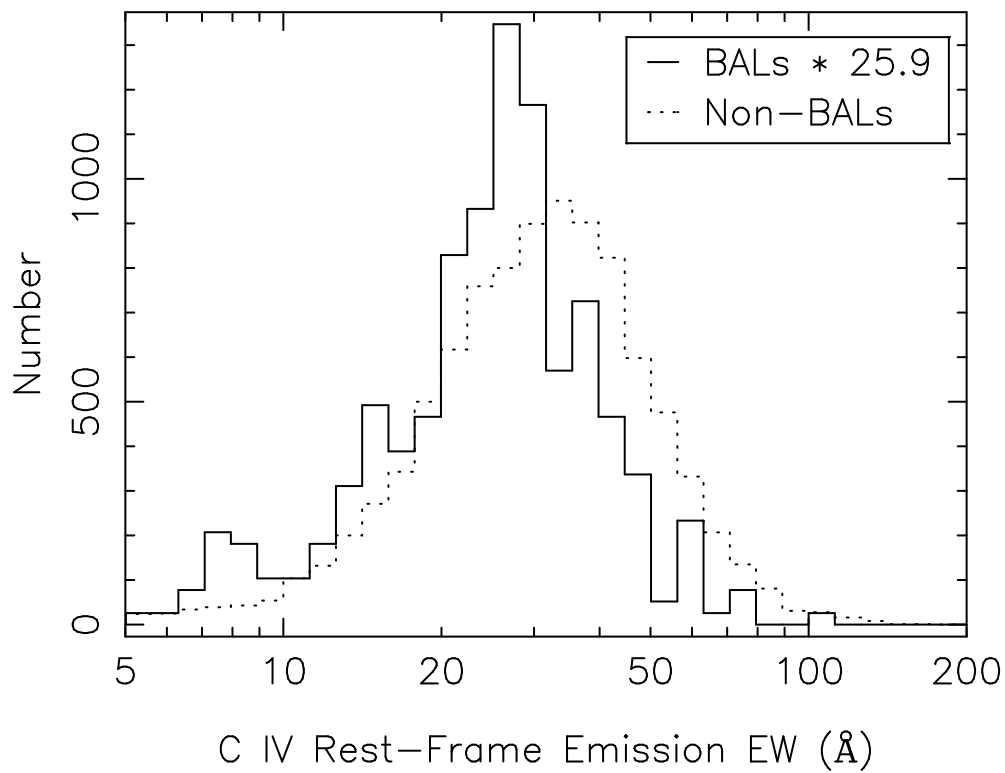}
      \caption{\label{plotBALCatEmEWFig}\ion{C}{4}~$\lambda$1549.5 model emission line EW for BAL QSOs (solid histogram) and non-BAL QSOs (dotted histogram).  The BAL distribution has been scaled by a constant for easy comparison.  We considered only QSOs with $z \ge 1.68$ and $SN_{1700} > 9$ to ensure that the entire \ion{C}{4} absorption region (out to --25,000~km~s$^{-1}$) was available and that spectra had high S/N.  The $x$-axis is logarithmic.}
   \end{center}
\end{figure}

\begin{figure} [ht]
  \begin{center}
      \includegraphics[width=4in, angle=270]{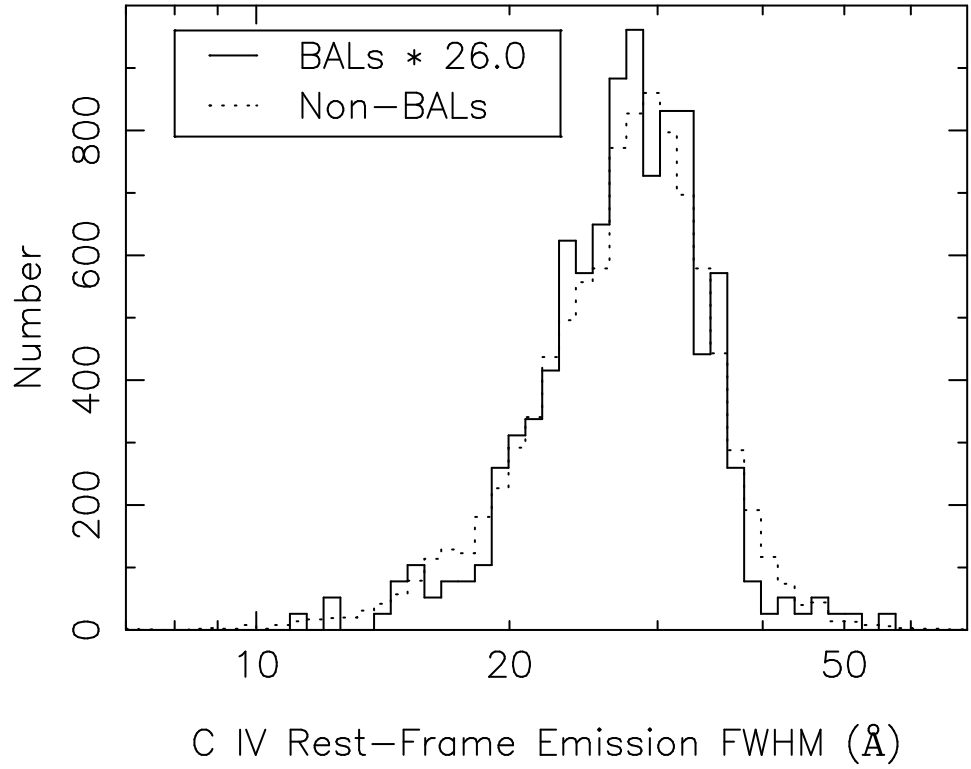}
      \caption{\label{plotBALCatEmFWHMFig}Same as Figure~\ref{plotBALCatEmEWFig}, but showing the \ion{C}{4} emission line FWHM.}
   \end{center}
\end{figure}

Figure~\ref{plotBALCatEmEWFig} shows the distribution of \ion{C}{4} emission equivalent width (EW) for BAL and non-BAL QSOs in our subsamples, while Figure~\ref{plotBALCatEmFWHMFig} shows the distributions of \ion{C}{4} FWHM.  The distribution of EWs is clearly different between the two groups, with BAL QSOs having weaker \ion{C}{4} emission EWs.  A Kolmogorov-Smirnov test confirms the difference at $>99.99$\% confidence.  On the other hand, we find no visual or statistical evidence that the distribution of FWHMs differs.  We caution in this case that we have fit line cores and not accounted for (often weakly-constrained) emission features in the red and/or blue wings of the line.  If the emission in the line wings or line emission from other ions differs between BALs and non-BALs \citep[e.g.,][]{rrhsvfykb03}, we would not be sensitive to these effects.

For non-BAL QSOs, the \ion{C}{4} emission EW is related to the UV luminosity according to the well-known Baldwin effect \citep{b77}, with $\log(EW) \equiv A_B  + \beta \log(L_{UV})$ and $\beta \approx -0.2$ \citep[e.g.,][]{krk90, wvbb06}.  If BAL QSOs are derived from a parent population with higher intrinsic UV luminosities than we observe (perhaps because excess UV absorption reddens BAL QSOs; \S\ref{bALQSOReddeningSec}), and if the Baldwin effect in BAL QSOs is based on the {\it intrinsic} (e.g., pre-reddened) luminosity, then the Baldwin effect would lead to weaker emission EWs in BAL QSOs compared to non-BAL QSOs of the same {\it observed} UV luminosity.

\begin{figure} [ht]
  \begin{center}
      \includegraphics[width=4in, angle=270]{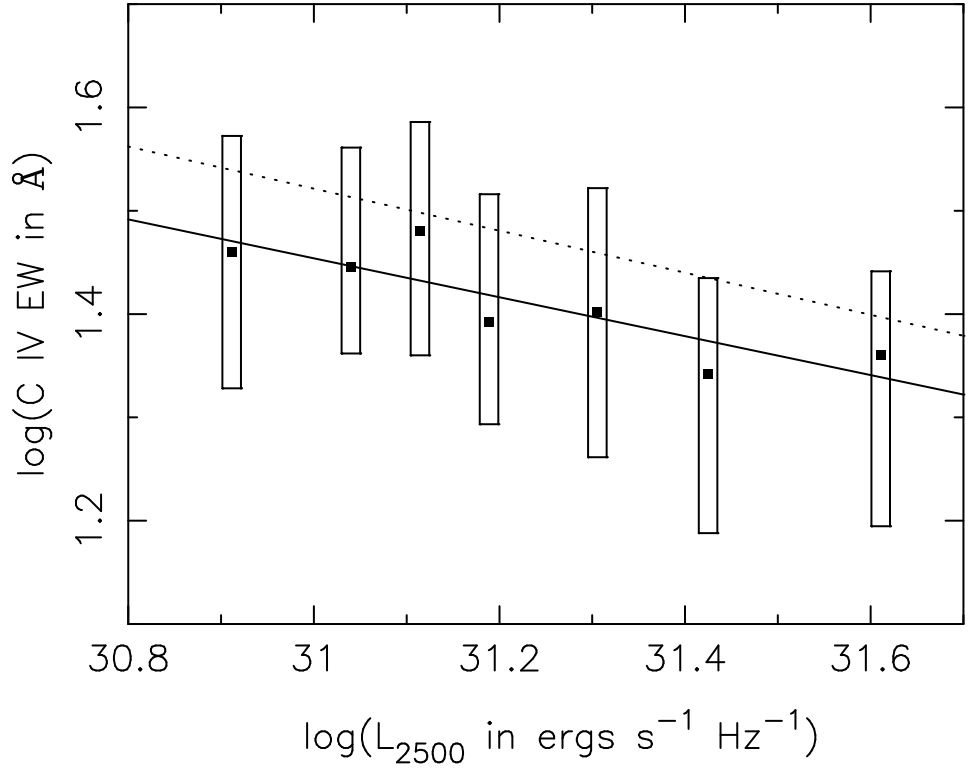}
      \caption{\label{bALVsNBBaldwinFig}Filled squares indicate the median $\log(L_{2500})$ and median $\log(C IV EW)$ for luminosity-ordered subsamples of 50 BAL QSOs with \ion{C}{4} $BI_0 > 0$ and $SN_{1700} > 9$.  The rectangles surrounding the filled squares indicate the range of the central two quartiles in EW.  The solid line indicates the best fit to these points.  The dotted line indicates a fit to a similarly-constructed set of subsamples of non-BAL QSOs.}
   \end{center}
\end{figure}

We have measured the Baldwin effect relation for BAL and non-BAL QSOs separately.  To do this, we ordered by $L_{2500}$ the samples of BAL and non-BAL QSOs used to generate Figure~\ref{plotBALCatEmEWFig} and divided them into groups of size 50 (BALs) or 200 (non-BALs).  For each group, we record the median $L_{2500}$ and \ion{C}{4}~EW.  The points in Figure~\ref{bALVsNBBaldwinFig} show these values for the sample of BAL QSOs.  Fitting these points for the BAL QSOs, we find $A_B = 7.291 \pm 1.40$ and $\beta = -0.188 \pm 0.045$; this fit is shown in the figure with a solid line.  For non-BAL QSOs, we find $A_B = 7.830 \pm 0.437$ and $\beta = -0.203 \pm 0.014$; this fit is shown as a dotted line for comparison.  For clarity, we do not plot the non-BAL data points; they are well-represented by the fit line.  Both logarithmic slopes are consistent with those previously found for the Baldwin effect, but the normalization for the BAL sample is smaller.  Fitting the median EW values reduces the influence of a significant number of outlying sources with extremely weak \ion{C}{4} emission.  If we fit directly the EW and $L_{2500}$ values for the full samples, we obtain similar results but with slightly steeper logarithmic slopes of $\beta =-0.24 \pm 0.01$ and $\beta = -0.27 \pm 0.05$, respectively.

If BAL continuum emission is absorbed compared to non-BAL QSOs, but the \ion{C}{4} emission is unaffected by absorption, we would expect BAL QSOs to have generally {\it larger} emission EWs than non-BAL QSOs, contrary to what we observe in Figure~\ref{plotBALCatEmEWFig}.  If both the line and continuum emission are equally absorbed, the EWs would be unchanged, but continuum luminosities would be fainter.  This would shift leftward the trend of EW against $L_{2500}$ plotted in Figure~\ref{bALVsNBBaldwinFig}.  However, a rather large shift is required to explain the discrepancy between BAL and non-BAL fit lines in the figure:  $L_{2500}$ would have to change by a factor of $\approx$2.  This is much larger than typically observed for BAL QSOs (\S\ref{bALQSOReddeningSec}, \S\ref{bALFracSec}).  While BAL QSOs exhibit a Baldwin effect, it appears that this effect alone does not account for the difference observed between BAL and non-BAL QSO \ion{C}{4} emission EWs.

Since we have only considered BAL QSOs that do not have strong BAL features obscuring the emission line regions, it is not obvious why BAL QSOs should have significantly different \ion{C}{4} emission line properties from non-BAL QSOs.  It may be that the BAL absorber or other absorbers associated with the BAL outflow along the line of sight partially obscure the broad emission line region (BELR).  The low-velocity absorption profile would have to be broad and smooth so that significant deficits in the line emission profile were not detected as BALs.  More complex scenarios may also be constructed to account for the effects of viewing, at different orientation angles, the anisotropic emission line radiation from the disk and the BAL outflow \citep[e.g.,][]{kv98}.  Additional orientation-dependent effects may be introduced as the emission is absorbed and scattered by material (including the BAL outflow itself) in the QSO nucleus.

Alternately, the decreased emission in BAL QSOs could be due to ionization effects.  For example, if the ionization state of the emission line region were lower for BAL QSOs than for non-BAL QSOs, perhaps due to the presence of additional absorbing material in BAL QSOs that blocks ionizing photons from the emission line region, then emission from highly-ionized species such as \ion{C}{4} could be inhibited in such cases.  In fact, a preliminary analysis of our data indicates that the ratio of \ion{C}{4} to \ion{C}{3}] emission EWs is significantly ($>$99.99\% confidence) lower for BALs than for non-BAL QSOs, which could indicate that the BELR gas is at a lower ionization state in BAL QSOs.  However, this effect could be explained by other factors besides ionization differences between BAL and non-BAL QSOs.  Further study is therefore needed to determine the extent and physical implications of any differences between broad line emission in BAL and non-BAL QSOs.

\subsection{X-Ray Properties of BAL QSOs\label{xRayBALQSOSec}}

The UV and \mbox{X-ray} luminosities of QSOs are correlated \citep[e.g.,][]{at82}.  Recent studies have carefully quantified the UV/\mbox{X-ray} luminosity relation across $\approx$5 orders of magnitude in UV luminosity \citep[e.g.,][]{sbsvv05, ssbaklsv06, jbssscg07}.  This relation is usually expressed in terms of the logarithm of the ratio between the monchromatic luminosities $L_{\nu}$ at 2~keV and 2500~\AA\ with the parameter
\begin{eqnarray}
\alpha_{OX} &\equiv& 0.3838 \log\biggl(\frac{L_{\nu}(2~\rm{keV})}{L_{\nu}(2500~\mathring{\rm A})}\biggr)\label{aOXDefnEqn}.
\end{eqnarray}

For typical QSOs, $\alpha_{OX}$ is a function of $L_{2500}$, the monochromatic luminosity at 2500~\AA.  \citet{jbssscg07} found
\begin{eqnarray}
\alpha_{OX}(L_{2500}) &=& (-0.140 \pm 0.007) \log(L_{2500}) + (2.705 \pm 0.212).\label{justAOXEqn}
\end{eqnarray}

Equation~\ref{justAOXEqn} enables quantitative measurement of how the \mbox{X-ray} luminosity of a given QSO differs from that expected for a typical QSO of the same UV luminosity.  This difference is parameterized by
\begin{eqnarray}
\Delta\alpha_{OX} &\equiv& \alpha_{OX} - \alpha_{OX}(L_{2500}),\label{dAOXDefnEqn}
\end{eqnarray}
where $\alpha_{OX}$ is the value observed for a particular source observed to have a UV luminosity of $L_{2500}$, and $\alpha_{OX}(L_{2500})$ is determined for a typical QSO from Equation~\ref{justAOXEqn}.  For example, $\Delta\alpha_{OX} = -0.4$ corresponds to \mbox{X-ray} weakness by a factor of $\approx$11 with respect to typical QSOs of the same UV luminosity, and $\Delta\alpha_{OX} = -0.9$ corresponds to \mbox{X-ray} weakness by a factor of $\approx$220.

Many SDSS QSOs have been observed in targeted or (typically) serendipitous observations with {\it Chandra} or {\it XMM-Newton}.  As part of a study to determine the \mbox{X-ray} properties of {\it non}-BAL QSOs, we implemented semi-automated processes to identify such sources and obtain \mbox{X-ray} flux densities from data available in the {\it Chandra} and {\it XMM-Newton} archives.  These procedures are described in detail in \citet{gbs08}.  Briefly, we determined which SDSS QSOs fell on {\it Chandra} ACIS or {\it XMM-Newton} MOS/$pn$ chips and reduced these data to obtain source and background spectra in each case.  We then used the Cash statistic to fit a broken power law to each unbinned spectrum, with the power law break set at (rest-frame) 2~keV.  From this fit, we determined the flux density at 2~keV, $F_{\nu}(2~{\rm keV})$.  We determined upper and lower limits on $F_{\nu}$ by adjusting the power law norm and re-fitting the remaining spectral parameters until the Cash statistic $C$ changed by $\Delta C = 1$.

We used Poisson statistics to determine whether a source was detected (based on expected background count rates) at $>$99\% confidence in either the observed-frame full (0.5--8~keV), soft (0.5--2~keV) or hard (2--8~keV) bands.  High angular resolution is essential to ``resolve away'' the \mbox{X-ray} background and maximize the rate of source detections.  The excellent resolution and astrometric precision of the {\it Chandra} ACIS instrument allowed a high source detection rate even for short exposure times and high off-axis angles.  For this reason, in cases where a source was observed multiple times, we selected (in an unbiased way) the longest {\it Chandra} ACIS exposure as most representative of source properties.  If {\it Chandra} ACIS observations were not available, we selected the longest {\it XMM-Newton} MOS camera observation to determine \mbox{X-ray} fluxes.  We prefer the MOS to the $pn$ camera because the MOS is more effective at ``resolving away'' the background, resulting in a superior detection fraction for these faint sources.

QSOs with strong radio emission are well known to be relatively \mbox{X-ray} bright \citep[e.g.,][]{wtgz87, blvsbbwg00}.  We wish to identify such sources so that we can understand the emission and absorption processes of BAL QSOs without the added complexity of enhanced \mbox{X-ray} emission that may be associated with a radio jet.  For sources observed in the FIRST radio survey \citep{bwh95}, we obtain the 1.4~GHz core flux densities from the DR5 QSO catalog.  For sources that were not covered by the FIRST survey, we used 1.4~GHz flux densities obtained from the NVSS survey \citep{ccgyptb98}.  We estimate the monochromatic luminosity at 5~GHz assuming the radio flux follows a power law with a spectral index $\alpha = -0.8$.  We then calculate the radio-loudness parameter \citep[e.g.,][]{sw80, ksssg89},
\begin{eqnarray}
\log(R^*) &\equiv& \log\biggl( \frac{L_{\nu}({\rm~5~GHz})}{L_{2500}} \biggr).\label{rStarEqn}
\end{eqnarray}
We classify sources with $\log(R^*) \ge 1$ as ``radio-loud'' and all other sources as ``radio-quiet.''  Values of $\log(R^*)$ (or the upper limit on $\log(R^*)$) are given in Table~\ref{xInfoTab}.  In some cases, the radio observations may not be sufficiently sensitive to reach to $\log(R^*) = 1$, but even so these cases are small in number and are not extremely radio-bright (see the discussion in \citet{gbs08}.  Throughout this section, we only consider radio-quiet sources.  A detailed study of the \mbox{X-ray} properties of radio-loud BAL QSOs is being conducted separately (Miller et al., in preparation).

In this work, we consider only sources that are within 10$\arcmin$ of the telescope pointing.  At larger off-axis angles, the angular resolution is significantly worse, leading to greater uncertainty in background estimation and a larger fraction of non-detections.

\subsubsection{X-Ray Weakness\label{xWeakSec}}
\begin{figure} [ht]
  \begin{center}
      \includegraphics[width=4in, angle=270]{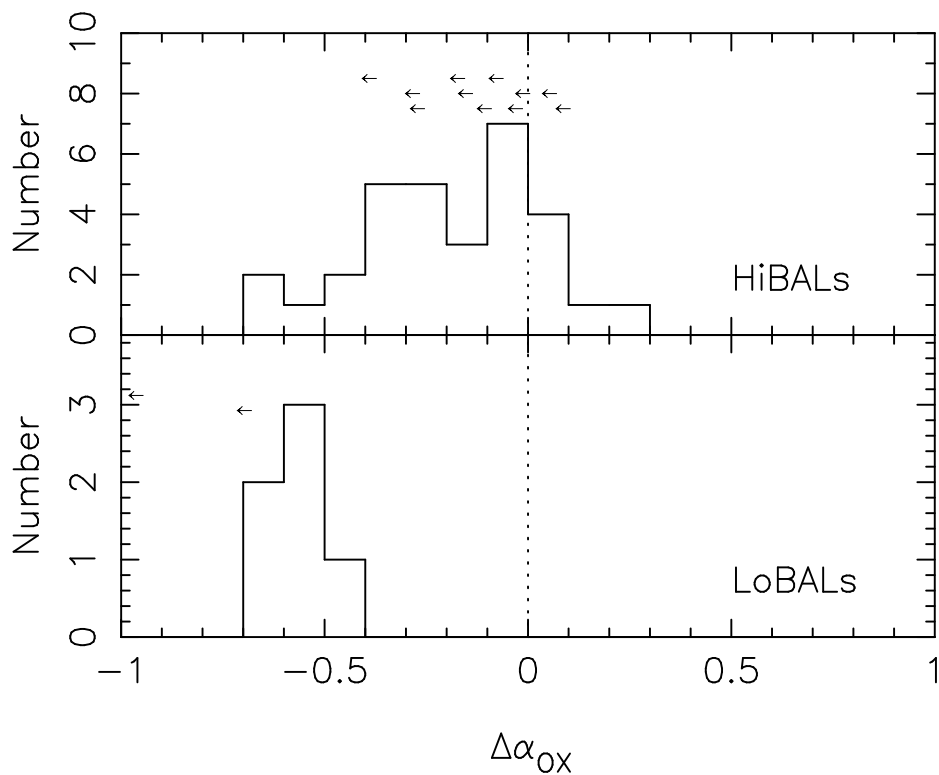}
      \caption{\label{hiLoBALDAOXFig}({\it Top panel:})  The distribution of $\Delta\alpha_{OX}$ (defined in \S\ref{xRayBALQSOSec}) for HiBAL QSOs.  Upper limits for sources that were not detected in \mbox{X-rays} are plotted with arrows.  The arrows are spaced vertically for visual clarity, but only the $x$-axis value is meaningful as an indication of the upper limit value.  Each arrow represents a single source.  The vertical dotted line indicates $\Delta\alpha_{OX} = 0$ for clarity.  ({\it Bottom panel:})  Same as top panel, but for LoBAL QSOs.}
   \end{center}
\end{figure}

Compared to typical QSOs of the same UV luminosity, BAL QSOs are unusually \mbox{X-ray} weak \citep[e.g.,][]{gsahfbftm95, blw00} due to strong \mbox{X-ray} absorption \citep[e.g.,][and references therein]{gamwe01, gblemwi01, gbcg02}.  Figure~\ref{hiLoBALDAOXFig} shows the distribution of $\Delta\alpha_{OX}$ for the 42 radio-quiet HiBAL (top panel) and 8 LoBAL (bottom panel) QSOs in our study.  The relatively high fraction of \mbox{X-ray} detections indicates that the \mbox{X-ray} emission of SDSS BAL QSOs is reasonably well-matched to the sensitivities of modern \mbox{X-ray} observatories.  As previous studies have indicated \citep[e.g.,][]{gbcpgs06}, LoBAL QSOs are found to be significantly \mbox{X-ray} weaker than HiBAL QSOs at 99.99\% confidence, according to a Gehan test implemented in the Astronomy Survival Analysis (ASURV) software package \citep[e.g.,][]{if90, lif92}.  The extreme \mbox{X-ray} weakness of LoBAL QSOs is presumably due to stronger \mbox{X-ray} absorption.

\subsubsection{X-Ray Brightness and BAL Properties\label{xWeakBALPropSec}}

\citet{gbcpgs06} examined the UV and \mbox{X-ray} properties of 35 BAL QSOs with optical/UV spectroscopy in the LBQS.  They found that most UV BAL properties were not strongly correlated with \mbox{X-ray} brightness in their sample.  The exception was $v_{max}$, which correlated with $\Delta\alpha_{OX}$ at $\approx$99.7\% confidence.  Motivated by their study, we have tested for correlations between BAL properties and $\Delta\alpha_{OX}$ in our larger sample of DR5 BAL QSOs with {\it Chandra} and/or {\it XMM-Newton} coverage.  We have limited our study to consider only radio-quiet QSOs that are confirmed HiBALs; our results are therefore independent of the effects of enhanced \mbox{X-ray} emission associated with radio jets, any physical factors which may differ between HiBAL and LoBAL QSOs, and also any biases inherent in the \ion{Mg}{2} BAL detection process (discussed in \S\ref{obsDataSec}).

\begin{figure} [ht]
  \begin{center}
      \includegraphics[scale=0.75, angle=270]{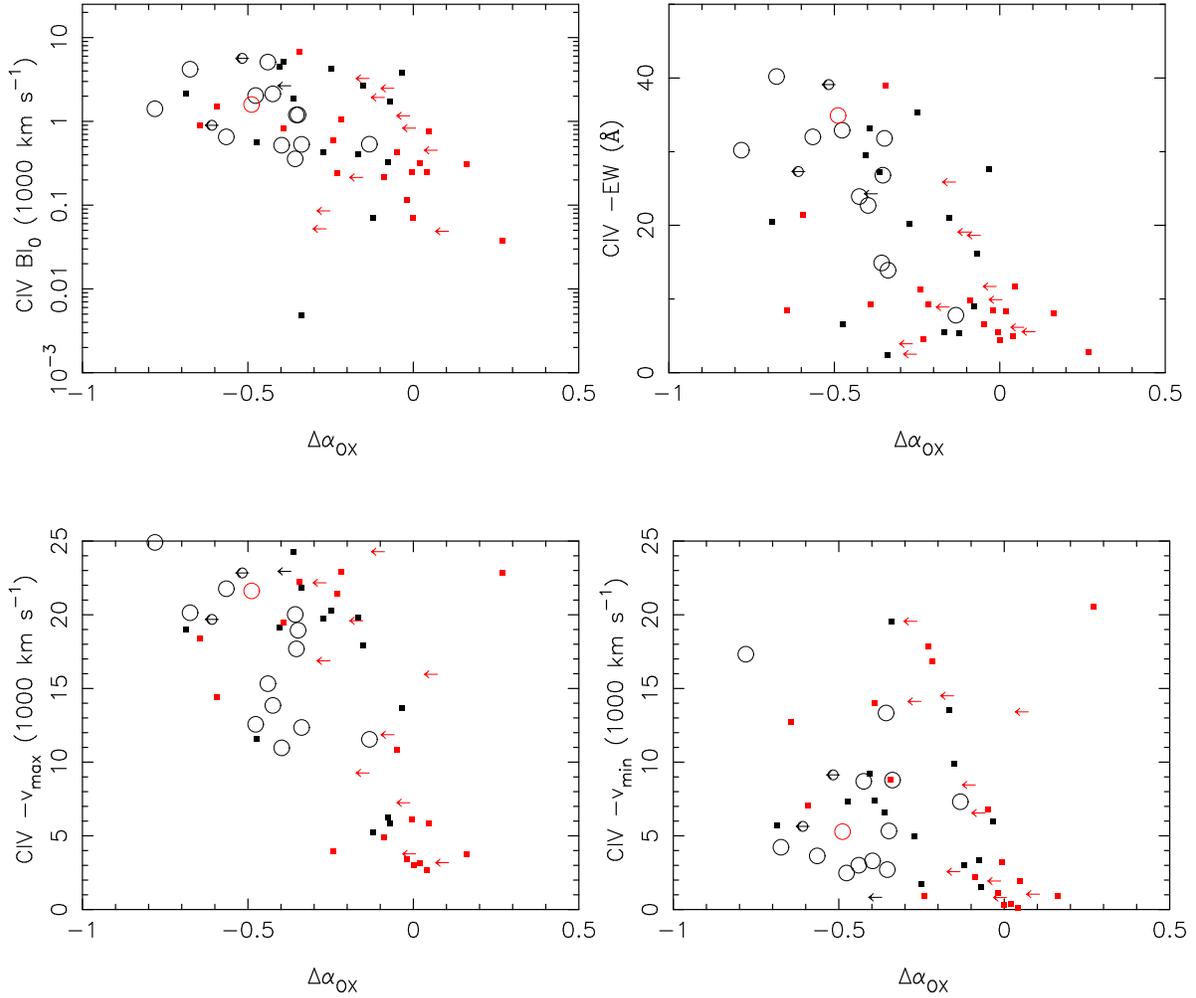}
      \caption{\label{bALCatCIVAllVsDAOXcFig}({\it Top left:)}  \ion{C}{4} $BI_0$ plotted against the relative \mbox{X-ray} brightness $\Delta\alpha_{OX}$ for HiBALs observed in \mbox{X-rays} with {\it Chandra} or {\it XMM-Newton}.  Filled squares represent sources from this catalog that were detected in \mbox{X-rays}, while arrows represent sources from this catalog that were not detected in \mbox{X-rays}.  Open circles represent sources from \citet{gbcpgs06} that are not included in this catalog and were detected in \mbox{X-rays}.  Smaller circles with arrows represent sources from \citet{gbcpgs06} that are not included in this catalog and were not detected in \mbox{X-rays}.  Black points have $L_{2500} \ge 1.44\times 10^{31}$~erg~s$^{-1}$, while red points have lower values of $L_{2500}$.  ({\it Top right:}) the same, but plotting \ion{C}{4} BAL absorption EW against $\Delta\alpha_{OX}$.  ({\it Bottom left:}) the same, but plotting \ion{C}{4} BAL absorption $v_{max}$ against $\Delta\alpha_{OX}$.  ({\it Bottom right:}) the same, but plotting \ion{C}{4} BAL absorption $v_{min}$ against $\Delta\alpha_{OX}$.}
   \end{center}
\end{figure}

We have plotted the \ion{C}{4} BAL $BI_0$, $EW_0$, $v_{max}$, and $v_{min}$ against $\Delta\alpha_{OX}$ in Figure~\ref{bALCatCIVAllVsDAOXcFig}.  In each panel, we have plotted data from this catalog with filled squares for \mbox{X-ray} detected sources, or arrows indicating 1$\sigma$ upper limits, in cases where the source was not detected in \mbox{X-rays}.  We have also plotted HiBALs from \citet{gbcpgs06} that were not included in the SDSS DR5 QSO catalog.  The \citet{gbcpgs06} non-SDSS HiBALs are plotted with open circles (for detected sources) or circles containing arrows (indicating upper limits for non-detected sources).  For these sources, we have measured \ion{C}{4} $BI_0$ and $v_{min}$ from the LBQS spectra using the procedures of this catalog.  As the figures show, the SDSS BALs extend the range of the \citet{gbcpgs06} study to include weaker BALs at lower maximum outflow velocities, and such sources tend to be less \mbox{X-ray} weak, i.e., they tend to have \mbox{X-ray} luminosities more similar to those of non-BAL QSOs.  We have coded the points in Figure~\ref{bALCatCIVAllVsDAOXcFig} so that black points represent sources with higher UV luminosities, having $L_{2500} \ge 1.44\times 10^{31}$~erg~s$^{-1}$.  Red points represent less UV-luminous sources.  While the less UV-luminous sources can be found with a wide range of BAL properties, they preferentially have weaker BAL features (i.e., lower values of $BI_0$ and $|EW|$) and are not as X-ray weak as the sources with higher values of $L_{2500}$.

Using Kendall's Tau test as implemented in ASURV, we have tested for correlations between \ion{C}{4} BAL properties and $\Delta\alpha_{OX}$ for the QSOs in this catalog alone, and also for the combined sample with the \citet{gbcpgs06} sources included.  In cases where a \citet{gbcpgs06} source was observed in the SDSS, we used the measurements from our current study.  The results of our correlation tests are shown in Table~\ref{bALCorrTestsTab}.  We find correlations at $>$95\% confidence between $BI_0$, $EW_0$, $v_{max}$, and $v_{min}$ for HiBALs in our sample alone; with the inclusion of the \citet{gbcpgs06} HiBAL sources, the significance rises to $>99.9$\% in some cases.  $f_{deep}$ does not show significant evidence for correlation with $\Delta\alpha_{OX}$.  While the correlation between $v_{min}$ and $\Delta\alpha_{OX}$ is formally significant at high confidence, Figure~\ref{bALCatCIVAllVsDAOXcFig} suggests that this is primarily due to a lack of high $|v_{min}|$ for relatively \mbox{X-ray} bright sources.  As Table~\ref{bALCorrTestsTab} shows, we find no significant correlations between BAL properties and the UV luminosity, $L_{\nu}(2500~\mathring{\rm A})$, alone.  However, we previously observed (in \S\ref{vDistSec}, using a much larger sample) that BAL QSOs with higher UV luminosities are more likely to have higher $|v_{max}|$ values.

Taken together, these correlations suggest that the \mbox{X-ray} absorber plays an important role in the acceleration of BAL outflows.  The depth of a BAL trough is likely determined by the fraction of the UV emission region covered by the BAL outflow at a given velocity \citep{ablgwbd99}.  This geometry may be relatively independent of the interior \mbox{X-ray} absorber, and thus it is not surprising that $f_{deep}$ is unrelated to $\Delta\alpha_{OX}$.  However, the BAL $BI$ and $EW$ are strongly dependent on the BAL width, and the correlations between $BI_0$, $EW$, $v_{max}$ and $\Delta\alpha_{OX}$ are reasonable in a scenario where the acceleration of the BAL, and therefore its velocity width, are influenced by the \mbox{X-ray} absorber which shields the BAL outflow from overionization and thereby enables efficient line-driven radiative acceleration.

\begin{figure} [ht]
  \begin{center}
      \includegraphics[width=4in, angle=270]{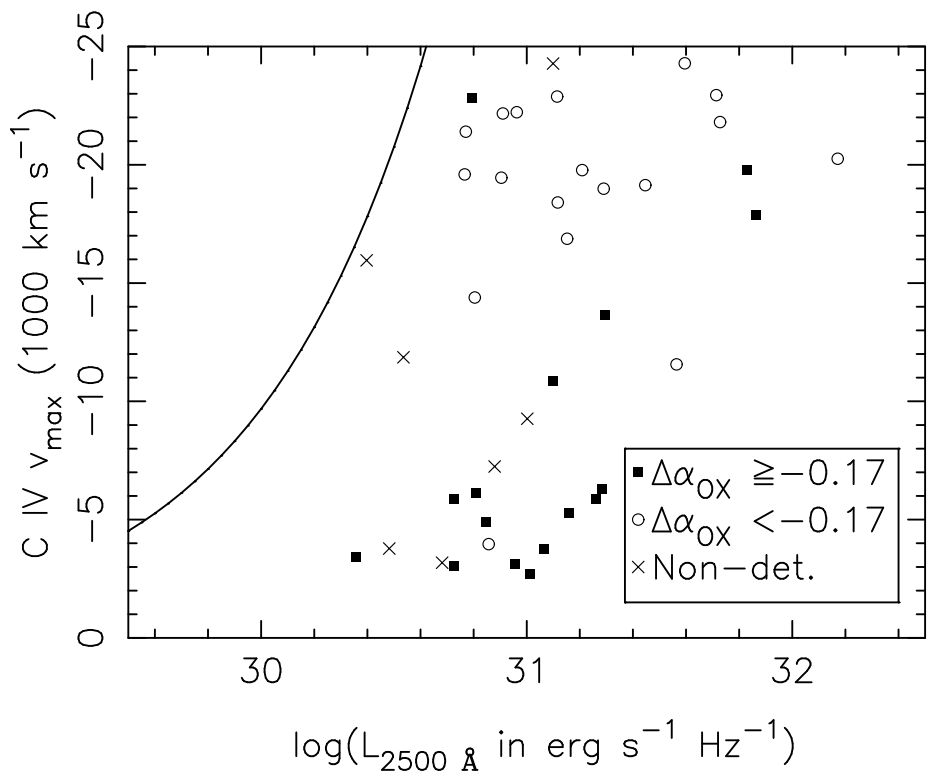}
      \caption{\label{envelopePlotFig}Plot of \ion{C}{4} $v_{max}$ against $L_{2500}$ for known HiBALs at $z \ge 1.68$.  Solid squares represent \mbox{X-ray} detected sources with $\Delta\alpha_{OX}$ above the median $\Delta\alpha_{OX}$ value.  Open circles represent \mbox{X-ray} detected sources with $\Delta\alpha_{OX}$ below the median value or undetected sources with upper limits below the median.  Crosses represent the remaining undetected sources.  The solid curve represents the ``envelope'' from Figure~7 of \citet{gbcssv07}.}
   \end{center}
\end{figure}

Recent studies have identified an upper ``envelope'' to absorber outflow velocities that is dependent on UV luminosity \citep{lb02, gbcssv07}.  This envelope may be interpreted as a terminal velocity that a QSO of a given luminosity can support for a radiatively-driven wind.  In Figure~\ref{envelopePlotFig}, we plot the \ion{C}{4} $v_{max}$ and $L_{2500}$ values for 42 QSOs that are known to be HiBALs (at $z \ge 1.68$) that also have \mbox{X-ray} data available, together with the envelope (solid line) derived by \citet{gbcssv07}.  We divide our sources into \mbox{X-ray} brighter or \mbox{X-ray} weaker subsamples according to whether their $\Delta\alpha_{OX}$ values fall above or below the median value of $-0.17$ for detected sources.  The plot shows that \mbox{X-ray} weaker sources with a range of a factor of $\approx$20 in $L_{2500}$ tend to have higher outflow velocities (as can also be seen from Figure~\ref{bALCatCIVAllVsDAOXcFig}) and generally lie closer to the envelope than do \mbox{X-ray} brighter sources.  Additional analysis is required to determine the nature of any relation between \mbox{X-ray} properties and BAL terminal velocities, but Figure~\ref{envelopePlotFig} demonstrates that the \mbox{X-ray} absorption mechanism (assumed to be the cause of \mbox{X-ray} weakness) is important to the process of BAL acceleration.

\section{CONCLUSIONS\label{concSec}}

We have presented a catalog of BAL properties for spectra of 5039 BAL QSOs in the SDSS DR5 QSO catalog and also for spectra of DR5 QSOs obtained in additional epochs that are available in the DR6 public database.  We have analyzed the UV absorption properties and also the \mbox{X-ray} properties of radio-quiet BAL QSOs that have been observed in targeted or serendipitous {\it Chandra} and/or {\it XMM-Newton} observations.

A preliminary analysis of the data in this catalog indicates that:
\begin{enumerate}
\item{BAL QSOs show stronger UV reddening than non-BAL QSOs (\S\ref{bALQSOReddeningSec}), in agreement with the findings of previous studies.  LoBAL QSOs are even redder than HiBAL QSOs, and also have a broader distribution of $R_{red}$, the ratio of flux densities at 1400 and 2500~\AA.}
\item{Estimates of the intrinsic BAL fraction are complicated by low S/N and survey selection effects.  This fraction is $\approx$16.4\% for traditional BALs, with some uncertainty due to the assumptions made to determine what fraction of BAL QSOs drop out of magnitude-limited surveys.  A preliminary analysis of a sample corrected for S/N and survey selection effects finds no dependence of the BAL fraction or distribution of $BI_0$ on UV luminosity for the relatively limited luminosity range in our sample.}
\item{BAL QSOs with higher UV luminosities are more likely to have a high $|v_{max}|$ than lower-luminosity QSOs (\S\ref{vDistSec}), as expected for a scenario in which radiation pressure accelerates BAL outflows.}
\item{While \ion{Si}{4} trough profiles are complicated by unmodeled emission at high velocities, there is some evidence that \ion{Si}{4} BAL troughs tend to appear in a narrower velocity range than those of \ion{C}{4} (\S\ref{vDistSec}).  Because \ion{Si}{4} is found at slightly lower ionization stages than \ion{C}{4}, this trend may reveal ionization-dependent structure in BAL outflows.}
\item{BAL QSOs and non-BAL SDSS QSOs both follow Baldwin effect relations between \ion{C}{4} emission and UV luminosity with similar logarithmic slopes.  However, BAL QSOs tend to have weaker \ion{C}{4} line emission than do non-BAL QSOs (\S\ref{cIVEmSec}).  This effect may be due to additional, unmodeled absorption of broad emission lines, or to the obstruction of ionizing \mbox{X-ray} photons from the BELR gas.}
\item{As previous studies have found, radio-quiet HiBAL QSOs are generally relatively \mbox{X-ray} weak compared to non-BAL QSOs, and radio-quiet LoBAL QSOs are even more \mbox{X-ray} weak than HiBAL QSOs are (\S\ref{xWeakSec}).  The \mbox{X-ray} weakness is attributed to strong \mbox{X-ray} absorption that is necessary to shield the BAL outflow from over-ionization.}
\item{In general, a larger degree of relative \mbox{X-ray} weakness is associated with stronger \ion{C}{4} BAL absorption and faster outflow velocities in radio-quiet, HiBAL QSOs (\S\ref{xWeakBALPropSec}).  These trends are consistent with a scenario in which strong \mbox{X-ray} absorption enables and enhances the effectiveness of radiation pressure to accelerate the BAL outflow.}
\end{enumerate}

In the near future, a web page will be established to track information relevant to this work.  Until that time, please contact R.~Gibson at the email address at the top of this document.  After 2008 December, R.~Gibson may be contacted at the University of Washington.

\acknowledgements
We gratefully acknowledge support from NASA LTSA grant NAG5-13035 (RRG, WNB, DPS), NASA grant NNX07AQ57G (RRG, WNB), NSF AST06--07634 (DPS, DVB), NSERC (PBH), SAO GO6-7107X (SFA), and GO5-6112X (SFA).  LJ and XF acknowledge support from NSF Grants AST 03-07384 and AST 08-06861 and a Packard Fellowship for Science and Engineering.

Funding for the SDSS and SDSS-II has been provided by the Alfred P. Sloan Foundation, the Participating Institutions, the National Science Foundation, the U.S. Department of Energy, the National Aeronautics and Space Administration, the Japanese Monbukagakusho, the Max Planck Society, and the Higher Education Funding Council for England.  The SDSS Web site \hbox{is {\tt http://www.sdss.org/}.}

The SDSS is managed by the Astrophysical Research Consortium (ARC) for the Participating Institutions.  The Participating Institutions are the American Museum of Natural History, Astrophysical Institute of Potsdam, University of Basel, Cambridge University, Case Western Reserve University, University of Chicago, Drexel University, Fermilab, the Institute for Advanced Study, the Japan Participation Group, Johns Hopkins University, the Joint Institute for Nuclear Astrophysics, the Kavli Institute for Particle Astrophysics and Cosmology, the Korean Scientist Group, the Chinese Academy of Sciences (LAMOST), Los Alamos National Laboratory, the Max-Planck-Institute for Astronomy (MPIA), the Max-Planck-Institute for Astrophysics (MPA), New Mexico State University, Ohio State University, University of Pittsburgh, University of Portsmouth, Princeton University, the United States Naval Observatory, and the University of Washington.

Most of the data analysis for this project was performed using the ISIS platform \citep{hd00}.  We thank the referee for helpful comments that have improved the quality of this work.





\bibliographystyle{apj}
\bibliography{apj-jour,bibliography}

\begin{thebibliography}{68}
\expandafter\ifx\csname natexlab\endcsname\relax\def\natexlab#1{#1}\fi

\bibitem[{{Adelman-McCarthy} {et~al.}(2007){Adelman-McCarthy}, {Ag{\"u}eros},
  {Allam}, {Anderson}, {Anderson}, {Annis}, {Bahcall}, {Bailer-Jones},
  {Baldry}, {Barentine}, {Beers}, {Belokurov}, {Berlind}, {Bernardi},
  {Blanton}, {Bochanski}, {Boroski}, {Bramich}, {Brewington}, {Brinchmann},
  {Brinkmann}, {Brunner}, {Budav{\'a}ri}, {Carey}, {Carliles}, {Carr},
  {Castander}, {Connolly}, {Cool}, {Cunha}, {Csabai}, {Dalcanton}, {Doi},
  {Eisenstein}, {Evans}, {Evans}, {Fan}, {Finkbeiner}, {Friedman}, {Frieman},
  {Fukugita}, {Gillespie}, {Gilmore}, {Glazebrook}, {Gray}, {Grebel}, {Gunn},
  {de Haas}, {Hall}, {Harvanek}, {Hawley}, {Hayes}, {Heckman}, {Hendry},
  {Hennessy}, {Hindsley}, {Hirata}, {Hogan}, {Hogg}, {Holtzman}, {Ichikawa},
  {Ichikawa}, {Ivezi{\'c}}, {Jester}, {Johnston}, {Jorgensen}, {Juri{\'c}},
  {Kauffmann}, {Kent}, {Kleinman}, {Knapp}, {Kniazev}, {Kron}, {Krzesinski},
  {Kuropatkin}, {Lamb}, {Lampeitl}, {Lee}, {Leger}, {Lima}, {Lin}, {Long},
  {Loveday}, {Lupton}, {Mandelbaum}, {Margon}, {Mart{\'{\i}}nez-Delgado},
  {Matsubara}, {McGehee}, {McKay}, {Meiksin}, {Munn}, {Nakajima}, {Nash},
  {Neilsen}, {Newberg}, {Nichol}, {Nieto-Santisteban}, {Nitta}, {Oyaizu},
  {Okamura}, {Ostriker}, {Padmanabhan}, {Park}, {Peoples}, {Pier}, {Pope},
  {Pourbaix}, {Quinn}, {Raddick}, {Re Fiorentin}, {Richards}, {Richmond},
  {Rix}, {Rockosi}, {Schlegel}, {Schneider}, {Scranton}, {Seljak}, {Sheldon},
  {Shimasaku}, {Silvestri}, {Smith}, {Smol{\v c}i{\'c}}, {Snedden}, {Stebbins},
  {Stoughton}, {Strauss}, {SubbaRao}, {Suto}, {Szalay}, {Szapudi}, {Szkody},
  {Tegmark}, {Thakar}, {Tremonti}, {Tucker}, {Uomoto}, {Vanden Berk},
  {Vandenberg}, {Vidrih}, {Vogeley}, {Voges}, {Vogt}, {Weinberg}, {West},
  {White}, {Wilhite}, {Yanny}, {Yocum}, {York}, {Zehavi}, {Zibetti}, \&
  {Zucker}}]{a-m+07}
{Adelman-McCarthy}, J.~K., {Ag{\"u}eros}, M.~A., {Allam}, S.~S., {Anderson},
  K.~S.~J., {Anderson}, S.~F., {Annis}, J., {Bahcall}, N.~A., {Bailer-Jones},
  C.~A.~L., {Baldry}, I.~K., {Barentine}, J.~C., {Beers}, T.~C., {Belokurov},
  V., {Berlind}, A., {Bernardi}, M., {Blanton}, M.~R., {Bochanski}, J.~J.,
  {Boroski}, W.~N., {Bramich}, D.~M., {Brewington}, H.~J., {Brinchmann}, J.,
  {Brinkmann}, J., {Brunner}, R.~J., {Budav{\'a}ri}, T., {Carey}, L.~N.,
  {Carliles}, S., {Carr}, M.~A., {Castander}, F.~J., {Connolly}, A.~J., {Cool},
  R.~J., {Cunha}, C.~E., {Csabai}, I., {Dalcanton}, J.~J., {Doi}, M.,
  {Eisenstein}, D.~J., {Evans}, M.~L., {Evans}, N.~W., {Fan}, X., {Finkbeiner},
  D.~P., {Friedman}, S.~D., {Frieman}, J.~A., {Fukugita}, M., {Gillespie}, B.,
  {Gilmore}, G., {Glazebrook}, K., {Gray}, J., {Grebel}, E.~K., {Gunn}, J.~E.,
  {de Haas}, E., {Hall}, P.~B., {Harvanek}, M., {Hawley}, S.~L., {Hayes}, J.,
  {Heckman}, T.~M., {Hendry}, J.~S., {Hennessy}, G.~S., {Hindsley}, R.~B.,
  {Hirata}, C.~M., {Hogan}, C.~J., {Hogg}, D.~W., {Holtzman}, J.~A.,
  {Ichikawa}, S.-i., {Ichikawa}, T., {Ivezi{\'c}}, {\v Z}., {Jester}, S.,
  {Johnston}, D.~E., {Jorgensen}, A.~M., {Juri{\'c}}, M., {Kauffmann}, G.,
  {Kent}, S.~M., {Kleinman}, S.~J., {Knapp}, G.~R., {Kniazev}, A.~Y., {Kron},
  R.~G., {Krzesinski}, J., {Kuropatkin}, N., {Lamb}, D.~Q., {Lampeitl}, H.,
  {Lee}, B.~C., {Leger}, R.~F., {Lima}, M., {Lin}, H., {Long}, D.~C.,
  {Loveday}, J., {Lupton}, R.~H., {Mandelbaum}, R., {Margon}, B.,
  {Mart{\'{\i}}nez-Delgado}, D., {Matsubara}, T., {McGehee}, P.~M., {McKay},
  T.~A., {Meiksin}, A., {Munn}, J.~A., {Nakajima}, R., {Nash}, T., {Neilsen},
  Jr., E.~H., {Newberg}, H.~J., {Nichol}, R.~C., {Nieto-Santisteban}, M.,
  {Nitta}, A., {Oyaizu}, H., {Okamura}, S., {Ostriker}, J.~P., {Padmanabhan},
  N., {Park}, C., {Peoples}, J.~J., {Pier}, J.~R., {Pope}, A.~C., {Pourbaix},
  D., {Quinn}, T.~R., {Raddick}, M.~J., {Re Fiorentin}, P., {Richards}, G.~T.,
  {Richmond}, M.~W., {Rix}, H.-W., {Rockosi}, C.~M., {Schlegel}, D.~J.,
  {Schneider}, D.~P., {Scranton}, R., {Seljak}, U., {Sheldon}, E., {Shimasaku},
  K., {Silvestri}, N.~M., {Smith}, J.~A., {Smol{\v c}i{\'c}}, V., {Snedden},
  S.~A., {Stebbins}, A., {Stoughton}, C., {Strauss}, M.~A., {SubbaRao}, M.,
  {Suto}, Y., {Szalay}, A.~S., {Szapudi}, I., {Szkody}, P., {Tegmark}, M.,
  {Thakar}, A.~R., {Tremonti}, C.~A., {Tucker}, D.~L., {Uomoto}, A., {Vanden
  Berk}, D.~E., {Vandenberg}, J., {Vidrih}, S., {Vogeley}, M.~S., {Voges}, W.,
  {Vogt}, N.~P., {Weinberg}, D.~H., {West}, A.~A., {White}, S.~D.~M.,
  {Wilhite}, B., {Yanny}, B., {Yocum}, D.~R., {York}, D.~G., {Zehavi}, I.,
  {Zibetti}, S., \& {Zucker}, D.~B. 2007, \apjs, 172, 634

\bibitem[{{Arav} {et~al.}(1999{\natexlab{a}}){Arav}, {Becker},
  {Laurent-Muehleisen}, {Gregg}, {White}, {Brotherton}, \& {de
  Kool}}]{ablgwbd99}
{Arav}, N., {Becker}, R.~H., {Laurent-Muehleisen}, S.~A., {Gregg}, M.~D.,
  {White}, R.~L., {Brotherton}, M.~S., \& {de Kool}, M. 1999{\natexlab{a}},
  \apj, 524, 566

\bibitem[{{Arav} {et~al.}(1999{\natexlab{b}}){Arav}, {Korista}, {de Kool},
  {Junkkarinen}, \& {Begelman}}]{akdjb99}
{Arav}, N., {Korista}, K.~T., {de Kool}, M., {Junkkarinen}, V.~T., \&
  {Begelman}, M.~C. 1999{\natexlab{b}}, \apj, 516, 27

\bibitem[{{Avni} \& {Tananbaum}(1982)}]{at82}
{Avni}, Y. \& {Tananbaum}, H. 1982, \apjl, 262, L17

\bibitem[{{Baldwin}(1977)}]{b77}
{Baldwin}, J.~A. 1977, \apj, 214, 679

\bibitem[{{Barlow}(1993)}]{b93}
{Barlow}, T.~A. 1993, PhD thesis, AA(California Univ.)

\bibitem[{{Becker} {et~al.}(2000){Becker}, {White}, {Gregg}, {Brotherton},
  {Laurent-Muehleisen}, \& {Arav}}]{bwgbla00}
{Becker}, R.~H., {White}, R.~L., {Gregg}, M.~D., {Brotherton}, M.~S.,
  {Laurent-Muehleisen}, S.~A., \& {Arav}, N. 2000, \apj, 538, 72

\bibitem[{{Becker} {et~al.}(1995){Becker}, {White}, \& {Helfand}}]{bwh95}
{Becker}, R.~H., {White}, R.~L., \& {Helfand}, D.~J. 1995, \apj, 450, 559

\bibitem[{{Brandt} {et~al.}(2000){Brandt}, {Laor}, \& {Wills}}]{blw00}
{Brandt}, W.~N., {Laor}, A., \& {Wills}, B.~J. 2000, \apj, 528, 637

\bibitem[{{Brinkmann} {et~al.}(2000){Brinkmann}, {Laurent-Muehleisen}, {Voges},
  {Siebert}, {Becker}, {Brotherton}, {White}, \& {Gregg}}]{blvsbbwg00}
{Brinkmann}, W., {Laurent-Muehleisen}, S.~A., {Voges}, W., {Siebert}, J.,
  {Becker}, R.~H., {Brotherton}, M.~S., {White}, R.~L., \& {Gregg}, M.~D. 2000,
  \aap, 356, 445

\bibitem[{{Brotherton} {et~al.}(2001){Brotherton}, {Tran}, {Becker}, {Gregg},
  {Laurent-Muehleisen}, \& {White}}]{btbglw01}
{Brotherton}, M.~S., {Tran}, H.~D., {Becker}, R.~H., {Gregg}, M.~D.,
  {Laurent-Muehleisen}, S.~A., \& {White}, R.~L. 2001, \apj, 546, 775

\bibitem[{{Cardelli} {et~al.}(1989){Cardelli}, {Clayton}, \& {Mathis}}]{ccm89}
{Cardelli}, J.~A., {Clayton}, G.~C., \& {Mathis}, J.~S. 1989, \apj, 345, 245

\bibitem[{{Condon} {et~al.}(1998){Condon}, {Cotton}, {Greisen}, {Yin},
  {Perley}, {Taylor}, \& {Broderick}}]{ccgyptb98}
{Condon}, J.~J., {Cotton}, W.~D., {Greisen}, E.~W., {Yin}, Q.~F., {Perley},
  R.~A., {Taylor}, G.~B., \& {Broderick}, J.~J. 1998, \aj, 115, 1693

\bibitem[{{Dai} {et~al.}(2008){Dai}, {Shankar}, \& {Sivakoff}}]{dss08}
{Dai}, X., {Shankar}, F., \& {Sivakoff}, G.~R. 2008, \apj, 672, 108

\bibitem[{{Fukugita} {et~al.}(1996){Fukugita}, {Ichikawa}, {Gunn}, {Doi},
  {Shimasaku}, \& {Schneider}}]{figdss96}
{Fukugita}, M., {Ichikawa}, T., {Gunn}, J.~E., {Doi}, M., {Shimasaku}, K., \&
  {Schneider}, D.~P. 1996, \aj, 111, 1748

\bibitem[{{Gallagher} {et~al.}(2002){Gallagher}, {Brandt}, {Chartas}, \&
  {Garmire}}]{gbcg02}
{Gallagher}, S.~C., {Brandt}, W.~N., {Chartas}, G., \& {Garmire}, G.~P. 2002,
  \apj, 567, 37

\bibitem[{{Gallagher} {et~al.}(2006){Gallagher}, {Brandt}, {Chartas},
  {Priddey}, {Garmire}, \& {Sambruna}}]{gbcpgs06}
{Gallagher}, S.~C., {Brandt}, W.~N., {Chartas}, G., {Priddey}, R., {Garmire},
  G.~P., \& {Sambruna}, R.~M. 2006, \apj, 644, 709

\bibitem[{{Gallagher} {et~al.}(2001){Gallagher}, {Brandt}, {Laor}, {Elvis},
  {Mathur}, {Wills}, \& {Iyomoto}}]{gblemwi01}
{Gallagher}, S.~C., {Brandt}, W.~N., {Laor}, A., {Elvis}, M., {Mathur}, S.,
  {Wills}, B.~J., \& {Iyomoto}, N. 2001, \apj, 546, 795

\bibitem[{{Ganguly} {et~al.}(2007){Ganguly}, {Brotherton}, {Cales}, {Scoggins},
  {Shang}, \& {Vestergaard}}]{gbcssv07}
{Ganguly}, R., {Brotherton}, M.~S., {Cales}, S., {Scoggins}, B., {Shang}, Z.,
  \& {Vestergaard}, M. 2007, \apj, 665, 990

\bibitem[{{Gibson} {et~al.}(2008{\natexlab{a}}){Gibson}, {Brandt}, \&
  {Schneider}}]{gbs08}
{Gibson}, R.~R., {Brandt}, W.~N., \& {Schneider}, D.~P. 2008{\natexlab{a}},
  \apj, 685, 773

\bibitem[{{Gibson} {et~al.}(2008{\natexlab{b}}){Gibson}, {Brandt}, {Schneider},
  \& {Gallagher}}]{gbsg08}
{Gibson}, R.~R., {Brandt}, W.~N., {Schneider}, D.~P., \& {Gallagher}, S.~C.
  2008{\natexlab{b}}, \apj, 675, 985

\bibitem[{{Green} {et~al.}(2001){Green}, {Aldcroft}, {Mathur}, {Wilkes}, \&
  {Elvis}}]{gamwe01}
{Green}, P.~J., {Aldcroft}, T.~L., {Mathur}, S., {Wilkes}, B.~J., \& {Elvis},
  M. 2001, \apj, 558, 109

\bibitem[{{Green} {et~al.}(1995){Green}, {Schartel}, {Anderson}, {Hewett},
  {Foltz}, {Brinkmann}, {Fink}, {Truemper}, \& {Margon}}]{gsahfbftm95}
{Green}, P.~J., {Schartel}, N., {Anderson}, S.~F., {Hewett}, P.~C., {Foltz},
  C.~B., {Brinkmann}, W., {Fink}, H., {Truemper}, J., \& {Margon}, B. 1995,
  \apj, 450, 51

\bibitem[{{Gunn} {et~al.}(1998){Gunn}, {Carr}, {Rockosi}, {Sekiguchi}, {Berry},
  {Elms}, {de Haas}, {Ivezi{\'c}}, {Knapp}, {Lupton}, {Pauls}, {Simcoe},
  {Hirsch}, {Sanford}, {Wang}, {York}, {Harris}, {Annis}, {Bartozek},
  {Boroski}, {Bakken}, {Haldeman}, {Kent}, {Holm}, {Holmgren}, {Petravick},
  {Prosapio}, {Rechenmacher}, {Doi}, {Fukugita}, {Shimasaku}, {Okada}, {Hull},
  {Siegmund}, {Mannery}, {Blouke}, {Heidtman}, {Schneider}, {Lucinio}, \&
  {Brinkman}}]{gcrsbehiklpshswyhabbbhkhhpprdfsohsmbhslb98}
{Gunn}, J.~E., {Carr}, M., {Rockosi}, C., {Sekiguchi}, M., {Berry}, K., {Elms},
  B., {de Haas}, E., {Ivezi{\'c}}, {\v Z}., {Knapp}, G., {Lupton}, R., {Pauls},
  G., {Simcoe}, R., {Hirsch}, R., {Sanford}, D., {Wang}, S., {York}, D.,
  {Harris}, F., {Annis}, J., {Bartozek}, L., {Boroski}, W., {Bakken}, J.,
  {Haldeman}, M., {Kent}, S., {Holm}, S., {Holmgren}, D., {Petravick}, D.,
  {Prosapio}, A., {Rechenmacher}, R., {Doi}, M., {Fukugita}, M., {Shimasaku},
  K., {Okada}, N., {Hull}, C., {Siegmund}, W., {Mannery}, E., {Blouke}, M.,
  {Heidtman}, D., {Schneider}, D., {Lucinio}, R., \& {Brinkman}, J. 1998, \aj,
  116, 3040

\bibitem[{{Gunn} {et~al.}(2006){Gunn}, {Siegmund}, {Mannery}, {Owen}, {Hull},
  {Leger}, {Carey}, {Knapp}, {York}, {Boroski}, {Kent}, {Lupton}, {Rockosi},
  {Evans}, {Waddell}, {Anderson}, {Annis}, {Barentine}, {Bartoszek}, {Bastian},
  {Bracker}, {Brewington}, {Briegel}, {Brinkmann}, {Brown}, {Carr},
  {Czarapata}, {Drennan}, {Dombeck}, {Federwitz}, {Gillespie}, {Gonzales},
  {Hansen}, {Harvanek}, {Hayes}, {Jordan}, {Kinney}, {Klaene}, {Kleinman},
  {Kron}, {Kresinski}, {Lee}, {Limmongkol}, {Lindenmeyer}, {Long}, {Loomis},
  {McGehee}, {Mantsch}, {Neilsen}, {Neswold}, {Newman}, {Nitta}, {Peoples},
  {Pier}, {Prieto}, {Prosapio}, {Rivetta}, {Schneider}, {Snedden}, \&
  {Wang}}]{gunn+06}
{Gunn}, J.~E., {Siegmund}, W.~A., {Mannery}, E.~J., {Owen}, R.~E., {Hull},
  C.~L., {Leger}, R.~F., {Carey}, L.~N., {Knapp}, G.~R., {York}, D.~G.,
  {Boroski}, W.~N., {Kent}, S.~M., {Lupton}, R.~H., {Rockosi}, C.~M., {Evans},
  M.~L., {Waddell}, P., {Anderson}, J.~E., {Annis}, J., {Barentine}, J.~C.,
  {Bartoszek}, L.~M., {Bastian}, S., {Bracker}, S.~B., {Brewington}, H.~J.,
  {Briegel}, C.~I., {Brinkmann}, J., {Brown}, Y.~J., {Carr}, M.~A.,
  {Czarapata}, P.~C., {Drennan}, C.~C., {Dombeck}, T., {Federwitz}, G.~R.,
  {Gillespie}, B.~A., {Gonzales}, C., {Hansen}, S.~U., {Harvanek}, M., {Hayes},
  J., {Jordan}, W., {Kinney}, E., {Klaene}, M., {Kleinman}, S.~J., {Kron},
  R.~G., {Kresinski}, J., {Lee}, G., {Limmongkol}, S., {Lindenmeyer}, C.~W.,
  {Long}, D.~C., {Loomis}, C.~L., {McGehee}, P.~M., {Mantsch}, P.~M.,
  {Neilsen}, Jr., E.~H., {Neswold}, R.~M., {Newman}, P.~R., {Nitta}, A.,
  {Peoples}, J.~J., {Pier}, J.~R., {Prieto}, P.~S., {Prosapio}, A., {Rivetta},
  C., {Schneider}, D.~P., {Snedden}, S., \& {Wang}, S.-i. 2006, \aj, 131, 2332

\bibitem[{{Hall} {et~al.}(2002){Hall}, {Anderson}, {Strauss}, {York},
  {Richards}, {Fan}, {Knapp}, {Schneider}, {Vanden Berk}, {Geballe}, {Bauer},
  {Becker}, {Davis}, {Rix}, {Nichol}, {Bahcall}, {Brinkmann}, {Brunner},
  {Connolly}, {Csabai}, {Doi}, {Fukugita}, {Gunn}, {Haiman}, {Harvanek},
  {Heckman}, {Hennessy}, {Inada}, {Ivezi{\'c}}, {Johnston}, {Kleinman},
  {Krolik}, {Krzesinski}, {Kunszt}, {Lamb}, {Long}, {Lupton}, {Miknaitis},
  {Munn}, {Narayanan}, {Neilsen}, {Newman}, {Nitta}, {Okamura}, {Pentericci},
  {Pier}, {Schlegel}, {Snedden}, {Szalay}, {Thakar}, {Tsvetanov}, {White}, \&
  {Zheng}}]{h+02}
{Hall}, P.~B., {Anderson}, S.~F., {Strauss}, M.~A., {York}, D.~G., {Richards},
  G.~T., {Fan}, X., {Knapp}, G.~R., {Schneider}, D.~P., {Vanden Berk}, D.~E.,
  {Geballe}, T.~R., {Bauer}, A.~E., {Becker}, R.~H., {Davis}, M., {Rix}, H.-W.,
  {Nichol}, R.~C., {Bahcall}, N.~A., {Brinkmann}, J., {Brunner}, R.,
  {Connolly}, A.~J., {Csabai}, I., {Doi}, M., {Fukugita}, M., {Gunn}, J.~E.,
  {Haiman}, Z., {Harvanek}, M., {Heckman}, T.~M., {Hennessy}, G.~S., {Inada},
  N., {Ivezi{\'c}}, {\v Z}., {Johnston}, D., {Kleinman}, S., {Krolik}, J.~H.,
  {Krzesinski}, J., {Kunszt}, P.~Z., {Lamb}, D.~Q., {Long}, D.~C., {Lupton},
  R.~H., {Miknaitis}, G., {Munn}, J.~A., {Narayanan}, V.~K., {Neilsen}, E.,
  {Newman}, P.~R., {Nitta}, A., {Okamura}, S., {Pentericci}, L., {Pier}, J.~R.,
  {Schlegel}, D.~J., {Snedden}, S., {Szalay}, A.~S., {Thakar}, A.~R.,
  {Tsvetanov}, Z., {White}, R.~L., \& {Zheng}, W. 2002, \apjs, 141, 267

\bibitem[{{Hewett} \& {Foltz}(2003)}]{hf03}
{Hewett}, P.~C. \& {Foltz}, C.~B. 2003, \aj, 125, 1784

\bibitem[{{Hewett} {et~al.}(1995){Hewett}, {Foltz}, \& {Chaffee}}]{hfc95}
{Hewett}, P.~C., {Foltz}, C.~B., \& {Chaffee}, F.~H. 1995, \aj, 109, 1498

\bibitem[{{Houck} \& {Denicola}(2000)}]{hd00}
{Houck}, J.~C. \& {Denicola}, L.~A. 2000, in ASP Conf. Ser. 216: Astronomical
  Data Analysis Software and Systems IX, ed. N.~{Manset}, C.~{Veillet}, \&
  D.~{Crabtree}, 591--+

\bibitem[{{Isobe} \& {Feigelson}(1990)}]{if90}
{Isobe}, T. \& {Feigelson}, E.~D. 1990, in Bulletin of the American
  Astronomical Society, Vol.~22, Bulletin of the American Astronomical Society,
  917--918

\bibitem[{{Ivezi{\'c}} {et~al.}(2004){Ivezi{\'c}}, {Lupton}, {Schlegel},
  {Boroski}, {Adelman-McCarthy}, {Yanny}, {Kent}, {Stoughton}, {Finkbeiner},
  {Padmanabhan}, {Rockosi}, {Gunn}, {Knapp}, {Strauss}, {Richards},
  {Eisenstein}, {Nicinski}, {Kleinman}, {Krzesinski}, {Newman}, {Snedden},
  {Thakar}, {Szalay}, {Munn}, {Smith}, {Tucker}, \&
  {Lee}}]{ilsbayksfprgksrenkknstsmstl04}
{Ivezi{\'c}}, {\v Z}., {Lupton}, R.~H., {Schlegel}, D., {Boroski}, B.,
  {Adelman-McCarthy}, J., {Yanny}, B., {Kent}, S., {Stoughton}, C.,
  {Finkbeiner}, D., {Padmanabhan}, N., {Rockosi}, C.~M., {Gunn}, J.~E.,
  {Knapp}, G.~R., {Strauss}, M.~A., {Richards}, G.~T., {Eisenstein}, D.,
  {Nicinski}, T., {Kleinman}, S.~J., {Krzesinski}, J., {Newman}, P.~R.,
  {Snedden}, S., {Thakar}, A.~R., {Szalay}, A., {Munn}, J.~A., {Smith}, J.~A.,
  {Tucker}, D., \& {Lee}, B.~C. 2004, Astronomische Nachrichten, 325, 583

\bibitem[{{Just} {et~al.}(2007){Just}, {Brandt}, {Shemmer}, {Steffen},
  {Schneider}, {Chartas}, \& {Garmire}}]{jbssscg07}
{Just}, D.~W., {Brandt}, W.~N., {Shemmer}, O., {Steffen}, A.~T., {Schneider},
  D.~P., {Chartas}, G., \& {Garmire}, G.~P. 2007, \apj, 665, 1004

\bibitem[{{Kellermann} {et~al.}(1989){Kellermann}, {Sramek}, {Schmidt},
  {Shaffer}, \& {Green}}]{ksssg89}
{Kellermann}, K.~I., {Sramek}, R., {Schmidt}, M., {Shaffer}, D.~B., \& {Green},
  R. 1989, \aj, 98, 1195

\bibitem[{{Kinney} {et~al.}(1990){Kinney}, {Rivolo}, \& {Koratkar}}]{krk90}
{Kinney}, A.~L., {Rivolo}, A.~R., \& {Koratkar}, A.~P. 1990, \apj, 357, 338

\bibitem[{{Knigge} {et~al.}(2008){Knigge}, {Scaringi}, {Goad}, \&
  {Cottis}}]{ksgc08}
{Knigge}, C., {Scaringi}, S., {Goad}, M.~R., \& {Cottis}, C.~E. 2008, ArXiv
  e-prints, 802

\bibitem[{{Krolik} \& {Voit}(1998)}]{kv98}
{Krolik}, J.~H. \& {Voit}, G.~M. 1998, \apjl, 497, L5+

\bibitem[{{Laor} \& {Brandt}(2002)}]{lb02}
{Laor}, A. \& {Brandt}, W.~N. 2002, \apj, 569, 641

\bibitem[{{Lavalley} {et~al.}(1992){Lavalley}, {Isobe}, \& {Feigelson}}]{lif92}
{Lavalley}, M., {Isobe}, T., \& {Feigelson}, E. 1992, in ASP Conf. Ser. 25:
  Astronomical Data Analysis Software and Systems I, ed. D.~M. {Worrall},
  C.~{Biemesderfer}, \& J.~{Barnes}, 245--+

\bibitem[{{Lundgren} {et~al.}(2007){Lundgren}, {Wilhite}, {Brunner}, {Hall},
  {Schneider}, {York}, {Vanden Berk}, \& {Brinkmann}}]{lwbhsyvb07}
{Lundgren}, B.~F., {Wilhite}, B.~C., {Brunner}, R.~J., {Hall}, P.~B.,
  {Schneider}, D.~P., {York}, D.~G., {Vanden Berk}, D.~E., \& {Brinkmann}, J.
  2007, \apj, 656, 73

\bibitem[{{Lupton} {et~al.}(1999){Lupton}, {Gunn}, \& {Szalay}}]{lgs99}
{Lupton}, R.~H., {Gunn}, J.~E., \& {Szalay}, A.~S. 1999, \aj, 118, 1406

\bibitem[{{Murray} {et~al.}(1995){Murray}, {Chiang}, {Grossman}, \&
  {Voit}}]{mcgv95}
{Murray}, N., {Chiang}, J., {Grossman}, S.~A., \& {Voit}, G.~M. 1995, \apj,
  451, 498

\bibitem[{{O'Donnell}(1994)}]{o94}
{O'Donnell}, J.~E. 1994, \apj, 422, 158

\bibitem[{{Pei}(1992)}]{p92}
{Pei}, Y.~C. 1992, \apj, 395, 130

\bibitem[{{Pier} {et~al.}(2003){Pier}, {Munn}, {Hindsley}, {Hennessy}, {Kent},
  {Lupton}, \& {Ivezi{\'c}}}]{pmhkli03}
{Pier}, J.~R., {Munn}, J.~A., {Hindsley}, R.~B., {Hennessy}, G.~S., {Kent},
  S.~M., {Lupton}, R.~H., \& {Ivezi{\'c}}, {\v Z}. 2003, \aj, 125, 1559

\bibitem[{{Reichard} {et~al.}(2003){Reichard}, {Richards}, {Hall}, {Schneider},
  {Vanden Berk}, {Fan}, {York}, {Knapp}, \& {Brinkmann}}]{rrhsvfykb03}
{Reichard}, T.~A., {Richards}, G.~T., {Hall}, P.~B., {Schneider}, D.~P.,
  {Vanden Berk}, D.~E., {Fan}, X., {York}, D.~G., {Knapp}, G.~R., \&
  {Brinkmann}, J. 2003, \aj, 126, 2594

\bibitem[{{Richards}(2006)}]{r06}
{Richards}, G.~T. 2006, ArXiv Astrophysics e-prints

\bibitem[{{Richards} {et~al.}(2002){Richards}, {Fan}, {Newberg}, {Strauss},
  {Vanden Berk}, {Schneider}, {Yanny}, {Boucher}, {Burles}, {Frieman}, {Gunn},
  {Hall}, {Ivezi{\'c}}, {Kent}, {Loveday}, {Lupton}, {Rockosi}, {Schlegel},
  {Stoughton}, {SubbaRao}, \& {York}}]{rfnsvsybbfghikllrsssy02}
{Richards}, G.~T., {Fan}, X., {Newberg}, H.~J., {Strauss}, M.~A., {Vanden
  Berk}, D.~E., {Schneider}, D.~P., {Yanny}, B., {Boucher}, A., {Burles}, S.,
  {Frieman}, J.~A., {Gunn}, J.~E., {Hall}, P.~B., {Ivezi{\'c}}, {\v Z}.,
  {Kent}, S., {Loveday}, J., {Lupton}, R.~H., {Rockosi}, C.~M., {Schlegel},
  D.~J., {Stoughton}, C., {SubbaRao}, M., \& {York}, D.~G. 2002, \aj, 123, 2945

\bibitem[{{Schlegel} {et~al.}(1998){Schlegel}, {Finkbeiner}, \&
  {Davis}}]{sfd98}
{Schlegel}, D.~J., {Finkbeiner}, D.~P., \& {Davis}, M. 1998, \apj, 500, 525

\bibitem[{{Schmidt} \& {Green}(1983)}]{sg83}
{Schmidt}, M. \& {Green}, R.~F. 1983, \apj, 269, 352

\bibitem[{{Schneider} {et~al.}(2007){Schneider}, {Hall}, {Richards}, {Strauss},
  {Vanden Berk}, {Anderson}, {Brandt}, {Fan}, {Jester}, {Gray}, {Gunn},
  {SubbaRao}, {Thakar}, {Stoughton}, {Szalay}, {Yanny}, {York}, {Bahcall},
  {Barentine}, {Blanton}, {Brewington}, {Brinkmann}, {Brunner}, {Castander},
  {Csabai}, {Frieman}, {Fukugita}, {Harvanek}, {Hogg}, {Ivezi{\'c}}, {Kent},
  {Kleinman}, {Knapp}, {Kron}, {Krzesi{\'n}ski}, {Long}, {Lupton}, {Nitta},
  {Pier}, {Saxe}, {Shen}, {Snedden}, {Weinberg}, \& {Wu}}]{s+07}
{Schneider}, D.~P., {Hall}, P.~B., {Richards}, G.~T., {Strauss}, M.~A., {Vanden
  Berk}, D.~E., {Anderson}, S.~F., {Brandt}, W.~N., {Fan}, X., {Jester}, S.,
  {Gray}, J., {Gunn}, J.~E., {SubbaRao}, M.~U., {Thakar}, A.~R., {Stoughton},
  C., {Szalay}, A.~S., {Yanny}, B., {York}, D.~G., {Bahcall}, N.~A.,
  {Barentine}, J., {Blanton}, M.~R., {Brewington}, H., {Brinkmann}, J.,
  {Brunner}, R.~J., {Castander}, F.~J., {Csabai}, I., {Frieman}, J.~A.,
  {Fukugita}, M., {Harvanek}, M., {Hogg}, D.~W., {Ivezi{\'c}}, {\v Z}., {Kent},
  S.~M., {Kleinman}, S.~J., {Knapp}, G.~R., {Kron}, R.~G., {Krzesi{\'n}ski},
  J., {Long}, D.~C., {Lupton}, R.~H., {Nitta}, A., {Pier}, J.~R., {Saxe},
  D.~H., {Shen}, Y., {Snedden}, S.~A., {Weinberg}, D.~H., \& {Wu}, J. 2007,
  \aj, 134, 102

\bibitem[{{Schneider} {et~al.}(2005){Schneider}, {Hall}, {Richards}, {Vanden
  Berk}, {Anderson}, {Fan}, {Jester}, {Stoughton}, {Strauss}, {SubbaRao},
  {Brandt}, {Gunn}, {Yanny}, {Bahcall}, {Barentine}, {Blanton}, {Boroski},
  {Brewington}, {Brinkmann}, {Brunner}, {Csabai}, {Doi}, {Eisenstein},
  {Frieman}, {Fukugita}, {Gray}, {Harvanek}, {Heckman}, {Ivezi{\'c}}, {Kent},
  {Kleinman}, {Knapp}, {Kron}, {Krzesinski}, {Long}, {Loveday}, {Lupton},
  {Margon}, {Munn}, {Neilsen}, {Newberg}, {Newman}, {Nichol}, {Nitta}, {Pier},
  {Rockosi}, {Saxe}, {Schlegel}, {Snedden}, {Szalay}, {Thakar}, {Uomoto},
  {Voges}, \& {York}}]{s+05}
{Schneider}, D.~P., {Hall}, P.~B., {Richards}, G.~T., {Vanden Berk}, D.~E.,
  {Anderson}, S.~F., {Fan}, X., {Jester}, S., {Stoughton}, C., {Strauss},
  M.~A., {SubbaRao}, M., {Brandt}, W.~N., {Gunn}, J.~E., {Yanny}, B.,
  {Bahcall}, N.~A., {Barentine}, J.~C., {Blanton}, M.~R., {Boroski}, W.~N.,
  {Brewington}, H.~J., {Brinkmann}, J., {Brunner}, R., {Csabai}, I., {Doi}, M.,
  {Eisenstein}, D.~J., {Frieman}, J.~A., {Fukugita}, M., {Gray}, J.,
  {Harvanek}, M., {Heckman}, T.~M., {Ivezi{\'c}}, {\v Z}., {Kent}, S.,
  {Kleinman}, S.~J., {Knapp}, G.~R., {Kron}, R.~G., {Krzesinski}, J., {Long},
  D.~C., {Loveday}, J., {Lupton}, R.~H., {Margon}, B., {Munn}, J.~A.,
  {Neilsen}, E.~H., {Newberg}, H.~J., {Newman}, P.~R., {Nichol}, R.~C.,
  {Nitta}, A., {Pier}, J.~R., {Rockosi}, C.~M., {Saxe}, D.~H., {Schlegel},
  D.~J., {Snedden}, S.~A., {Szalay}, A.~S., {Thakar}, A.~R., {Uomoto}, A.,
  {Voges}, W., \& {York}, D.~G. 2005, \aj, 130, 367

\bibitem[{{Smith} {et~al.}(2002){Smith}, {Tucker}, {Kent}, {Richmond},
  {Fukugita}, {Ichikawa}, {Ichikawa}, {Jorgensen}, {Uomoto}, {Gunn}, {Hamabe},
  {Watanabe}, {Tolea}, {Henden}, {Annis}, {Pier}, {McKay}, {Brinkmann}, {Chen},
  {Holtzman}, {Shimasaku}, \& {York}}]{stkrfiijughwthapmbchsy02}
{Smith}, J.~A., {Tucker}, D.~L., {Kent}, S., {Richmond}, M.~W., {Fukugita}, M.,
  {Ichikawa}, T., {Ichikawa}, S.-i., {Jorgensen}, A.~M., {Uomoto}, A., {Gunn},
  J.~E., {Hamabe}, M., {Watanabe}, M., {Tolea}, A., {Henden}, A., {Annis}, J.,
  {Pier}, J.~R., {McKay}, T.~A., {Brinkmann}, J., {Chen}, B., {Holtzman}, J.,
  {Shimasaku}, K., \& {York}, D.~G. 2002, \aj, 123, 2121

\bibitem[{{Sprayberry} \& {Foltz}(1992)}]{sf92}
{Sprayberry}, D. \& {Foltz}, C.~B. 1992, \apj, 390, 39

\bibitem[{{Sramek} \& {Weedman}(1980)}]{sw80}
{Sramek}, R.~A. \& {Weedman}, D.~W. 1980, \apj, 238, 435

\bibitem[{{Steffen} {et~al.}(2006){Steffen}, {Strateva}, {Brandt}, {Alexander},
  {Koekemoer}, {Lehmer}, {Schneider}, \& {Vignali}}]{ssbaklsv06}
{Steffen}, A.~T., {Strateva}, I., {Brandt}, W.~N., {Alexander}, D.~M.,
  {Koekemoer}, A.~M., {Lehmer}, B.~D., {Schneider}, D.~P., \& {Vignali}, C.
  2006, \aj, 131, 2826

\bibitem[{{Stoughton} {et~al.}(2002){Stoughton}, {Lupton}, {Bernardi},
  {Blanton}, {Burles}, {Castander}, {Connolly}, {Eisenstein}, {Frieman},
  {Hennessy}, {Hindsley}, {Ivezi{\'c}}, {Kent}, {Kunszt}, {Lee}, {Meiksin},
  {Munn}, {Newberg}, {Nichol}, {Nicinski}, {Pier}, {Richards}, {Richmond},
  {Schlegel}, {Smith}, {Strauss}, {SubbaRao}, {Szalay}, {Thakar}, {Tucker},
  {Vanden Berk}, {Yanny}, {Adelman}, {Anderson}, {Anderson}, {Annis},
  {Bahcall}, {Bakken}, {Bartelmann}, {Bastian}, {Bauer}, {Berman},
  {B{\"o}hringer}, {Boroski}, {Bracker}, {Briegel}, {Briggs}, {Brinkmann},
  {Brunner}, {Carey}, {Carr}, {Chen}, {Christian}, {Colestock}, {Crocker},
  {Csabai}, {Czarapata}, {Dalcanton}, {Davidsen}, {Davis}, {Dehnen},
  {Dodelson}, {Doi}, {Dombeck}, {Donahue}, {Ellman}, {Elms}, {Evans}, {Eyer},
  {Fan}, {Federwitz}, {Friedman}, {Fukugita}, {Gal}, {Gillespie}, {Glazebrook},
  {Gray}, {Grebel}, {Greenawalt}, {Greene}, {Gunn}, {de Haas}, {Haiman},
  {Haldeman}, {Hall}, {Hamabe}, {Hansen}, {Harris}, {Harris}, {Harvanek},
  {Hawley}, {Hayes}, {Heckman}, {Helmi}, {Henden}, {Hogan}, {Hogg}, {Holmgren},
  {Holtzman}, {Huang}, {Hull}, {Ichikawa}, {Ichikawa}, {Johnston}, {Kauffmann},
  {Kim}, {Kimball}, {Kinney}, {Klaene}, {Kleinman}, {Klypin}, {Knapp},
  {Korienek}, {Krolik}, {Kron}, {Krzesi{\'n}ski}, {Lamb}, {Leger},
  {Limmongkol}, {Lindenmeyer}, {Long}, {Loomis}, {Loveday}, {MacKinnon},
  {Mannery}, {Mantsch}, {Margon}, {McGehee}, {McKay}, {McLean}, {Menou},
  {Merelli}, {Mo}, {Monet}, {Nakamura}, {Narayanan}, {Nash}, {Neilsen},
  {Newman}, {Nitta}, {Odenkirchen}, {Okada}, {Okamura}, {Ostriker}, {Owen},
  {Pauls}, {Peoples}, {Peterson}, {Petravick}, {Pope}, {Pordes}, {Postman},
  {Prosapio}, {Quinn}, {Rechenmacher}, {Rivetta}, {Rix}, {Rockosi}, {Rosner},
  {Ruthmansdorfer}, {Sandford}, {Schneider}, {Scranton}, {Sekiguchi}, {Sergey},
  {Sheth}, {Shimasaku}, {Smee}, {Snedden}, {Stebbins}, {Stubbs}, {Szapudi},
  {Szkody}, {Szokoly}, {Tabachnik}, {Tsvetanov}, {Uomoto}, {Vogeley}, {Voges},
  {Waddell}, {Walterbos}, {Wang}, {Watanabe}, {Weinberg}, {White}, {White},
  {Wilhite}, {Wolfe}, {Yasuda}, {York}, {Zehavi}, \& {Zheng}}]{s+02}
{Stoughton}, C., {Lupton}, R.~H., {Bernardi}, M., {Blanton}, M.~R., {Burles},
  S., {Castander}, F.~J., {Connolly}, A.~J., {Eisenstein}, D.~J., {Frieman},
  J.~A., {Hennessy}, G.~S., {Hindsley}, R.~B., {Ivezi{\'c}}, {\v Z}., {Kent},
  S., {Kunszt}, P.~Z., {Lee}, B.~C., {Meiksin}, A., {Munn}, J.~A., {Newberg},
  H.~J., {Nichol}, R.~C., {Nicinski}, T., {Pier}, J.~R., {Richards}, G.~T.,
  {Richmond}, M.~W., {Schlegel}, D.~J., {Smith}, J.~A., {Strauss}, M.~A.,
  {SubbaRao}, M., {Szalay}, A.~S., {Thakar}, A.~R., {Tucker}, D.~L., {Vanden
  Berk}, D.~E., {Yanny}, B., {Adelman}, J.~K., {Anderson}, Jr., J.~E.,
  {Anderson}, S.~F., {Annis}, J., {Bahcall}, N.~A., {Bakken}, J.~A.,
  {Bartelmann}, M., {Bastian}, S., {Bauer}, A., {Berman}, E., {B{\"o}hringer},
  H., {Boroski}, W.~N., {Bracker}, S., {Briegel}, C., {Briggs}, J.~W.,
  {Brinkmann}, J., {Brunner}, R., {Carey}, L., {Carr}, M.~A., {Chen}, B.,
  {Christian}, D., {Colestock}, P.~L., {Crocker}, J.~H., {Csabai}, I.,
  {Czarapata}, P.~C., {Dalcanton}, J., {Davidsen}, A.~F., {Davis}, J.~E.,
  {Dehnen}, W., {Dodelson}, S., {Doi}, M., {Dombeck}, T., {Donahue}, M.,
  {Ellman}, N., {Elms}, B.~R., {Evans}, M.~L., {Eyer}, L., {Fan}, X.,
  {Federwitz}, G.~R., {Friedman}, S., {Fukugita}, M., {Gal}, R., {Gillespie},
  B., {Glazebrook}, K., {Gray}, J., {Grebel}, E.~K., {Greenawalt}, B.,
  {Greene}, G., {Gunn}, J.~E., {de Haas}, E., {Haiman}, Z., {Haldeman}, M.,
  {Hall}, P.~B., {Hamabe}, M., {Hansen}, B., {Harris}, F.~H., {Harris}, H.,
  {Harvanek}, M., {Hawley}, S.~L., {Hayes}, J.~J.~E., {Heckman}, T.~M.,
  {Helmi}, A., {Henden}, A., {Hogan}, C.~J., {Hogg}, D.~W., {Holmgren}, D.~J.,
  {Holtzman}, J., {Huang}, C.-H., {Hull}, C., {Ichikawa}, S.-I., {Ichikawa},
  T., {Johnston}, D.~E., {Kauffmann}, G., {Kim}, R.~S.~J., {Kimball}, T.,
  {Kinney}, E., {Klaene}, M., {Kleinman}, S.~J., {Klypin}, A., {Knapp}, G.~R.,
  {Korienek}, J., {Krolik}, J., {Kron}, R.~G., {Krzesi{\'n}ski}, J., {Lamb},
  D.~Q., {Leger}, R.~F., {Limmongkol}, S., {Lindenmeyer}, C., {Long}, D.~C.,
  {Loomis}, C., {Loveday}, J., {MacKinnon}, B., {Mannery}, E.~J., {Mantsch},
  P.~M., {Margon}, B., {McGehee}, P., {McKay}, T.~A., {McLean}, B., {Menou},
  K., {Merelli}, A., {Mo}, H.~J., {Monet}, D.~G., {Nakamura}, O., {Narayanan},
  V.~K., {Nash}, T., {Neilsen}, Jr., E.~H., {Newman}, P.~R., {Nitta}, A.,
  {Odenkirchen}, M., {Okada}, N., {Okamura}, S., {Ostriker}, J.~P., {Owen}, R.,
  {Pauls}, A.~G., {Peoples}, J., {Peterson}, R.~S., {Petravick}, D., {Pope},
  A., {Pordes}, R., {Postman}, M., {Prosapio}, A., {Quinn}, T.~R.,
  {Rechenmacher}, R., {Rivetta}, C.~H., {Rix}, H.-W., {Rockosi}, C.~M.,
  {Rosner}, R., {Ruthmansdorfer}, K., {Sandford}, D., {Schneider}, D.~P.,
  {Scranton}, R., {Sekiguchi}, M., {Sergey}, G., {Sheth}, R., {Shimasaku}, K.,
  {Smee}, S., {Snedden}, S.~A., {Stebbins}, A., {Stubbs}, C., {Szapudi}, I.,
  {Szkody}, P., {Szokoly}, G.~P., {Tabachnik}, S., {Tsvetanov}, Z., {Uomoto},
  A., {Vogeley}, M.~S., {Voges}, W., {Waddell}, P., {Walterbos}, R., {Wang},
  S.-i., {Watanabe}, M., {Weinberg}, D.~H., {White}, R.~L., {White}, S.~D.~M.,
  {Wilhite}, B., {Wolfe}, D., {Yasuda}, N., {York}, D.~G., {Zehavi}, I., \&
  {Zheng}, W. 2002, \aj, 123, 485

\bibitem[{{Strateva} {et~al.}(2005){Strateva}, {Brandt}, {Schneider}, {Vanden
  Berk}, \& {Vignali}}]{sbsvv05}
{Strateva}, I.~V., {Brandt}, W.~N., {Schneider}, D.~P., {Vanden Berk}, D.~G.,
  \& {Vignali}, C. 2005, \aj, 130, 387

\bibitem[{{Tolea} {et~al.}(2002){Tolea}, {Krolik}, \& {Tsvetanov}}]{tkt02}
{Tolea}, A., {Krolik}, J.~H., \& {Tsvetanov}, Z. 2002, \apjl, 578, L31

\bibitem[{{Trump} {et~al.}(2006){Trump}, {Hall}, {Reichard}, {Richards},
  {Schneider}, {Vanden Berk}, {Knapp}, {Anderson}, {Fan}, {Brinkman},
  {Kleinman}, \& {Nitta}}]{thrrsvkafbkn06}
{Trump}, J.~R., {Hall}, P.~B., {Reichard}, T.~A., {Richards}, G.~T.,
  {Schneider}, D.~P., {Vanden Berk}, D.~E., {Knapp}, G.~R., {Anderson}, S.~F.,
  {Fan}, X., {Brinkman}, J., {Kleinman}, S.~J., \& {Nitta}, A. 2006, \apjs,
  165, 1

\bibitem[{{Turnshek} {et~al.}(1988){Turnshek}, {Grillmair}, {Foltz}, \&
  {Weymann}}]{tfgw88}
{Turnshek}, D.~A., {Grillmair}, C.~J., {Foltz}, C.~B., \& {Weymann}, R.~J.
  1988, \apj, 325, 651

\bibitem[{{Vanden Berk} {et~al.}(2001){Vanden Berk}, {Richards}, {Bauer},
  {Strauss}, {Schneider}, {Heckman}, {York}, {Hall}, {Fan}, {Knapp},
  {Anderson}, {Annis}, {Bahcall}, {Bernardi}, {Briggs}, {Brinkmann}, {Brunner},
  {Burles}, {Carey}, {Castander}, {Connolly}, {Crocker}, {Csabai}, {Doi},
  {Finkbeiner}, {Friedman}, {Frieman}, {Fukugita}, {Gunn}, {Hennessy},
  {Ivezi{\'c}}, {Kent}, {Kunszt}, {Lamb}, {Leger}, {Long}, {Loveday}, {Lupton},
  {Meiksin}, {Merelli}, {Munn}, {Newberg}, {Newcomb}, {Nichol}, {Owen}, {Pier},
  {Pope}, {Rockosi}, {Schlegel}, {Siegmund}, {Smee}, {Snir}, {Stoughton},
  {Stubbs}, {SubbaRao}, {Szalay}, {Szokoly}, {Tremonti}, {Uomoto}, {Waddell},
  {Yanny}, \& {Zheng}}]{v+01}
{Vanden Berk}, D.~E., {Richards}, G.~T., {Bauer}, A., {Strauss}, M.~A.,
  {Schneider}, D.~P., {Heckman}, T.~M., {York}, D.~G., {Hall}, P.~B., {Fan},
  X., {Knapp}, G.~R., {Anderson}, S.~F., {Annis}, J., {Bahcall}, N.~A.,
  {Bernardi}, M., {Briggs}, J.~W., {Brinkmann}, J., {Brunner}, R., {Burles},
  S., {Carey}, L., {Castander}, F.~J., {Connolly}, A.~J., {Crocker}, J.~H.,
  {Csabai}, I., {Doi}, M., {Finkbeiner}, D., {Friedman}, S., {Frieman}, J.~A.,
  {Fukugita}, M., {Gunn}, J.~E., {Hennessy}, G.~S., {Ivezi{\'c}}, {\v Z}.,
  {Kent}, S., {Kunszt}, P.~Z., {Lamb}, D.~Q., {Leger}, R.~F., {Long}, D.~C.,
  {Loveday}, J., {Lupton}, R.~H., {Meiksin}, A., {Merelli}, A., {Munn}, J.~A.,
  {Newberg}, H.~J., {Newcomb}, M., {Nichol}, R.~C., {Owen}, R., {Pier}, J.~R.,
  {Pope}, A., {Rockosi}, C.~M., {Schlegel}, D.~J., {Siegmund}, W.~A., {Smee},
  S., {Snir}, Y., {Stoughton}, C., {Stubbs}, C., {SubbaRao}, M., {Szalay},
  A.~S., {Szokoly}, G.~P., {Tremonti}, C., {Uomoto}, A., {Waddell}, P.,
  {Yanny}, B., \& {Zheng}, W. 2001, \aj, 122, 549

\bibitem[{{Verner} {et~al.}(1995){Verner}, {Ferland}, {Korista}, \&
  {Yakovlev}}]{vfky95}
{Verner}, D.~A., {Ferland}, G.~J., {Korista}, K.~T., \& {Yakovlev}, D.~G. 1995,
  Bulletin of the American Astronomical Society, 27, 859

\bibitem[{{Verner} {et~al.}(1996){Verner}, {Verner}, \& {Ferland}}]{vvf96}
{Verner}, D.~A., {Verner}, E.~M., \& {Ferland}, G.~J. 1996, Atomic Data and
  Nuclear Data Tables, 64, 1

\bibitem[{{Voit} {et~al.}(1993){Voit}, {Weymann}, \& {Korista}}]{vwk93}
{Voit}, G.~M., {Weymann}, R.~J., \& {Korista}, K.~T. 1993, \apj, 413, 95

\bibitem[{{Weymann} {et~al.}(1991){Weymann}, {Morris}, {Foltz}, \&
  {Hewett}}]{wmfh91}
{Weymann}, R.~J., {Morris}, S.~L., {Foltz}, C.~B., \& {Hewett}, P.~C. 1991,
  \apj, 373, 23

\bibitem[{{Wilhite} {et~al.}(2006){Wilhite}, {Vanden Berk}, {Brunner}, \&
  {Brinkmann}}]{wvbb06}
{Wilhite}, B.~C., {Vanden Berk}, D.~E., {Brunner}, R.~J., \& {Brinkmann}, J.~V.
  2006, \apj, 641, 78

\bibitem[{{Worrall} {et~al.}(1987){Worrall}, {Tananbaum}, {Giommi}, \&
  {Zamorani}}]{wtgz87}
{Worrall}, D.~M., {Tananbaum}, H., {Giommi}, P., \& {Zamorani}, G. 1987, \apj,
  313, 596

\bibitem[{{York} {et~al.}(2000){York}, {Adelman}, {Anderson}, {Anderson},
  {Annis}, {Bahcall}, {Bakken}, {Barkhouser}, {Bastian}, {Berman}, {Boroski},
  {Bracker}, {Briegel}, {Briggs}, {Brinkmann}, {Brunner}, {Burles}, {Carey},
  {Carr}, {Castander}, {Chen}, {Colestock}, {Connolly}, {Crocker}, {Csabai},
  {Czarapata}, {Davis}, {Doi}, {Dombeck}, {Eisenstein}, {Ellman}, {Elms},
  {Evans}, {Fan}, {Federwitz}, {Fiscelli}, {Friedman}, {Frieman}, {Fukugita},
  {Gillespie}, {Gunn}, {Gurbani}, {de Haas}, {Haldeman}, {Harris}, {Hayes},
  {Heckman}, {Hennessy}, {Hindsley}, {Holm}, {Holmgren}, {Huang}, {Hull},
  {Husby}, {Ichikawa}, {Ichikawa}, {Ivezi{\'c}}, {Kent}, {Kim}, {Kinney},
  {Klaene}, {Kleinman}, {Kleinman}, {Knapp}, {Korienek}, {Kron}, {Kunszt},
  {Lamb}, {Lee}, {Leger}, {Limmongkol}, {Lindenmeyer}, {Long}, {Loomis},
  {Loveday}, {Lucinio}, {Lupton}, {MacKinnon}, {Mannery}, {Mantsch}, {Margon},
  {McGehee}, {McKay}, {Meiksin}, {Merelli}, {Monet}, {Munn}, {Narayanan},
  {Nash}, {Neilsen}, {Neswold}, {Newberg}, {Nichol}, {Nicinski}, {Nonino},
  {Okada}, {Okamura}, {Ostriker}, {Owen}, {Pauls}, {Peoples}, {Peterson},
  {Petravick}, {Pier}, {Pope}, {Pordes}, {Prosapio}, {Rechenmacher}, {Quinn},
  {Richards}, {Richmond}, {Rivetta}, {Rockosi}, {Ruthmansdorfer}, {Sandford},
  {Schlegel}, {Schneider}, {Sekiguchi}, {Sergey}, {Shimasaku}, {Siegmund},
  {Smee}, {Smith}, {Snedden}, {Stone}, {Stoughton}, {Strauss}, {Stubbs},
  {SubbaRao}, {Szalay}, {Szapudi}, {Szokoly}, {Thakar}, {Tremonti}, {Tucker},
  {Uomoto}, {Vanden Berk}, {Vogeley}, {Waddell}, {Wang}, {Watanabe},
  {Weinberg}, {Yanny}, \& {Yasuda}}]{y+00}
{York}, D.~G., {Adelman}, J., {Anderson}, Jr., J.~E., {Anderson}, S.~F.,
  {Annis}, J., {Bahcall}, N.~A., {Bakken}, J.~A., {Barkhouser}, R., {Bastian},
  S., {Berman}, E., {Boroski}, W.~N., {Bracker}, S., {Briegel}, C., {Briggs},
  J.~W., {Brinkmann}, J., {Brunner}, R., {Burles}, S., {Carey}, L., {Carr},
  M.~A., {Castander}, F.~J., {Chen}, B., {Colestock}, P.~L., {Connolly}, A.~J.,
  {Crocker}, J.~H., {Csabai}, I., {Czarapata}, P.~C., {Davis}, J.~E., {Doi},
  M., {Dombeck}, T., {Eisenstein}, D., {Ellman}, N., {Elms}, B.~R., {Evans},
  M.~L., {Fan}, X., {Federwitz}, G.~R., {Fiscelli}, L., {Friedman}, S.,
  {Frieman}, J.~A., {Fukugita}, M., {Gillespie}, B., {Gunn}, J.~E., {Gurbani},
  V.~K., {de Haas}, E., {Haldeman}, M., {Harris}, F.~H., {Hayes}, J.,
  {Heckman}, T.~M., {Hennessy}, G.~S., {Hindsley}, R.~B., {Holm}, S.,
  {Holmgren}, D.~J., {Huang}, C.-h., {Hull}, C., {Husby}, D., {Ichikawa},
  S.-I., {Ichikawa}, T., {Ivezi{\'c}}, {\v Z}., {Kent}, S., {Kim}, R.~S.~J.,
  {Kinney}, E., {Klaene}, M., {Kleinman}, A.~N., {Kleinman}, S., {Knapp},
  G.~R., {Korienek}, J., {Kron}, R.~G., {Kunszt}, P.~Z., {Lamb}, D.~Q., {Lee},
  B., {Leger}, R.~F., {Limmongkol}, S., {Lindenmeyer}, C., {Long}, D.~C.,
  {Loomis}, C., {Loveday}, J., {Lucinio}, R., {Lupton}, R.~H., {MacKinnon}, B.,
  {Mannery}, E.~J., {Mantsch}, P.~M., {Margon}, B., {McGehee}, P., {McKay},
  T.~A., {Meiksin}, A., {Merelli}, A., {Monet}, D.~G., {Munn}, J.~A.,
  {Narayanan}, V.~K., {Nash}, T., {Neilsen}, E., {Neswold}, R., {Newberg},
  H.~J., {Nichol}, R.~C., {Nicinski}, T., {Nonino}, M., {Okada}, N., {Okamura},
  S., {Ostriker}, J.~P., {Owen}, R., {Pauls}, A.~G., {Peoples}, J., {Peterson},
  R.~L., {Petravick}, D., {Pier}, J.~R., {Pope}, A., {Pordes}, R., {Prosapio},
  A., {Rechenmacher}, R., {Quinn}, T.~R., {Richards}, G.~T., {Richmond}, M.~W.,
  {Rivetta}, C.~H., {Rockosi}, C.~M., {Ruthmansdorfer}, K., {Sandford}, D.,
  {Schlegel}, D.~J., {Schneider}, D.~P., {Sekiguchi}, M., {Sergey}, G.,
  {Shimasaku}, K., {Siegmund}, W.~A., {Smee}, S., {Smith}, J.~A., {Snedden},
  S., {Stone}, R., {Stoughton}, C., {Strauss}, M.~A., {Stubbs}, C., {SubbaRao},
  M., {Szalay}, A.~S., {Szapudi}, I., {Szokoly}, G.~P., {Thakar}, A.~R.,
  {Tremonti}, C., {Tucker}, D.~L., {Uomoto}, A., {Vanden Berk}, D., {Vogeley},
  M.~S., {Waddell}, P., {Wang}, S.-i., {Watanabe}, M., {Weinberg}, D.~H.,
  {Yanny}, B., \& {Yasuda}, N. 2000, \aj, 120, 1579

\end{thebibliography}

\clearpage

\pagestyle{empty}
\begin{deluxetable}{cccccccccccccc}
\rotate
\tabletypesize{\tiny}
\tablecolumns{14}
\tablewidth{0pc}
\tablecaption{DR5 BAL Catalog\tablenotemark{a}\label{bALCatTab}}
\tablehead{
\colhead{DR5 Name} & 
\colhead{RA} & 
\colhead{Dec} & 
\colhead{$z$} & 
\colhead{\ion{Si}{4}} & 
\colhead{\ion{Si}{4}} & 
\colhead{\ion{Si}{4}} & 
\colhead{\ion{Si}{4}} & 
\colhead{\ion{Si}{4}} & 
\colhead{\ion{Si}{4}} & 
\colhead{\ion{C}{4}} & 
\colhead{\ion{C}{4}} & 
\colhead{\ion{C}{4}} & 
\colhead{\ion{C}{4}}\\
\colhead{} & 
\colhead{(J2000)} & 
\colhead{(J2000)} & 
\colhead{} & 
\colhead{$BI$ (km~s$^{-1}$)} & 
\colhead{$BI_0$ (km~s$^{-1}$)} & 
\colhead{$EW_0$ (\AA)} & 
\colhead{$v_{min}$ (km~s$^{-1}$)} & 
\colhead{$v_{max}$ (km~s$^{-1}$)} & 
\colhead{$f_{deep}$} & 
\colhead{$BI$ (km~s$^{-1}$)} & 
\colhead{$BI_0$ (km~s$^{-1}$)} & 
\colhead{$EW_0$ (\AA)} & 
\colhead{$v_{min}$ (km~s$^{-1}$)}
}
\startdata
$000013.80-005446.8$ &    0.057505 & $  -0.913004$ & 1.84 & $0$ & $276.6$ & $ -6.2$ & $-2249$ & $-4872$ & $0.58$ &  $0$ & $404.3$ & $ -9.9$ & $ -358$ \\
$000038.65+011426.3$ &    0.161078 & $   1.240640$ & 1.84 & $1733.2$ & $1733.2$ & $-15.9$ & $-6778$ & $-10295$ & $0.75$ &  $2144.7$ & $2144.7$ & $-19.2$ & $-5923$ \\
$000046.41+011420.8$ &    0.193415 & $   1.239118$ & 3.76 & $801.3$ & $801.3$ & $ -8.2$ & $-7157$ & $-10537$ & $0.61$ &  $5803.2$ & $5803.2$ & $-44.6$ & $-4232$ \\
$000051.56+001202.5$ &    0.214856 & $   0.200710$ & 3.88 & $1008.6$ & $1435.0$ & $-12.4$ & $-2423$ & $-6701$ & $0.74$ &  $2170.4$ & $2759.8$ & $-22.7$ & $-2258$ \\
$000056.89-010409.7$ &    0.237056 & $  -1.069386$ & 2.11 & $842.1$ & $1504.2$ & $-12.5$ & $-2214$ & $-6700$ & $0.65$ &  $1355.5$ & $2353.8$ & $-20.7$ & $-1497$ \\
\tablehead{
\colhead{DR5 Name} &
\colhead{\ion{C}{4}} & 
\colhead{\ion{C}{4}} & 
\colhead{\ion{Al}{3}} & 
\colhead{\ion{Al}{3}} & 
\colhead{\ion{Al}{3}} & 
\colhead{\ion{Al}{3}} & 
\colhead{\ion{Al}{3}} & 
\colhead{\ion{Al}{3}} & 
\colhead{\ion{Mg}{2}} & 
\colhead{\ion{Mg}{2}} & 
\colhead{\ion{Mg}{2}} & 
\colhead{\ion{Mg}{2}} & 
\colhead{\ion{Mg}{2}}\\
\colhead{} &
\colhead{$v_{max}$ (km~s$^{-1}$)} & 
\colhead{$f_{deep}$} & 
\colhead{$BI$ (km~s$^{-1}$)} & 
\colhead{$BI_0$ (km~s$^{-1}$)} & 
\colhead{$EW_0$ (\AA)} & 
\colhead{$v_{min}$ (km~s$^{-1}$)} & 
\colhead{$v_{max}$(km~s$^{-1}$)} & 
\colhead{$f_{deep}$} & 
\colhead{$BI$ (km~s$^{-1}$)} & 
\colhead{$BI_0$ (km~s$^{-1}$)} & 
\colhead{$EW_0$ (\AA)} & 
\colhead{$v_{min}$ (km~s$^{-1}$)} & 
\colhead{$v_{max}$ (km~s$^{-1}$)}
}\tablebreak
$000013.80-005446.8$ & $-3326$ & $0.77$ &  $0$ & $0$ & $0$ & $$ & $$ & $$ &  $0$ & $0$ & $0$ & $$ & $$ \\
$000038.65+011426.3$ & $-10544$ & $0.85$ &  $0$ & $0$ & $0$ & $$ & $$ & $$ &  $0$ & $0$ & $0$ & $$ & $$ \\
$000046.41+011420.8$ & $-19529$ & $0.78$ &  $0$ & $0$ & $0$ & $$ & $$ & $$ &  $$ & $$ & $$ & $$ & $$ \\
$000051.56+001202.5$ & $-8054$ & $0.86$ &  $0$ & $0$ & $0$ & $$ & $$ & $$ &  $$ & $$ & $$ & $$ & $$ \\
$000056.89-010409.7$ & $-8605$ & $0.56$ &  $0$ & $0$ & $0$ & $$ & $$ & $$ &  $0$ & $0$ & $0$ & $$ & $$ \\
\tablehead{
\colhead{DR5 Name} & 
\colhead{\ion{Mg}{2}} & 
\colhead{\ion{Si}{4}} & 
\colhead{\ion{C}{4}} & 
\colhead{\ion{Al}{3}} & 
\colhead{\ion{Mg}{2}} & 
\colhead{\ion{Si}{4}} & 
\colhead{\ion{C}{4}} & 
\colhead{\ion{Al}{3}} & 
\colhead{\ion{Mg}{2}} & 
\colhead{\ion{C}{4}} & 
\colhead{$SN_{1700}$} & 
\colhead{$\log(F_{1400})$} & 
\colhead{$\log(F_{2500})$}\tablenotemark{b} \\
\colhead{} &
\colhead{$f_{deep}$} & 
\colhead{EmL} & 
\colhead{EmL} & 
\colhead{EmL} & 
\colhead{EmL} & 
\colhead{BMBB} & 
\colhead{BMBB} & 
\colhead{BMBB} & 
\colhead{BMBB} & 
\colhead{BWA} & 
\colhead{} & 
\colhead{erg~cm$^{-2}$~s$^{-1}$~Hz$^{-1}$} & 
\colhead{erg~cm$^{-2}$~s$^{-1}$~Hz$^{-1}$}
}\tablebreak
$000013.80-005446.8$ & $$ &  1 & 1 & 0 & 0 &  0 & 0 &  &  &  1 & $  3.7$ & $-27.55$ & $-27.38$\\
$000038.65+011426.3$ & $$ &  0 & 0 & 0 & 0 &  0 & 0 &  &  &  0 & $  2.9$ & $-27.70$ & $-27.49$\\
$000046.41+011420.8$ & $$ &  1 & 1 & 0 & 0 &  0 & 0 &  &  &  0 & $  4.2$ & $-27.37$ & $$\\
$000051.56+001202.5$ & $$ &  0 & 1 & 0 & 0 &  0 & 0 &  &  &  0 & $  4.6$ & $-27.45$ & $$\\
$000056.89-010409.7$ & $$ &  1 & 0 & 0 & 0 &  0 & 0 &  &  &  0 & $  4.7$ & $-27.51$ & $-27.15$\\
\enddata
\tablenotetext{a}{The full version of this table is available in the electronic edition online.}
\tablenotetext{b}{For redshifts $z \la 0.72$, the SDSS spectrum does not extend below (rest-frame) 2200~\AA.  Consequently, the UV continuum estimate is less certain, as the 2200--3000~\AA\ region may have significant emission and absorption features that obscure the underlying continuum.}
\end{deluxetable}
\clearpage

\clearpage
\begin{deluxetable}{ccccccccccccccc}
\rotate
\tabletypesize{\tiny}
\tablecolumns{15}
\tablewidth{0pc}
\tablecaption{DR5 BAL Catalog Duplicate Spectra\tablenotemark{a}\label{bALCatDupTab}}
\tablehead{
\colhead{DR5 Name} & 
\colhead{RA} & 
\colhead{Dec} & 
\colhead{$z$} &
\colhead{MJD} & 
\colhead{Plate} & 
\colhead{Fiber} & 
\colhead{\ion{Si}{4}} & 
\colhead{\ion{Si}{4}} & 
\colhead{\ion{Si}{4}} & 
\colhead{\ion{Si}{4}} & 
\colhead{\ion{Si}{4}} & 
\colhead{\ion{Si}{4}} & 
\colhead{\ion{C}{4}} &
\colhead{}\\
\colhead{} & 
\colhead{(J2000)} & 
\colhead{(J2000)} & 
\colhead{} &
\colhead{} & 
\colhead{} & 
\colhead{} & 
\colhead{$BI$ (km~s$^{-1}$)} & 
\colhead{$BI_0$ (km~s$^{-1}$)} & 
\colhead{$EW_0$ (\AA)} & 
\colhead{$v_{min}$(km~s$^{-1}$)} & 
\colhead{$v_{max}$(km~s$^{-1}$)} & 
\colhead{$f_{deep}$} & 
\colhead{$BI$ (km~s$^{-1}$)} & 
\colhead{}
}
\startdata	
$000056.89-010409.7$ &    0.237056 & $  -1.069386$ & 2.11 & 52203 & 0685 & 194 &  $1233.8$ & $1885.3$ & $-14.1$ & $-2145$ & $-8632$ & $0.36$ &  $1238.2$  &\\
$001130.55+005550.7$ &    2.877324 & $   0.930756$ & 2.31 & 52519 & 0686 & 603 &  $259.4$ & $259.4$ & $ -5.0$ & $-2955$ & $-6612$ & $0.17$ &  $245.2$  &\\
$001130.55+005550.7$ &    2.877324 & $   0.930756$ & 2.31 & 52518 & 0687 & 339 &  $ 76.6$ & $ 76.6$ & $ -3.3$ & $-4404$ & $-6888$ & $0.03$ &  $285.8$  &\\
$001438.28-010750.1$ &    3.659519 & $  -1.130607$ & 1.82 & 52518 & 0687 & 249 &  $0$ & $0$ & $0$ & $$ & $$ & $$ &  $571.5$\\
$001824.95+001525.8$ &    4.603987 & $   0.257193$ & 2.43 & 51816 & 0390 & 386 &  $1697.3$ & $2522.0$ & $-18.9$ & $-1841$ & $-7776$ & $0.76$ &  $4371.1$  &\\
\tablehead{
\colhead{DR5 Name} &  
\colhead{\ion{C}{4}} & 
\colhead{\ion{C}{4}} & 
\colhead{\ion{C}{4}} & 
\colhead{\ion{C}{4}} & 
\colhead{\ion{C}{4}} & 
\colhead{\ion{Al}{3}} & 
\colhead{\ion{Al}{3}} & 
\colhead{\ion{Al}{3}} & 
\colhead{\ion{Al}{3}} & 
\colhead{\ion{Al}{3}} & 
\colhead{\ion{Al}{3}} & 
\colhead{\ion{Mg}{2}} & 
\colhead{\ion{Mg}{2}} &
\colhead{}\\
\colhead{} &
\colhead{$BI_0$ (km~s$^{-1}$)} & 
\colhead{$EW_0$ (\AA)} & 
\colhead{$v_{min}$ (km~s$^{-1}$)} & 
\colhead{$v_{max}$ (km~s$^{-1}$)} & 
\colhead{$f_{deep}$} & 
\colhead{$BI$ (km~s$^{-1}$)} & 
\colhead{$BI_0$ (km~s$^{-1}$)} & 
\colhead{$EW_0$ (\AA)} & 
\colhead{$v_{min}$ (km~s$^{-1}$)} & 
\colhead{$v_{max}$ (km~s$^{-1}$)} & 
\colhead{$f_{deep}$} & 
\colhead{$BI$ (km~s$^{-1}$)} & 
\colhead{$BI_0$ (km~s$^{-1}$)} & 
\colhead{}
}\tablebreak
$000056.89-010409.7$ & $1988.4$ & $-19.4$ & $-1773$ & $-8260$ & $0.59$ &  $0$ & $0$ & $0$ & $$ & $$ & $$ &  $0$ & $0$  &\\
$001130.55+005550.7$ & $866.8$ & $-11.7$ & $-1409$ & $-6516$ & $0.38$ &  $0$ & $0$ & $0$ & $$ & $$ & $$ &  $0$ & $0$  &\\
$001130.55+005550.7$ & $943.4$ & $-12.2$ & $-1409$ & $-6378$ & $0.44$ &  $0$ & $0$ & $0$ & $$ & $$ & $$ &  $0$ & $0$ &\\
$001438.28-010750.1$ & $571.5$ & $-10.6$ & $-3153$ & $-6052$ & $0.67$ &  $0$ & $0$ & $0$ & $$ & $$ & $$ &  $0$ & $0$\\ 
$001824.95+001525.8$ & $5849.7$ & $-39.2$ & $-1538$ & $-12162$ & $0.82$ &  $  9.7$ & $  9.7$ & $ -3.2$ & $-3551$ & $-5691$ & $0.00$ &  $0$ & $0$  &\\
\tablehead{
\colhead{DR5 Name} &  
\colhead{\ion{Mg}{2}} & 
\colhead{\ion{Mg}{2}} & 
\colhead{\ion{Mg}{2}} & 
\colhead{\ion{Mg}{2}} & 
\colhead{\ion{Si}{4}} & 
\colhead{\ion{C}{4}} & 
\colhead{\ion{Al}{3}} & 
\colhead{\ion{Mg}{2}} & 
\colhead{\ion{Si}{4}} & 
\colhead{\ion{C}{4}} & 
\colhead{\ion{Al}{3}} & 
\colhead{\ion{Mg}{2}} & 
\colhead{\ion{C}{4}} & 
\colhead{$SN_{1700}$} \\ 
\colhead{} &
\colhead{$EW_0$ (\AA)} & 
\colhead{$v_{min}$ (km~s$^{-1}$)} & 
\colhead{$v_{max}$ (km~s$^{-1}$)} & 
\colhead{$f_{deep}$} & 
\colhead{EmL} & 
\colhead{EmL} & 
\colhead{EmL} & 
\colhead{EmL} & 
\colhead{BMBB} & 
\colhead{BMBB} & 
\colhead{BMBB} & 
\colhead{BMBB} & 
\colhead{BWA} & 
\colhead{}
}
\tablebreak
$000056.89-010409.7$ & $0$ & $$ & $$ & $$ &  1 & 1 & 0 & 0 &  0 & 0 &  &  &  0 &   4.4\\
$001130.55+005550.7$ & $0$ & $$ & $$ & $$ &  1 & 1 & 0 & 0 &  0 & 0 &  &  &  0 &   9.5\\
$001130.55+005550.7$ & $943.4$ & $0$ & $$ & $$ & $$ &  1 & 1 & 0 & 0 &  0 & 0 &    &  0 &   9.8\\
$001438.28-010750.1$ & $0$ & $$ & $$ & $$ &  0 & 1 & 0 & 0 &   & 0 &  &  &  0 &   7.0\\
$001824.95+001525.8$ & $0$ & $0$ & $$ & $$ & $$ &  0 & 0 & 0 & 0 &  0 & 0 & 0   &  0 &   4.0\\
\enddata
\tablenotetext{a}{The full version of this table is available in the electronic edition online.}
\end{deluxetable}
\clearpage
\pagestyle{plaintop}
\begin{deluxetable}{lrrr}
\tabletypesize{\scriptsize}
\tablecolumns{4}
\tablewidth{0pc}
\tablecaption{DR5 Non-BALs with \ion{C}{4} {\tt BlueWingAbs} Flag Set\label{nonBALCIVBWATab}}
\tablehead{\colhead{DR5 Name} & \colhead{RA} & \colhead{Dec} & \colhead{$z$}}
\startdata
$000058.77+003510.5$ &    0.244895 & $   0.586257$ & 1.65\\
$000124.23-093349.8$ &    0.350965 & $  -9.563839$ & 1.63\\
$000839.31-005336.8$ &    2.163799 & $  -0.893578$ & 2.62\\
$002018.74-104851.3$ &    5.078088 & $ -10.814263$ & 1.64\\
$003554.17-001353.0$ &    8.975741 & $  -0.231411$ & 2.78\\
$003841.36-004254.4$ &    9.672349 & $  -0.715120$ & 1.55\\
$004218.71+003237.1$ &   10.577999 & $   0.543640$ & 3.06\\
$005844.81-003934.1$ &   14.686740 & $  -0.659494$ & 1.60\\
$011202.98-103616.5$ &   18.012429 & $ -10.604590$ & 2.06\\
$011705.18+152931.7$ &   19.271614 & $  15.492161$ & 1.85\\
$011918.15-095239.0$ &   19.825643 & $  -9.877519$ & 1.78\\
$012802.66+000040.9$ &   22.011116 & $   0.011362$ & 1.87\\
$020755.76+011159.4$ &   31.982337 & $   1.199843$ & 2.07\\
$022430.17-004131.1$ &   36.125713 & $  -0.691996$ & 1.67\\
$022534.09+000347.8$ &   36.392059 & $   0.063296$ & 1.73\\
$022849.70-004550.4$ &   37.207116 & $  -0.764002$ & 2.54\\
$025103.47-064719.7$ &   42.764498 & $  -6.788819$ & 1.94\\
$030311.30+002051.3$ &   45.797098 & $   0.347600$ & 1.76\\
$030758.75+004259.0$ &   46.994819 & $   0.716413$ & 2.52\\
$032505.57-003445.5$ &   51.273226 & $  -0.579311$ & 2.50\\
$034000.92+001944.6$ &   55.003860 & $   0.329071$ & 2.67\\
$073157.88+365557.8$ &  112.991170 & $  36.932735$ & 2.14\\
$073453.93+422443.7$ &  113.724745 & $  42.412157$ & 1.90\\
$074136.49+385847.6$ &  115.402070 & $  38.979891$ & 1.87\\
$080024.41+160152.5$ &  120.101736 & $  16.031277$ & 1.65\\
$080937.80+421029.3$ &  122.407505 & $  42.174819$ & 1.67\\
$081145.52+052435.3$ &  122.939672 & $   5.409811$ & 3.57\\
$082107.11+360303.5$ &  125.279636 & $  36.050996$ & 1.66\\
$083314.24+482335.4$ &  128.309371 & $  48.393181$ & 1.60\\
$084951.46+383548.4$ &  132.464419 & $  38.596792$ & 2.43\\
$085507.83+061117.3$ &  133.782626 & $   6.188150$ & 3.18\\
$090351.08+535101.0$ &  135.962844 & $  53.850293$ & 2.86\\
$090539.53+330923.4$ &  136.414726 & $  33.156514$ & 2.82\\
$091002.53+422100.5$ &  137.510549 & $  42.350159$ & 2.08\\
$091053.58+563119.9$ &  137.723266 & $  56.522218$ & 2.36\\
$091342.48+372603.3$ &  138.427000 & $  37.434263$ & 2.13\\
$091520.41+041557.4$ &  138.835061 & $   4.265964$ & 1.92\\
$091545.38+083625.7$ &  138.939124 & $   8.607161$ & 1.61\\
$091741.60+594557.3$ &  139.423349 & $  59.765919$ & 2.20\\
$091835.40+051544.6$ &  139.647508 & $   5.262410$ & 1.89\\
$092316.08+325504.2$ &  140.817004 & $  32.917851$ & 1.64\\
$092550.48+264920.2$ &  141.460340 & $  26.822298$ & 2.07\\
$092830.60+101406.9$ &  142.127503 & $  10.235262$ & 2.80\\
$092940.75+321033.0$ &  142.419819 & $  32.175839$ & 1.78\\
$093330.53+362736.8$ &  143.377235 & $  36.460226$ & 1.61\\
$093433.86+565509.6$ &  143.641090 & $  56.919338$ & 1.90\\
$093753.26+381625.3$ &  144.471948 & $  38.273704$ & 3.19\\
$095108.76+314705.8$ &  147.786534 & $  31.784968$ & 3.03\\
$095317.39+281926.2$ &  148.322488 & $  28.323965$ & 1.63\\
$095722.81+622334.9$ &  149.345077 & $  62.393040$ & 2.25\\
$095805.67+375029.0$ &  149.523629 & $  37.841405$ & 1.95\\
$100145.01+511611.7$ &  150.437550 & $  51.269942$ & 1.94\\
$100247.30+451648.5$ &  150.697102 & $  45.280166$ & 1.63\\
$100425.66+444406.7$ &  151.106926 & $  44.735208$ & 1.77\\
$101303.73+434606.3$ &  153.265570 & $  43.768418$ & 2.93\\
$102259.79+010123.6$ &  155.749157 & $   1.023223$ & 1.56\\
$103227.63+612853.6$ &  158.115135 & $  61.481566$ & 2.31\\
$103403.87+380248.4$ &  158.516153 & $  38.046790$ & 3.55\\
$104157.94+473329.6$ &  160.491423 & $  47.558243$ & 1.64\\
$104212.55+404122.8$ &  160.552321 & $  40.689669$ & 1.74\\
$105216.05+485619.1$ &  163.066878 & $  48.938658$ & 2.24\\
$105332.47+293737.6$ &  163.385319 & $  29.627115$ & 2.08\\
$105623.71+384155.0$ &  164.098806 & $  38.698626$ & 1.85\\
$110249.51+013640.6$ &  165.706317 & $   1.611299$ & 1.76\\
$110312.88+120855.9$ &  165.803696 & $  12.148876$ & 1.58\\
$112249.43+002047.5$ &  170.705996 & $   0.346535$ & 2.57\\
$112745.14+633618.4$ &  171.938114 & $  63.605118$ & 1.56\\
$112813.92+494857.1$ &  172.058017 & $  49.815880$ & 1.55\\
$112906.44+064809.3$ &  172.276850 & $   6.802611$ & 3.22\\
$114817.70+135949.8$ &  177.073777 & $  13.997174$ & 1.87\\
$115701.92+562144.9$ &  179.258005 & $  56.362482$ & 2.17\\
$120720.58+644842.3$ &  181.835781 & $  64.811759$ & 1.85\\
$121005.00+125904.3$ &  182.520873 & $  12.984554$ & 1.59\\
$121538.55+340451.9$ &  183.910634 & $  34.081101$ & 2.20\\
$122145.26+412525.0$ &  185.438614 & $  41.423626$ & 2.03\\
$123437.37+062803.7$ &  188.655710 & $   6.467713$ & 2.12\\
$125215.26+475231.1$ &  193.063601 & $  47.875327$ & 1.61\\
$130140.58+544907.3$ &  195.419119 & $  54.818701$ & 1.75\\
$131234.56+011409.7$ &  198.144034 & $   1.236055$ & 2.07\\
$131936.87+114652.3$ &  199.903666 & $  11.781212$ & 2.23\\
$132707.05+313838.6$ &  201.779411 & $  31.644065$ & 2.81\\
$134556.93+612509.6$ &  206.487213 & $  61.419352$ & 1.77\\
$135249.81-031354.3$ &  208.207550 & $  -3.231758$ & 4.75\\
$135306.34+113804.7$ &  208.276448 & $  11.634643$ & 1.62\\
$140030.56+594229.7$ &  210.127351 & $  59.708277$ & 1.92\\
$141438.04+373357.1$ &  213.658535 & $  37.565878$ & 1.57\\
$141609.15+502925.5$ &  214.038133 & $  50.490434$ & 1.73\\
$142342.68+572410.5$ &  215.927859 & $  57.402930$ & 1.66\\
$142901.69+441358.6$ &  217.257070 & $  44.232945$ & 1.62\\
$143236.18+234914.4$ &  218.150757 & $  23.820679$ & 2.94\\
$143911.28+563855.3$ &  219.797039 & $  56.648710$ & 1.64\\
$144650.53+470814.3$ &  221.710566 & $  47.137309$ & 1.69\\
$144844.81+494834.8$ &  222.186714 & $  49.809694$ & 2.07\\
$145432.54+343523.9$ &  223.635601 & $  34.589985$ & 1.61\\
$145609.59+354908.5$ &  224.039977 & $  35.819048$ & 1.61\\
$145647.53+533539.5$ &  224.198044 & $  53.594306$ & 1.62\\
$150220.46+465233.5$ &  225.585282 & $  46.875985$ & 4.26\\
$150344.91+005026.8$ &  225.937156 & $   0.840782$ & 1.88\\
$151555.26+605156.0$ &  228.980261 & $  60.865565$ & 1.86\\
$152500.12+414336.8$ &  231.250508 & $  41.726898$ & 1.89\\
$152952.38+553346.8$ &  232.468265 & $  55.563016$ & 2.94\\
$154725.59+534814.5$ &  236.856625 & $  53.804052$ & 1.87\\
$155218.09+045635.2$ &  238.075379 & $   4.943133$ & 1.57\\
$160329.14+090218.2$ &  240.871427 & $   9.038390$ & 4.74\\
$162246.27+290054.1$ &  245.692808 & $  29.015031$ & 2.15\\
$162802.82+301219.1$ &  247.011771 & $  30.205308$ & 2.44\\
$163045.20+345924.5$ &  247.688340 & $  34.990159$ & 3.36\\
$163955.00+310048.6$ &  249.979187 & $  31.013525$ & 1.72\\
$164207.25+354306.0$ &  250.530227 & $  35.718337$ & 2.33\\
$164547.63+232824.6$ &  251.448488 & $  23.473525$ & 1.86\\
$165444.63+384808.6$ &  253.685981 & $  38.802401$ & 1.73\\
$210255.16-064112.2$ &  315.729853 & $  -6.686746$ & 2.11\\
$212957.88+100347.2$ &  322.491177 & $  10.063128$ & 3.90\\
$215002.70+011343.8$ &  327.511251 & $   1.228854$ & 3.27\\
$220116.75+125636.4$ &  330.319831 & $  12.943448$ & 2.92\\
$221347.32+001928.4$ &  333.447208 & $   0.324570$ & 2.31\\
$222339.21+003942.1$ &  335.913409 & $   0.661708$ & 2.96\\
$235833.49-110532.7$ &  359.639563 & $ -11.092431$ & 1.98\\
\enddata
\end{deluxetable}
\clearpage

\begin{deluxetable}{lrrrccc}
\tabletypesize{\scriptsize}
\tablecolumns{7}
\tablewidth{0pc}
\tablecaption{BALs with Absorption from Other Ions\label{contamBALTab}}
\tablehead{\colhead{DR5 Name} & \colhead{RA} & \colhead{Dec} & \colhead{$z$} & \colhead{\ion{Si}{4} Contam?}\tablenotemark{a} & \colhead{\ion{Al}{3} Contam?}\tablenotemark{a} & \colhead{\ion{Mg}{2} Contam?}\tablenotemark{a}}
\startdata
$031856.62-060037.7$ &   49.735942 & $  -6.010475$ & 1.82 & 1 & 1 & 1\\
$074554.74+181817.0$ &  116.478093 & $  18.304736$ & 1.84 & 0 & 0 & 1\\
$074850.39+442439.0$ &  117.209995 & $  44.410834$ & 1.20 & 1 & 0 & 1\\
$080248.19+551328.9$ &  120.700793 & $  55.224706$ & 2.24 & 0 & 0 & 1\\
$080401.06+230922.5$ &  121.004436 & $  23.156270$ & 1.84 & 1 & 0 & 1\\
$080501.77+221409.7$ &  121.257378 & $  22.236030$ & 0.46 & 1 & 0 & 0\\
$081426.45+364713.5$ &  123.610221 & $  36.787098$ & 0.48 & 1 & 0 & 0\\
$081547.33+555154.0$ &  123.947212 & $  55.865010$ & 1.44 & 0 & 0 & 1\\
$083942.11+380526.3$ &  129.925466 & $  38.090666$ & 0.95 & 0 & 0 & 1\\
$084044.41+363327.8$ &  130.185078 & $  36.557738$ & 1.84 & 0 & 0 & 1\\
$084526.06+010332.2$ &  131.358610 & $   1.058945$ & 1.89 & 0 & 0 & 1\\
$084554.24+423003.5$ &  131.476038 & $  42.500991$ & 1.85 & 1 & 0 & 0\\
$085216.78+233916.8$ &  133.069948 & $  23.654686$ & 0.70 & 0 & 0 & 1\\
$085910.39+423911.2$ &  134.793316 & $  42.653126$ & 1.25 & 0 & 1 & 1\\
$090106.43+291128.5$ &  135.276799 & $  29.191262$ & 0.55 & 1 & 0 & 0\\
$090720.29+365050.1$ &  136.834576 & $  36.847253$ & 0.99 & 1 & 1 & 0\\
$092221.25+084312.8$ &  140.588543 & $   8.720224$ & 0.77 & 1 & 0 & 1\\
$093224.48+084008.1$ &  143.102039 & $   8.668921$ & 1.14 & 1 & 0 & 1\\
$093348.37+313335.2$ &  143.451569 & $  31.559805$ & 1.80 & 1 & 0 & 0\\
$094347.22+324314.2$ &  145.946760 & $  32.720627$ & 1.09 & 1 & 0 & 0\\
$101927.36+022521.4$ &  154.864039 & $   2.422628$ & 1.52 & 0 & 0 & 1\\
$102036.09+602339.0$ &  155.150382 & $  60.394179$ & 1.22 & 0 & 0 & 1\\
$102216.71+345929.6$ &  155.569661 & $  34.991574$ & 1.84 & 0 & 0 & 1\\
$102236.23+372241.4$ &  155.650962 & $  37.378175$ & 2.12 & 0 & 0 & 1\\
$102343.13+553132.3$ &  155.929735 & $  55.525657$ & 1.55 & 1 & 0 & 0\\
$102413.94+465355.3$ &  156.058095 & $  46.898705$ & 1.31 & 1 & 0 & 0\\
$103349.80+332812.0$ &  158.457509 & $  33.470003$ & 0.95 & 0 & 0 & 1\\
$103524.68+452008.4$ &  158.852840 & $  45.335676$ & 2.19 & 0 & 1 & 0\\
$103958.21+061119.7$ &  159.992549 & $   6.188832$ & 0.44 & 1 & 0 & 0\\
$112001.68+620459.6$ &  170.007008 & $  62.083230$ & 1.44 & 0 & 0 & 1\\
$112235.98+042228.2$ &  170.649927 & $   4.374512$ & 3.76 & 0 & 1 & 0\\
$112526.12+002901.3$ &  171.358859 & $   0.483701$ & 1.14 & 0 & 0 & 1\\
$112828.31+011337.9$ &  172.117966 & $   1.227217$ & 2.64 & 0 & 0 & 1\\
$113212.92+010441.3$ &  173.053845 & $   1.078150$ & 3.88 & 0 & 0 & 1\\
$113424.64+323802.4$ &  173.602671 & $  32.634015$ & 1.32 & 0 & 0 & 1\\
$113734.06+024159.3$ &  174.391920 & $   2.699808$ & 2.76 & 1 & 1 & 0\\
$114013.74+624156.3$ &  175.057256 & $  62.698984$ & 1.41 & 1 & 0 & 0\\
$114225.75+663237.4$ &  175.607303 & $  66.543737$ & 1.18 & 1 & 0 & 0\\
$115436.60+030006.3$ &  178.652517 & $   3.001775$ & 1.46 & 0 & 0 & 1\\
$120627.62+002335.3$ &  181.615115 & $   0.393165$ & 2.11 & 0 & 0 & 1\\
$123549.95+013252.6$ &  188.958127 & $   1.547953$ & 2.55 & 0 & 0 & 1\\
$123816.38+085146.8$ &  189.568260 & $   8.863020$ & 1.09 & 0 & 0 & 1\\
$124140.51+012228.4$ &  190.418794 & $   1.374580$ & 1.90 & 1 & 1 & 0\\
$124452.50+583427.6$ &  191.218759 & $  58.574354$ & 1.65 & 1 & 0 & 0\\
$125942.80+121312.6$ &  194.928337 & $  12.220168$ & 0.29 & 0 & 0 & 1\\
$130138.87+023137.4$ &  195.411975 & $   2.527074$ & 2.08 & 1 & 0 & 0\\
$131010.74-003007.2$ &  197.544790 & $  -0.502027$ & 0.55 & 1 & 0 & 0\\
$131935.61+152040.6$ &  199.898379 & $  15.344636$ & 0.45 & 1 & 0 & 0\\
$131957.70+283311.1$ &  199.990434 & $  28.553085$ & 1.40 & 0 & 0 & 1\\
$132139.86-004151.9$ &  200.416109 & $  -0.697774$ & 0.42 & 1 & 0 & 0\\
$134026.43+634433.1$ &  205.110163 & $  63.742535$ & 0.53 & 1 & 0 & 0\\
$134145.12-003631.0$ &  205.438034 & $  -0.608638$ & 0.52 & 1 & 0 & 0\\
$134417.40+125030.9$ &  206.072505 & $  12.841917$ & 1.42 & 1 & 0 & 0\\
$134818.03+423205.1$ &  207.075135 & $  42.534762$ & 0.40 & 1 & 0 & 0\\
$134934.14+245540.1$ &  207.392268 & $  24.927807$ & 1.97 & 1 & 0 & 0\\
$135246.37+423923.5$ &  208.193219 & $  42.656554$ & 1.40 & 1 & 0 & 0\\
$135421.82+380327.3$ &  208.590946 & $  38.057602$ & 2.21 & 1 & 0 & 0\\
$140419.53+342807.5$ &  211.081378 & $  34.468771$ & 1.62 & 0 & 1 & 1\\
$142010.28+604722.3$ &  215.042863 & $  60.789546$ & 1.92 & 0 & 0 & 1\\
$142216.85+091524.3$ &  215.570224 & $   9.256768$ & 1.41 & 1 & 1 & 1\\
$142321.84+063031.7$ &  215.841033 & $   6.508812$ & 1.06 & 1 & 0 & 0\\
$143015.56+634721.5$ &  217.564840 & $  63.789328$ & 1.97 & 1 & 0 & 0\\
$143156.30+432556.3$ &  217.984602 & $  43.432328$ & 1.34 & 1 & 0 & 0\\
$143751.16+530706.8$ &  219.463185 & $  53.118575$ & 0.53 & 1 & 0 & 0\\
$144002.24+371058.5$ &  220.009350 & $  37.182922$ & 0.40 & 0 & 0 & 1\\
$144406.69+331345.8$ &  221.027888 & $  33.229412$ & 3.58 & 1 & 0 & 0\\
$144424.55+013457.0$ &  221.102292 & $   1.582513$ & 1.79 & 1 & 0 & 0\\
$144800.14+404311.7$ &  222.000614 & $  40.719921$ & 1.82 & 0 & 0 & 1\\
$144838.74+563919.1$ &  222.161437 & $  56.655330$ & 1.83 & 0 & 0 & 1\\
$144849.30-013524.1$ &  222.205440 & $  -1.590050$ & 0.93 & 1 & 1 & 1\\
$145138.28+415401.0$ &  222.909522 & $  41.900296$ & 1.34 & 1 & 0 & 0\\
$145507.23+571651.8$ &  223.780126 & $  57.281080$ & 2.64 & 0 & 0 & 1\\
$145729.62+042655.7$ &  224.373450 & $   4.448831$ & 1.20 & 0 & 0 & 1\\
$150848.80+605551.9$ &  227.203334 & $  60.931090$ & 0.36 & 0 & 0 & 1\\
$154741.96+444340.0$ &  236.924871 & $  44.727790$ & 1.17 & 1 & 0 & 1\\
$155359.96+005641.4$ &  238.499834 & $   0.944841$ & 2.31 & 1 & 0 & 0\\
$155816.71+475645.5$ &  239.569657 & $  47.945992$ & 1.27 & 1 & 1 & 0\\
$160627.56+445503.2$ &  241.614869 & $  44.917573$ & 1.44 & 1 & 0 & 0\\
$160744.30+513736.2$ &  241.934600 & $  51.626749$ & 2.21 & 1 & 1 & 0\\
$161013.07+452850.3$ &  242.554494 & $  45.480643$ & 0.65 & 1 & 0 & 0\\
$163313.28+352050.8$ &  248.305352 & $  35.347462$ & 2.15 & 0 & 0 & 1\\
$165008.87+311734.7$ &  252.536981 & $  31.292988$ & 1.96 & 1 & 0 & 0\\
$170602.51+304424.8$ &  256.510462 & $  30.740228$ & 1.45 & 1 & 0 & 0\\
$170931.01+630357.2$ &  257.379217 & $  63.065902$ & 1.27 & 0 & 0 & 1\\
$171701.00+304357.6$ &  259.254169 & $  30.732683$ & 0.55 & 0 & 1 & 1\\
$172341.08+555340.5$ &  260.921168 & $  55.894597$ & 1.76 & 1 & 1 & 1\\
$210757.67-062010.6$ &  316.990293 & $  -6.336294$ & 2.23 & 0 & 0 & 1\\
$222700.62-075356.7$ &  336.752603 & $  -7.899091$ & 1.02 & 0 & 0 & 1\\
$223841.88+142154.9$ &  339.674541 & $  14.365252$ & 3.06 & 1 & 0 & 0\\
$224117.74-010218.6$ &  340.323945 & $  -1.038511$ & 1.82 & 0 & 0 & 1\\
$224207.43-093219.3$ &  340.530964 & $  -9.538706$ & 0.81 & 1 & 0 & 0\\
$235038.95+133816.3$ &  357.662311 & $  13.637877$ & 1.65 & 1 & 0 & 0\\
\enddata
\tablenotetext{a}{A ``1'' indicates visual evidence of probable BAL contamination.}
\end{deluxetable}
\clearpage

\clearpage
\begin{deluxetable}{lrrrrcrrrr}
\rotate
\tabletypesize{\scriptsize}
\tablecolumns{9}
\tablewidth{0pc}
\tablecaption{X-Ray Source Information\label{xInfoTab}\tablenotemark{a}}
\tablehead{\colhead{SDSS} & \colhead{ObsId\tablenotemark{b}} & \colhead{X-Ray} & \colhead{0.5--2 keV Counts} & \colhead{2--8 keV Counts} & \colhead{X-Ray} & \colhead{$\log(R^*)$} & \colhead{$\log(L_{\rm 2~keV})$\tablenotemark{f}} & \colhead{$\alpha_{OX}$\tablenotemark{g}} & \colhead{$\Delta\alpha_{OX}$\tablenotemark{g}} \\ \colhead{Source} & \colhead{} & \colhead{Exposure (ks)\tablenotemark{c}} & \colhead{Source/BG} & \colhead{Source/BG} & \colhead{Detected?\tablenotemark{e}} & \colhead{} & \colhead{} & \colhead{} & \colhead{}}
\startdata
$001130.55+005550.7$ & 0403760301 & 25.4 & 34/12.7 & 29/10.5 & 1 & $<0.70$ & 26.33 & $-1.94$(0.03) & $-0.25$(0.03)\\
$002331.21-011045.6$ & 4079 & 1.9 & 2/0.6 & 2/0.7 & 0 & $<1.32$ & $<$26.76 & $<$$-1.50$ & $<$$0.09$\\
$003135.56+003421.3$ & 2101 & 6.7 & 32/0.4 & 10/0.7 & 1 & $<0.75$ & 26.84 & $-1.71$(0.03) & $-0.03$(0.03)\\
$004527.68+143816.1$ & 6889 & 11.4 & 7/0.2 & 5/0.7 & 1 & $<0.54$ & 25.62 & $-2.37$(0.06) & $-0.62$(0.06)\\
$005025.13-005718.3$ & 4825 & 13.0 & 16/1.2 & 12/2.7 & 1 & $<1.13$ & 26.31 & $-1.81$(0.05) & $-0.17$(0.05)\\
$005355.15-000309.3$ & 4830 & 7.1 & 18/0.1 & 10/0.3 & 1 & $<0.56$ & 26.16 & $-1.94$(0.04) & $-0.27$(0.04)\\
$011227.60-011221.7$ & 4832 & 5.9 & 2/0.1 & 7/0.3 & 1 & $<0.50$ & 25.14 & $-2.36$(0.08) & $-0.69$(0.08)\\
$024230.65-000029.6$ & 344 & 47.4 & 12/3.0 & 20/2.6 & 1 & $<0.92$ & 24.88 & $-2.44$(0.06) & $-0.77$(0.06)\\
$024304.68+000005.4$ & 0111200201 & 39.0 & 84/14.5 & 49/10.3 & 1 & $<0.64$ & 26.72 & $-1.75$(0.02) & $-0.08$(0.02)\\
$030000.57+004828.0$ & 4145 & 4.7 & 1/0.7 & 0/0.5 & 0 & $<0.12$ & $<$24.92 & $<$$-2.33$ & $<$$-0.70$\\
$034313.77+010336.9$ & 5650 & 7.9 & 41/0.2 & 26/0.4 & 1 & $<0.73$ & 26.58 & $-1.85$(0.02) & $-0.16$(0.02)\\
$073739.96+384413.2$ & 6888 & 10.5 & 8/0.1 & 4/0.3 & 1 & $<0.10$ & 25.66 & $-2.23$(0.07) & $-0.53$(0.07)\\
$075627.77+445836.4$ & 3033 & 7.0 & 17/0.8 & 12/1.5 & 1 & $<0.79$ & 26.52 & $-1.78$(0.04) & $-0.12$(0.04)\\
$082859.97+500451.3$ & 0303550901 & 17.0 & 81/61.6 & 98/87.3 & 1 & $<1.15$ & 26.41 & $-1.76$(0.03) & $-0.12$(0.03)\\
$083104.90+532500.1$ & 5656 & 11.6 & 7/0.8 & 9/4.5 & 1 & $<0.39$ & 25.87 & $-2.19$(0.07) & $-0.47$(0.07)\\
$084044.41+363327.8$ & 817 & 4.1 & 0/0.7 & 8/0.4 & 1 & $0.11$ & 25.61 & $-2.27$(0.06) & $-0.56$(0.06)\\
$084327.88+361723.2$ & 532 & 8.1 & 5/2.7 & 6/5.8 & 0 & $<1.38$ & $<$26.38 & $<$$-1.58$ & $<$$-0.01$\\
$084538.66+342043.6$ & 818 & 4.2 & 40/0.2 & 12/0.3 & 1 & $<0.13$ & 26.89 & $-1.91$(0.03) & $-0.15$(0.03)\\
$090904.52-000234.5$ & 5703 & 1.3 & 0/0.1 & 2/0.1 & 1 & $<1.44$ & 26.28 & $-1.56$(0.12) & $-0.02$(0.12)\\
$091251.87+591600.7$ & 3034 & 9.8 & 1/0.2 & 1/0.2 & 0 & $<1.17$ & $<$25.55 & $<$$-1.91$ & $<$$-0.34$\\
$091400.95+410600.9$ & 509 & 9.1 & 2/2.4 & 3/3.7 & 0 & $<1.18$ & $<$26.14 & $<$$-1.78$ & $<$$-0.17$\\
$092011.33+495404.0$ & 6885 & 1.5 & 1/0.0 & 2/0.1 & 1 & $<0.16$ & 26.17 & $-2.08$(0.10) & $-0.36$(0.10)\\
$092138.45+301546.9$ & 0150620101 & 15.8 & 7/4.1 & 13/3.2 & 1 & $<0.60$ & 25.52 & $-2.14$(0.08) & $-0.49$(0.08)\\
$093514.71+033545.7$ & 5705 & 1.8 & 2/0.0 & 3/0.1 & 1 & $<0.10$ & 26.32 & $-2.08$(0.08) & $-0.34$(0.08)\\
$093918.07+355615.0$ & 0021740101 & 34.0 & 8/6.3 & 7/10.8 & 0 & $<0.79$ & $<$26.13 & $<$$-1.93$ & $<$$-0.27$\\
$094309.56+481140.5$ & 0201470101 & 50.1 & 8/5.8 & 7/5.2 & 0 & $<0.91$ & $<$25.94 & $<$$-1.91$ & $<$$-0.28$\\
$095110.56+393243.9$ & 0111290101 & 21.2 & 11/7.6 & 9/7.5 & 0 & $<1.23$ & $<$26.24 & $<$$-1.65$ & $<$$-0.08$\\
$095544.91+410755.0$ & 5759 & 40.2 & 11/0.4 & 11/0.8 & 1 & $<1.08$ & 25.07 & $-2.20$(0.04) & $-0.59$(0.04)\\
$095944.47+051158.3$ & 0204791101 & 15.3 & 11/18.8 & 16/8.1 & 1 & $<0.94$ & 26.29 & $-1.71$(0.04) & $-0.11$(0.04)\\
$100711.81+053208.9$ & 6882 & 1.0 & 9/0.0 & 6/0.0 & 1 & $<-0.18$ & 26.83 & $-2.05$(0.06) & $-0.25$(0.06)\\
$104233.86+010206.2$ & 4086 & 1.7 & 4/0.2 & 3/0.2 & 1 & $<0.72$ & 26.73 & $-1.74$(0.06) & $-0.07$(0.06)\\
$104705.07+590728.4$ & 5028 & 71.1 & 40/1.9 & 175/8.2 & 1 & $<0.64$ & 24.71 & $-1.88$(0.00) & $-0.44$(0.00)\\
$110637.16+522233.4$ & 0304071201 & 8.0 & 5/3.3 & 12/2.7 & 1 & $<0.98$ & 26.02 & $-1.86$(0.09) & $-0.24$(0.09)\\
$114111.61-014306.6$ & 4146 & 10.0 & 12/0.2 & 11/0.4 & 1 & $<0.42$ & 25.53 & $-2.11$(0.06) & $-0.47$(0.06)\\
$120522.18+443140.4$ & 4162 & 29.7 & 55/4.1 & 46/6.1 & 1 & $<0.77$ & 26.25 & $-1.87$(0.03) & $-0.22$(0.03)\\
$120550.19+020131.5$ & 5700 & 2.1 & 8/0.1 & 2/0.1 & 1 & $<0.16$ & 26.83 & $-1.92$(0.07) & $-0.17$(0.07)\\
$120626.17+151335.6$ & 4834 & 7.1 & 3/0.2 & 1/0.3 & 1 & $<0.68$ & 25.61 & $-2.08$(0.08) & $-0.44$(0.08)\\
$120924.07+103612.0$ & 1764 & 1.2 & 3/-0.0 & 0/-0.0 & 0 & $0.15$ & $<$25.25 & $<$$-1.98$ & $<$$-0.43$\\
$120945.11+454404.6$ & 0033540601 & 12.5 & 5/4.8 & 8/5.8 & 0 & $<0.71$ & $<$26.52 & $<$$-1.76$ & $<$$-0.11$\\
$121125.48+151851.5$ & 2104 & 4.9 & 4/0.1 & 2/0.1 & 1 & $<0.54$ & 25.95 & $-2.08$(0.07) & $-0.39$(0.07)\\
$121205.30+152005.8$ & 2104 & 4.9 & 21/1.0 & 5/2.2 & 1 & $<1.20$ & 27.31 & $-1.34$(0.04) & $0.27$(0.04)\\
$121440.27+142859.1$ & 2456 & 4.5 & 1/0.1 & 0/0.2 & 0 & $<0.31$ & $<$25.45 & $<$$-2.28$ & $<$$-0.59$\\
$121930.95+104700.1$ & 2106 & 4.6 & 1/0.2 & 5/0.2 & 1 & $<0.38$ & 25.74 & $-2.14$(0.07) & $-0.46$(0.07)\\
$122033.87+334312.0$ & 2118 & 3.1 & 132/0.0 & 53/0.2 & 1 & $<0.68$ & 27.59 & $-1.30$(0.02) & $0.33$(0.02)\\
$122637.02+013016.0$ & 0110990201 & 28.6 & 24/6.0 & 8/6.0 & 1 & $<0.66$ & 26.10 & $-1.88$(0.05) & $-0.25$(0.05)\\
$122654.39-005430.6$ & 4060 & 19.8 & 42/0.7 & 10/1.1 & 1 & $<0.68$ & 26.58 & $-1.89$(0.02) & $-0.19$(0.02)\\
$122708.29+012638.4$ & 0110990201 & 28.6 & 10/4.3 & 6/3.7 & 1 & $<1.00$ & 25.66 & $-2.01$(0.07) & $-0.39$(0.07)\\
$122726.90-001003.9$ & 4865 & 4.9 & 0/0.1 & 4/1.3 & 0 & $<0.86$ & $<$26.20 & $<$$-1.77$ & $<$$-0.16$\\
$130136.12+000157.9$ & 4862 & 4.9 & 4/0.6 & 4/0.9 & 1 & $<0.36$ & 25.97 & $-2.10$(0.09) & $-0.41$(0.09)\\
$133428.06-012349.0$ & 2110 & 5.1 & 3/0.0 & 3/0.2 & 1 & $0.72$ & 25.87 & $-2.23$(0.07) & $-0.50$(0.07)\\
$133639.40+514605.2$ & 0084190201 & 47.5 & 22/17.4 & 13/12.5 & 0 & $<1.03$ & $<$26.34 & $<$$-1.79$ & $<$$-0.15$\\
$142412.13+054033.6$ & 5177 & 3.0 & 7/0.1 & 3/0.7 & 1 & $<1.03$ & 26.61 & $-1.61$(0.10) & $-0.01$(0.10)\\
$142620.30+351712.1$ & 3619 & 4.7 & 15/0.6 & 6/0.8 & 1 & $<0.77$ & 26.85 & $-1.60$(0.05) & $0.04$(0.05)\\
$142640.83+332158.7$ & 4255 & 5.0 & 0/0.1 & 1/0.3 & 0 & $<0.93$ & $<$25.74 & $<$$-1.91$ & $<$$-0.32$\\
$142652.94+375359.9$ & 542 & 8.1 & 11/0.5 & 9/1.0 & 1 & $<1.05$ & 26.00 & $-1.83$(0.06) & $-0.23$(0.06)\\
$143031.78+322145.9$ & 4279 & 5.0 & 20/0.1 & 6/0.2 & 1 & $<0.96$ & 27.20 & $-1.48$(0.04) & $0.16$(0.04)\\
$143117.93+364706.0$ & 4126 & 3.9 & 7/0.0 & 3/1.0 & 1 & $<0.87$ & 26.68 & $-1.70$(0.06) & $-0.05$(0.06)\\
$143411.23+334015.3$ & 4236 & 4.6 & 10/0.0 & 4/0.2 & 1 & $<1.09$ & 26.57 & $-1.60$(0.06) & $0.00$(0.06)\\
$143513.89+484149.3$ & 0110930401 & 9.1 & 22/3.1 & 12/3.1 & 1 & $<0.91$ & 26.76 & $-1.61$(0.03) & $0.02$(0.03)\\
$143752.75+042854.5$ & 5704 & 1.5 & 1/0.0 & 0/0.0 & 0 & $<0.17$ & $<$26.18 & $<$$-2.12$ & $<$$-0.39$\\
$143853.36+354918.7$ & 3596 & 4.6 & 6/0.1 & 1/0.2 & 1 & $<0.89$ & 26.27 & $-1.73$(0.06) & $-0.13$(0.06)\\
$144027.00+032637.9$ & 4942 & 23.1 & 6/5.0 & 11/2.7 & 1 & $<0.87$ & 25.14 & $-2.30$(0.08) & $-0.64$(0.08)\\
$144625.48+025548.6$ & 0203050801 & 7.9 & 3/2.2 & 6/3.4 & 0 & $<0.98$ & $<$26.58 & $<$$-1.65$ & $<$$-0.03$\\
$144842.45+042403.1$ & 6886 & 8.9 & 7/0.2 & 10/0.3 & 1 & $<0.08$ & 25.41 & $-2.37$(0.08) & $-0.65$(0.08)\\
$145002.34+421505.2$ & 5717 & 4.3 & 1/0.0 & 1/0.7 & 1 & $<0.94$ & 25.82 & $-1.97$(0.13) & $-0.34$(0.13)\\
$154359.44+535903.2$ & 822 & 4.5 & 61/0.2 & 18/0.2 & 1 & $<0.21$ & 27.17 & $-1.81$(0.03) & $-0.05$(0.03)\\
$155259.33+551033.9$ & 0145450601 & 9.5 & 13/7.6 & 13/11.2 & 0 & $<1.33$ & $<$27.04 & $<$$-1.44$ & $<$$0.17$\\
$155338.19+551401.9$ & 0145450601 & 9.5 & 16/5.6 & 14/8.7 & 1 & $<0.76$ & 26.25 & $-1.81$(0.06) & $-0.18$(0.06)\\
$221511.93-004549.9$ & 4147 & 4.6 & 0/0.1 & 0/0.3 & 0 & $<-0.17$ & $<$24.73 & $<$$-2.71$ & $<$$-0.96$\\
$234810.54+010551.3$ & 861 & 23.2 & 65/2.4 & 15/4.0 & 1 & $<1.13$ & 26.68 & $-1.55$(0.02) & $0.05$(0.02)\\
$235251.16-001950.4$ & 2115 & 5.8 & 7/0.4 & 1/1.4 & 1 & $<1.15$ & 26.41 & $-1.70$(0.06) & $-0.09$(0.06)\\
$235253.51-002850.4$ & 2115 & 5.8 & 0/0.1 & 0/0.3 & 0 & $<0.68$ & $<$24.72 & $<$$-2.42$ & $<$$-0.78$\\
$235759.42-000420.5$ & 0303110801 & 9.5 & 8/5.1 & 6/4.9 & 0 & $<1.42$ & $<$26.49 & $<$$-1.50$ & $<$$0.05$\\
\enddata
\tablenotetext{a}{The full version of this table is available in the electronic edition online.}
\tablenotetext{b}{The {\it Chandra} or {\it XMM-Newton} observation identification number.  Ten-digit numbers correspond to {\it XMM-Newton} observations, while shorter numbers correspond to {\it Chandra} observations.}
\tablenotetext{c}{The effective \mbox{X-ray} exposure reported by the CIAO or SAS toolchains for the extraction of the source region.}
\tablenotetext{d}{The observed-frame soft (0.5--2~keV) and hard (2--8~keV) counts in the source region.}
\tablenotetext{e}{A ``1'' indicates an \mbox{X-ray} detection based on the technical criteria in \S\ref{xRayBALQSOSec}; ``0'' indicates no detection.}
\tablenotetext{f}{The monochromatic luminosity is in units of erg~s$^{-1}$~Hz$^{-1}$.}
\tablenotetext{g}{The error, assumed to be dominated by the error on \mbox{X-ray} luminosity, is reported in parentheses.}
\end{deluxetable}
\clearpage

\begin{deluxetable}{lrr}
\tablecolumns{3}
\tablewidth{0pc}
\tablecaption{Tests for HiBAL \ion{C}{4} Absorption Correlations with $\Delta\alpha_{OX}$\label{bALCorrTestsTab}}
\tablehead{\colhead{} & \colhead{Only DR5 HiBALs} & \colhead{With G06 HiBALs} \\ \colhead{} & \colhead{Confidence} & \colhead{Confidence}}
\startdata
$BI_0$ vs. $\Delta\alpha_{OX}$ & 98.91 & 99.97\\
$EW$ vs. $\Delta\alpha_{OX}$ & 96.89 & 99.99\\
$v_{max}$ vs. $\Delta\alpha_{OX}$ & 99.97 & $>$99.99\\
$v_{min}$ vs. $\Delta\alpha_{OX}$ & 99.50 & 98.88\\
$f_{deep}$ vs. $\Delta\alpha_{OX}$ & 78.93 & 81.70\\
$BI_0$ vs. $L_{\nu}(2500~\mathring{A})$ & 47.74 & 24.85\\
$EW$ vs. $L_{\nu}(2500~\mathring{A})$ & 32.75 & 2.20\\
$v_{max}$ vs. $L_{\nu}(2500~\mathring{A})$ & 31.16 & 54.28\\
$v_{min}$ vs. $L_{\nu}(2500~\mathring{A})$ & 67.60 & 67.60\\
$f_{deep}$ vs. $L_{\nu}(2500~\mathring{A})$ & 6.94 & 40.55\\
\enddata
\end{deluxetable}
\clearpage

\clearpage

\clearpage

\clearpage

\begin{figure} [ht]
  \begin{center}
      \includegraphics[width=4in, angle=270]{f7.ps}
      \caption{\label{bALCatBI0DistCIVMedSpecsFig}Median (solid line) and mean (dotted line) rest-frame spectra of the absorber transmission for \ion{C}{4} BALs with $BI_0 > 0$.  Spectra with \ion{C}{4} BALs were sorted by $BI_0$ and divided into four groups of 400 objects.  Each spectrum was divided by its emission model and rebinned to a common wavelength grid.  The spectra for each group are plotted in order of increasing $BI_0$.}
   \end{center}
\end{figure}

\clearpage

\clearpage

\clearpage

\clearpage

\clearpage

\clearpage

\end{document}